\documentclass[12pt]{article}
\usepackage[english]{babel}
\usepackage{multirow} 
\usepackage{graphicx}
\usepackage{xcolor}
\usepackage{amsmath,amsfonts,amssymb,graphicx}
\usepackage{appendix}
\usepackage{natbib}
\usepackage{apalike}
\usepackage{csquotes}
\usepackage{physics}
\usepackage{float}
\usepackage[left=1in,right=1in,top=1in,bottom=1in]{geometry}
\usepackage{hyperref}
\usepackage{subcaption}
\usepackage{bbm}
\usepackage{booktabs}
\usepackage{algorithm}
\usepackage{algorithmic}
\usepackage{setspace}
\usepackage{authblk} 
\usepackage{soul}

\newtheorem{theorem}{Theorem}

\newtheorem{remark}{Remark}
\newtheorem{corollary}{Corollary}
\newtheorem{proposition}{Proposition}
\AtBeginDocument{%
  \let\mathbb\relax
  \DeclareMathAlphabet{\mathbb}{U}{msb}{m}{n}%
}
\DeclareMathOperator*{\argmax}{argmax}

\title{A Cumulative Ordered Spike-and-Slab Prior for Adaptive Dimension Selection in Joint Latent Space Models}
\author{Bin Lv}
\author{Yincai Tang}
\author{Siliang Zhang\thanks{\texttt{Corresponding author: slzhang@fem.ecnu.edu.cn}}}
\affil{Key Laboratory of Advanced Theory and Application in Statistics and Data Science-MOE, School of Statistics, East China Normal University}
\date{}

\begin{document}
\setstretch{1.26}
\maketitle

\begin{abstract}

Network models are increasingly vital in psychometrics for analyzing relational data, which are often accompanied by high-dimensional node attributes. Joint latent space models (JLSM) provide an elegant framework for integrating these data sources by assuming a shared underlying latent representation; however, a persistent methodological challenge is determining the dimension of the latent space, as existing methods typically require pre-specification or rely on computationally intensive post-hoc procedures. 
{The key innovation of this work is a cumulative ordered spike-and-slab (COSS) prior, which we  incorporate within a Bayesian joint latent space modeling framework.}
This prior enables the latent dimension to be inferred automatically and simultaneously with all model parameters. We develop an efficient Markov Chain Monte Carlo (MCMC) algorithm for posterior computation.
Theoretically, we establish that the posterior distribution concentrates on the true latent dimension and that parameter estimates achieve Hellinger consistency at a near-optimal rate that adapts to the unknown dimensionality. Through extensive simulations and {three} real-data applications, we demonstrate the method's superior performance in both dimension recovery and parameter estimation. Our work offers a principled, computationally efficient, and theoretically grounded solution for adaptive dimension selection in psychometric network models.

\end{abstract}

\section{Introduction}\label{section:1}

Network analysis has become an essential tool in psychometrics for understanding the structure of relationships that underpins psychological phenomena \citep{epskamp2017generalized,epskamp2018estimating,jeon2021mapping,borsboom2021network}.
Its increasing adoption has spurred a rapid expansion of quantitative methods \citep{sweet2015incorporating, liu2018structural, sweet2020latent,che2021network,liu2021social}, from Bayesian graphical models for ordinal data \citep{marsman2025bayesian} to specialized latent space approaches for item response data \citep{jin2019doubly, jeon2021mapping}, alongside comprehensive surveys charting the progress \citep{sweet2016social, wang2021recent}. 

A frontier in this domain is the integration of network structures with the rich, often high-dimensional, attributes of the nodes themselves \citep{mccallum2000automating, leskovec2012learning,zhang2022joint}.
The interaction between these data sources is typically reciprocal. An individual's attributes influence their network position, while their network ties can shape their behaviors, a confounding that presents a modeling challenge \citep{fratiglioni2000influence,mcpherson2001birds, shalizi2011homophily, vanderweele2011sensitivity}.
This dynamic necessitates an integrated analytical framework, as modeling both data sources jointly is crucial for achieving a robust and comprehensive understanding of the underlying latent constructs {\citep{zhang2016community, zhang2022joint, wang2022modeling,wang2023joint}}.

Latent space models (LSMs) offer powerful and interpretable tools for this purpose \citep{hoff2002latent}. LSMs represent nodes as points in a low-dimensional geometric space, where proximity is proportional to the likelihood of a connection. This framework extends naturally to a \textit{joint} LSM, where a \textit{shared} latent space generates both the network ties and the node attributes {\citep{gollini2016joint, salter2017latent, wang2022modeling, wang2023joint}}. By unifying these data sources, JLSM can borrow statistical strength across them, yielding more robust and accurate estimates of the underlying structure than could be achieved by analyzing either data source in isolation.

Despite their power, LSMs face a fundamental and persistent challenge: the selection of the latent space dimension. This choice is not a mere technicality; it has implications for both statistical validity and substantive interpretation. An under-dimensioned model risks missing key structural features of the data (underfitting), while an over-dimensioned model is prone to capturing spurious noise (overfitting) and produces latent factors that are difficult to interpret. Standard practice involves fitting a series of models with different pre-specified dimensions and selecting the ``best'' one using post-hoc criteria like cross-validation or information criteria (e.g., AIC, BIC) {\citep{hoff2007modeling, Chen02012018, zhang2022joint, wang2022modeling,wang2023joint}}. These approaches are not only computationally prohibitive but also conceptually unsatisfying, as they separate the crucial task of dimension selection from the main inferential procedure.
Moreover, the dependency between data points in network data violates the assumptions of standard cross-validation, the effective number of parameters required for information criteria can be ambiguous, and the marginal log likelihood is difficult to calculate in the LSMs context \citep{spiegelhalter2002bayesian,vaida2005conditional,li2020network}.

A fully Bayesian solution involves placing a prior over the latent dimension and inferring it alongside model parameters. However, this approach is not straightforward. Early attempts using ordered shrinkage priors \citep{bhattacharya2011sparse, durante2014nonparametric} often induced excessive shrinkage and were sensitive to hyperparameters \citep{durante2017note}.
More recent methods have advanced this goal by employing sophisticated shrinkage techniques. For instance, rotational strategies have shown promise in factor analysis \citep{rovckova2016fast}, and spike-and-slab priors have been successfully applied to related problems \citep{ishwaran2005spike, guhaniyogi2020joint, guha2021bayesian} and to network eigenmodels \citep{loyal2025spike}.
While the cumulative shrinkage process prior has also emerged as a powerful tool in Gaussian linear factor models \citep{legramanti2020bayesian, schiavon2022generalized}, its empirical performance and theoretical properties are under-explored within the complex setting of joint network models.

{In this work, we propose a cumulative ordered spike-and-slab (COSS) prior that facilitates automatic inference of the effective latent dimensionality within Bayesian joint latent space models.}
This prior induces a stochastic ordering and progressively penalizes additional dimensions, allowing superfluous dimensions to fade during estimation. 
Our approach builds upon recent advancements in adaptive Bayesian shrinkage for latent variable models \citep{rovckova2016fast, legramanti2020bayesian, schiavon2022generalized, li2025sparse, loyal2025spike}.

{Our approach differs from previous JLSMs primarily in the treatment of latent dimensionality. 
Existing approaches typically determine the number of latent dimensions through cross-validation or post-hoc analysis \citep{wang2022modeling,wang2023joint,zhang2022joint}, which requires fitting multiple candidate models and can therefore be computationally demanding, especially for large networks. 
Moreover, cross-validation assumes that held-out observations are approximately independent, an assumption that is difficult to justify for network data, where nodes and edges are inherently dependent.
Beyond these general considerations, certain existing strategies address only part of the dimension-selection problem.
Both \cite{zhang2022joint} and Chapter~3 of \cite{wang2022modeling} select the latent dimensionality based solely on the network component without incorporating information from node attributes, while the criteria adopted in \cite{wang2023joint} and Chapter~2 of \cite{wang2022modeling} are designed for binary attributes only. 
Our framework addresses these limitations through a cumulative ordered spike-and-slab (COSS) prior that determines the effective latent dimensionality automatically, while accommodating mixed data types within the exponential family.}

{Apart from the treatment of latent dimensionality, the approaches also differ in computational framework and model specification. Our inner-product formulation for the
joint likelihood aligns with that of \cite{zhang2022joint}. 
In contrast, \cite{wang2023joint} and Chapter~2 of \cite{wang2022modeling} specify a distance-based latent space model estimated via variational EM, while Chapter~3 of \cite{wang2022modeling} places a multivariate normal distribution on the concatenated latent variables for the network and node attributes. Our contribution is thus not in
the likelihood specification but in the inferential framework: a fully Bayesian approach with an adaptive MCMC sampler that jointly estimates all model parameters and the latent dimensionality within a single posterior computation.}

The key contributions of our work are four-fold: (1) 
{We introduce the cumulative ordered spike-and-slab (COSS) prior into a fully Bayesian JLSM, establishing a coherent inferential framework that integrates parameter estimation and latent dimension selection without requiring multiple model fits.
(2) We develop an efficient adaptive MCMC algorithm for posterior inference, including a reversible-jump mechanism that expands or contracts the latent dimension during sampling, making the approach practical for real-world applications. (3) We provide theoretical guarantees for the proposed framework: the posterior latent dimension concentrates around the true dimension as the sample size grows, and the model parameters achieve a near-optimal Hellinger contraction rate that adapts to the unknown dimensionality.} (4) We demonstrate through extensive simulations and {three} real-data applications that our method outperforms existing approaches in dimension recovery, parameter estimation accuracy, and computational efficiency.

The remainder of the article is structured as follows. In Section 2, we present the joint latent space model. In Section 3, we develop the novel cumulative ordered spike-and-slab prior that enables adaptive dimension selection.
Section 4 details the MCMC algorithm for posterior computation. We then establish the theoretical underpinnings of our approach in Section 5, proving key asymptotic properties of the posterior.
The method's empirical performance is evaluated through extensive simulation studies in Section 6 and {three} real-data applications in Section 7. Finally, Section 8 provides a concluding discussion.

\section{The Joint Latent Space Model}\label{section:2}
  
Let $\mathcal{G} = (\mathcal{V}, \mathcal{E})$ be an undirected network on a set of $n$ nodes $\mathcal{V} = \{1, \dots, n\}$, where $\mathcal{E}$ is the set of edges. The network structure is represented by a symmetric adjacency matrix $\mathbf{A} \in \{0, 1\}^{n \times n}$, where $A_{ii'} = 1$ if an edge exists between nodes $i$ and $i'$, and $A_{ii'} = 0$ otherwise. In addition to the network, we observe a $q$-dimensional attribute vector $\boldsymbol{y}_i \in \mathbb{R}^q$ for each node $i$. These are collected in the matrix $\mathbf{Y} = [\boldsymbol{y}_1, \dots, \boldsymbol{y}_n]^\top \in \mathbb{R}^{n \times q}$.

We consider a joint latent space model that assumes the network $\mathbf{A}$ and the node attributes $\mathbf{Y}$ are generated from a common set of unobserved latent variables. Specifically, each node $i$ is associated with a latent vector $\boldsymbol{z}_i \in \mathbb{R}^k$. The collection of these positions is denoted by the matrix $\mathbf{Z} = [\boldsymbol{z}_1, \dots, \boldsymbol{z}_n]^\top \in \mathbb{R}^{n \times k}$. 
The latent dimension $k$ is unknown and a primary object of inference in our framework. 
The joint likelihood of the observed data $(\mathbf{A}, \mathbf{Y})$ is specified through two conditionally independent components given $\mathbf{Z}$.

\paragraph{Network Model}
For the adjacency matrix $\mathbf{A}$, we assume that the edges are conditionally independent given their latent positions. Specifically, for $1 \le i < i' \le n, A_{ii'}\sim \text{Bernoulli}(P_{ii'})$. We adopt the inner product formulation \citep{ma2020universal, zhang2022joint} for its flexibility in modeling network structures:
\begin{equation}\label{eq:network_model}
    \text{logit}(P_{ii'}) = \Theta^{A}_{ii'} = \alpha_i + \alpha_{i'} + \boldsymbol{z}_i^\top \boldsymbol{z}_{i'},
\end{equation}
where $\text{logit}(x) = \log\{x/(1-x)\}$. The parameters $\boldsymbol{\alpha} = (\alpha_1, \dots, \alpha_n)^\top$ capture node-specific degree heterogeneity; a larger $\alpha_i$ increases the propensity of node $i$ to form connections. The inner product term $\boldsymbol{z}_i^\top \boldsymbol{z}_{i'}$ captures homophily and facilitates transitivity, as nodes with similar latent positions (yielding a larger inner product) are more likely to connect.

\paragraph{Node Attribute Model}
{
For the node attribute matrix $\mathbf{Y}$, we adopt a generalized latent factor model framework \citep{skrondal2004generalized,chen2020structured} and assume that the entries $Y_{ij}$ are conditionally independent given $\mathbf{Z}$. Specifically, assume the distribution of $Y_{ij}$ belongs to the exponential family in canonical form: 
\begin{equation}\label{eq:pym}
    f(Y_{ij} \mid \boldsymbol{z}_i, \boldsymbol{\beta}_j, \gamma_j, \phi_j) = \exp\left\{\frac{Y_{ij}\Theta^{Y}_{ij} - b(\Theta^{Y}_{ij})}{\phi_j} + c(Y_{ij}, \phi_j)\right\}, \quad \Theta_{ij}^Y = \gamma_j + \boldsymbol{\beta}_j^\top \boldsymbol{z}_i,
\end{equation}
where $\Theta_{ij}^Y$ denotes the natural parameter, $\phi_j$ is the dispersion parameter, and $b(\cdot)$ and $c(\cdot)$ are known functions determined by the chosen distribution. Furthermore, let $\boldsymbol{\gamma} = (\gamma_1, \dots ,\gamma_q)^ \top\in \mathbb{R}^q$ collects the intercept terms, and $\mathbf{B} = [\boldsymbol{\beta}_1,\dots,\boldsymbol\beta_q]^\top \in\mathbb{R}^{q\times k}$ be the matrix of factor loadings.
This framework encompasses many common models as special cases. When $Y_{ij}$ is continuous, taking $b(\theta) = \theta^2/2$ recovers the linear factor model: $Y_{ij} \mid \boldsymbol{z}_i, \boldsymbol{\beta}_j, \gamma_j, \sigma_j^2 \sim \mathcal{N}(\gamma_j + \boldsymbol{z}_i^\top\boldsymbol{\beta}_j, \sigma_j^2)$. When $Y_{ij}$ is binary, taking $b(\theta) = \log(1 + e^\theta)$ yields the multidimensional item response theory (MIRT) model: $Y_{ij} \mid \boldsymbol{z}_i, \boldsymbol{\beta}_j, \gamma_j \sim \text{Bernoulli}\big( \mathrm{logit}^{-1}(\gamma_j + \boldsymbol{z}_i^\top\boldsymbol{\beta}_j) \big)$.}

Given the model specifications in (1) and (2), the joint likelihood of the observed data $(\mathbf{A},\mathbf{Y})$, conditional on the latent positions $\mathbf{Z}$ and the model parameters $ {\Xi = }\{\boldsymbol\alpha,\boldsymbol{\gamma},\mathbf{B},\boldsymbol{\phi}\}$ ($\boldsymbol{\phi}$ denotes the scale parameters, if applicable), is formulated based on the assumption of conditional independence between the network and the node attributes, given the shared latent structure. The joint likelihood is the product of the likelihoods from the two components:
\begin{equation}
    \begin{aligned}
p(\mathbf{A}, \mathbf{Y}\mid \mathbf{Z}, \Xi) &= p(\mathbf{A}\mid \mathbf{Z}, \boldsymbol\alpha) p(\mathbf{Y}\mid \mathbf{Z}, \boldsymbol\gamma, \mathbf{B}, \boldsymbol\phi)\\
&= \left[ \prod_{1 \le i < i'\le n} \frac{\exp(A_{ii'}\Theta_{ii'}^A)}{1+\exp(\Theta_{ii'}^A)} \right] \times \left[ \prod_{i=1}^n \prod_{j=1}^q \exp\left\{\frac{Y_{ij}\Theta^{Y}_{ij} - b(\Theta^{Y}_{ij})}{\phi_j} + c(Y_{ij}, \phi_j)\right\} \right].
    \end{aligned}
\label{eq:joint_likelihood}
\end{equation}
This unified framework facilitates information sharing across data sources, thereby enhancing the robustness of the inference on the underlying latent structure. However, the model is subject to identifiability challenges inherent to latent variable models, where different parameter configurations can produce the identical likelihood. Addressing these invariances is crucial for meaningful interpretation.

The first one is rotational invariance. Specifically, for any $k \times k$ orthogonal matrix $\mathbf{Q}$, the transformations $\mathbf{Z} \to \mathbf{Z}\mathbf{Q}$ and $\mathbf{B} \to \mathbf{B}\mathbf{Q}$ leave both the inner products $\mathbf{Z}\mathbf{Z}^\top$ and the linear predictors $\mathbf{Z}\mathbf{B}^\top$ in \eqref{eq:joint_likelihood} unchanged. This rotational ambiguity, which includes column permutations (label switching) and sign-flips, implies that the latent coordinates and loadings are not uniquely identifiable from the likelihood alone. We address this through a cumulative ordered spike-and-slab prior detailed in Section~\ref{section:coss_prior}. This prior induces a stochastic ordering on the variances of the latent dimensions, providing a soft identifiability constraint that breaks the symmetry and encourages an interpretable orientation aligned with decreasing explained variance.
Second, to fix the origin of the latent space, we specify a zero-mean prior for each latent position, such that $\mathbb{E}(\mathbf{z}_i) = \mathbf{0}$. 

We further remark that the assumption that models (\ref{eq:network_model}) and (\ref{eq:pym}) share identical latent variables $\mathbf{Z}$ can be relaxed, as stated in \cite{zhang2022joint}. {Suppose the network $\mathbf{A}$ is modeled by $\mathbf{Z}_{A}\in \mathbb{R}^{n\times k_A}$ via Eq.~(\ref{eq:network_model}), and the attributes $\mathbf{Y}$ are modeled by $\mathbf{Z}_{Y}\in\mathbb{R}^{n\times k_Y}$ via an exponential family model with natural parameter $\Theta_{ij}^Y = \gamma_j + \boldsymbol{z}_{Y,i}^\top\boldsymbol{\beta}_j.$
If the two latent spaces are related by an approximate linear transformation $\mathbf{Z}_{Y}\approx \mathbf{Z}_{A}\boldsymbol W$ for some transformation matrix $\mathbf{W}\in \mathbb{R}^{k_A\times k_Y}$, then $\Theta_{ij}^Y$ can be approximated as 
        $
            \Theta_{ij}^Y \approx  \gamma_j + \boldsymbol{z}_{A,i}^\top(\mathbf{W}\boldsymbol{\beta}_j) = \gamma_j + \boldsymbol{z}_{A,i}^\top\boldsymbol{\beta}_j^{\prime},
        $
where $\boldsymbol{\beta}_j^{\prime} = \mathbf{W}\boldsymbol{\beta}_j$.} Thus the joint model formulation remains a valid approximation even if the latent variables are not identical, provided one can be sufficiently described by another one through a linear mapping. In this context, $\mathbf{Z}$ serves as the shared latent representation modeling both components.

We adopt a fully Bayesian approach for parameter estimation, treating the latent variables $\mathbf{Z}$ and the parameters $\boldsymbol{\alpha}$, $\boldsymbol{\gamma}$, and $\mathbf{B}$ as random variables. We specify the following conventional prior distributions to the model parameters: $\boldsymbol{\alpha} \sim \mathcal{N}(\boldsymbol{0}, \sigma_{\alpha}^2 \mathbf{I}_n)$, $\boldsymbol{\gamma} \sim \mathcal{N}(\boldsymbol{0}, \sigma_{\gamma}^2 \mathbf{I}_q)$, and $\beta_{jh} \sim \mathcal{N}(0, \sigma_{B}^2)$ for $j = 1, \dots, q$ and $h = 1, \dots, k$. For the linear factor model, we place an inverse-gamma prior on the noise variances: $\sigma_j^2 \sim \text{IG}(a_\sigma, b_\sigma)$. The prior for the latent variables $\mathbf{Z}$, which is central to our goal of inferring the dimension $k$, is detailed in the next section.

\section{Adaptive Dimension Selection via Cumulative Shrinkage}\label{section:coss_prior}

A critical challenge in latent space modeling is the selection of the latent dimension $k$. A fixed, pre-specified $k$ risks either underfitting by missing important structural features or overfitting by modeling spurious noise. To address this, we develop a Bayesian procedure that infers the effective number of latent dimensions directly from the data. 
Our approach begins by positing a model with a large, fixed number of potential latent dimensions. We then introduce a novel shrinkage prior on the columns of the latent variable matrix $\mathbf{Z}$ that adaptively deactivates superfluous dimensions.

The key to our prior is the principle of ordered importance, where subsequent dimensions contribute progressively less explanatory power. We formalize this by constructing a non-homogeneous cumulative shrinkage process, which integrates a spike-and-slab framework with an ordered prior structure. This construction induces a stochastic ordering on the variance parameters of the latent dimensions, effectively shrinking the norms of unnecessary columns in $\mathbf{Z}$ towards zero while robustly retaining the signal in the active dimensions. We first define this process as a general tool for ordered variable selection before demonstrating its specific application to latent space inference.

\subsection{The Non-Homogeneous Spike-and-Slab Cumulative Shrinkage Process}

We first introduce the general construction of the prior, which combines elements of spike-and-slab priors \citep{ishwaran2005spike} and the cumulative shrinkage process \citep{legramanti2020bayesian}.  
Let $\{\theta_h\}_{h=1}^k$ be a sequence of parameters, which will later represent the variances of the latent columns.
We assign them a spike-and-slab prior where the probability of being drawn from the spike increases with the index $h$:
\begin{equation}\label{eq:sscsp}
\theta_h\mid\pi_h \sim (1 - \pi_h)P_{\text{slab}} + \pi_h P_{\text{spike}},
\end{equation}
where $P_{\text{slab}}$ is a distribution for active (non-zero) parameters and $P_{\text{spike}}$ is a distribution tightly concentrated at zero.
The key step lies in the construction of the spike probabilities $\pi_h$. To enforce cumulative shrinkage, we require the sequence of spike probabilities to be stochastically increasing: $\pi_1 \le \pi_2 \le \cdots \le \pi_k$. We achieve this by using a \textbf{stick-breaking process}:
\begin{equation}\label{eq:stickbreaking}
    \pi_h = \sum_{l=1}^{h} \omega_l, \quad \text{where} \quad \omega_l = v_l \prod_{m=1}^{l-1}(1 - v_m).
\end{equation}
The weights $\{\omega_l\}_{l=1}^k$ are constructed by sequentially breaking a stick of unit length. At each step $l$, a proportion $v_l$ of the \textit{remaining} stick is broken off, forming the weight $\omega_l$. To induce the desired ordering, we specify the proportions $\{v_l\}$ as:
\begin{equation*}
    v_1 \sim \text{Beta}(\kappa, 1), \quad v_l \sim \text{Beta}(a, 1) \text{ for } l = 2, \dots, k-1, \quad \text{and } v_k = 1.
\end{equation*}
The weights $\omega_l$ define a probability mass function over the indices $\{1, \dots, k\}$, and $\pi_h$ is the cumulative probability up to index $h$. The stick-breaking variables $v_l \in [0,1]$ determine how the probability mass is allocated. The process is \textit{non-homogeneous} because the distribution for the first break, governed by hyperparameter $\kappa > 0$, differs from that of subsequent breaks, governed by $a > 0$. By setting a large $\kappa$, we can encourage the first weight $\omega_1$ to be small, thus making the first parameter $\theta_1$ likely to be active (drawn from the slab). Conversely, larger values of $a$ push subsequent weights towards zero, increasing the spike probability $\pi_h$ more rapidly for larger $h$.

This construction leads to an ordering property, formalized below. Let the spike distribution be a point mass at zero, $P_{\text{spike}}=\delta_0$.

\begin{proposition}[Stochastic Ordering]\label{pro:ordering}
Let $\{\theta_h\}_{h=1}^k$ follow the prior defined by \eqref{eq:sscsp}--\eqref{eq:stickbreaking} with $P_{\text{spike}}=\delta_0$. Then for any $\epsilon > 0$, $\Pr(|\theta_{h+1}| \le \epsilon) > \Pr(|\theta_h| \le \epsilon)$ for all $1 \le h < k$.
\end{proposition}

Proposition \ref{pro:ordering} confirms that the prior makes successive parameters stochastically smaller. That is, $\theta_{h+1}$ is more likely to be near zero than $\theta_h$, formalizing the notion of ordered, adaptive shrinkage.
Apart from the stochastic ordering, we further have that this prior exhibits exponentially decaying tails, provided the hyperparameters are chosen appropriately. This is crucial for establishing adaptive posterior contraction rates, and we refer readers to Lemma S3 in the Supplementary Material for details.

\subsection{Prior Specification for the Latent Positions}
We now apply this framework to specify the prior for the latent position matrix $\mathbf{Z}$. We assume that all latent positions $z_{ih}$ in a given column $h$ share a common variance parameter $\theta_h$. The full hierarchical prior is:
\begin{equation}\label{eq:priorZ}
    \begin{aligned}
        z_{ih} \mid \theta_h &\sim \mathcal{N}(0, \theta_h), \quad \text{for } i=1, \dots, n, \\
        \theta_h \mid \pi_h &\sim (1 - \pi_h)\text{IG}(a_{\theta}, b_{\theta}) + \pi_h \delta_{\theta_0},
    \end{aligned}
\end{equation}
where the spike probabilities $\{\pi_h\}$ are generated by the non-homogeneous stick-breaking process in \eqref{eq:stickbreaking}. The slab distribution is an Inverse-Gamma, a conventional choice for variances that is computationally convenient. 
The spike component, $\delta_{\theta_0}$, is a point mass at a small positive value $\theta_0 > 0$, representing a near-zero variance. 
In practice, setting $\theta_0$ to a small, fixed positive value results in a continuous shrinkage prior, which can improve the mixing and stability of MCMC samplers by avoiding true zeros while still achieving effective dimension selection \citep{ishwaran2005spike,legramanti2020bayesian}. For subsequent theoretical analysis, we can consider the limiting case where the spike variance $\theta_0 \to 0$. 
We refer to the above hierarchical construction for the latent positions as the \textbf{Cumulative Ordered Spike-and-Slab (COSS) prior}.

A key feature of the COSS prior is the marginal distribution it induces on the latent positions $z_{ih}$ after integrating out the random variances $\theta_h$. This marginal is a two-component mixture:
$$z_{ih}\sim(1-\pi_h)t_{2a_\theta}(0,b_\theta/a_\theta) + \pi_h\mathcal{N}(0,\theta_0),$$
where $t_{2a_\theta}(0,b_\theta/a_\theta)$ denotes a Student's $t$-distribution with $2a_\theta$ degrees of freedom, mean zero, and scale parameter $\sqrt{b_\theta/a_\theta}$. The slab component is heavy-tailed, which provides robustness by preventing excessive shrinkage of genuinely large latent positions. The spike component is a narrow Gaussian, which strongly pulls irrelevant positions towards zero.

This prior on $\mathbf{Z}$ directly inherits the ordering property from Proposition~\ref{pro:ordering}, which is crucial for resolving the inherent column permutation invariance of the latent space likelihood.
\begin{corollary}[Stochastic Ordering of Latent Positions]\label{cor:orderz}
    Let $z_{ih}$ be drawn from the COSS prior defined in \eqref{eq:priorZ}. For any $\epsilon > 0$, the inequality $\Pr(|z_{i,h+1}| \le \epsilon) > \Pr(|z_{ih}| \le \epsilon)$ holds for all $1 \le h < k$, provided that the spike component is more concentrated around zero than the slab component.\footnote{A sufficient condition for this is that the variance of the spike is smaller than the variance of the slab $\theta_0 < b_\theta/a_\theta$, which is ensured by hyperparameter choice.}
\end{corollary}
{The columns of $\mathbf{Z}$ are thus stochastically ordered, with higher-indexed columns progressively shrunk toward zero. 
This ordering serves as a soft identifiability constraint. In standard latent space models, the columns of $\mathbf{Z}$ are exchangeable so that any permutation yields an equivalent representation. The COSS prior breaks this permutation invariance: reordering the columns would alter their prior probabilities, making permuted configurations less probable. Unlike hard constraints such as lower-triangular factor loadings, this soft ordering does not restrict the parameter space and therefore avoids the mixing difficulties that rigid constraints can introduce. At the same time, by concentrating prior variance in the leading columns, the COSS prior ensures that substantive latent structure is captured by the first few dimensions before additional dimensions are activated, providing a natural mechanism for adaptive dimension selection.
}

\section{Posterior Computation}\label{sec:estimation}

We develop an efficient Markov Chain Monte Carlo (MCMC) algorithm for posterior inference of the joint latent space model equipped with the cumulative shrinkage prior. The algorithm is based on a Gibbs sampler that leverages data augmentation and an adaptive truncation scheme to efficiently explore the posterior distribution of both the parameters and the latent dimension.
We first describe the sampler for a fixed truncation level $k$ and then introduce an adaptive scheme that allows $k$ to be adjusted dynamically during sampling, enhancing computational efficiency.

\subsection{Posterior Computation via Gibbs Sampling}

The primary challenges in posterior sampling arise from the non-conjugacy induced by the logistic network model (\ref{eq:network_model}) and the complex structure of the COSS prior (\ref{eq:priorZ}). We address these challenges using two key strategies.

To address the non-conjugacy induced by the logistic likelihood of the network component, we employ the Pólya-Gamma data augmentation scheme of \citet{polson2013bayesian}. This approach leverages the following integral representation of the logistic function:
\begin{equation}\label{eq:polya_gamma}
    \frac{(e^\phi)^a}{(1+e^\phi)^b} = 2^{-b} e^{\kappa\phi}\int_0^\infty e^{-\frac{\tau\phi^2}{2}}p(\tau;b,0)d\tau,
\end{equation}
where $\kappa = a-b/2$ and $p(\tau;b,0)$ denotes the density for the Pólya-Gamma random variable $\tau$.
The introduction of the Pólya-Gamma latent variables induces conditional conjugacy, enabling updates of the network parameters $\boldsymbol{\alpha}$ and latent positions $\mathbf{Z}$ from normal distributions within a Gibbs sampling framework.
The sampling procedure and its detailed derivation are provided in Algorithm~\ref{alg:gibbs1} and Section A of the Supplementary Material, respectively.

Next, in order to efficiently sample the parameters of the COSS prior, we introduce discrete random variable $\rho_h\in\{1,\ldots,k\}$ such that $\Pr(\rho_h = l \mid \boldsymbol{\omega}) = \omega_l$. Thus the conditional distribution for the variance parameter $\theta_h$ in \eqref{eq:priorZ} can be rewritten as:
\begin{equation}\label{eq:rho}
    \theta_h \mid \rho_h \sim \left\{1 - \mathbbm{1}(\rho_h \leq h)\right\} \text{IG}(a_{\theta},b_{\theta}) + \mathbbm{1}(\rho_h \leq h) \delta_{\theta_0},
\end{equation}
where $\mathbbm{1}(\cdot)$ denotes the indicator function. If $\rho_h \leq h$, the dimension is assigned to the spike; otherwise, it is assigned to the slab. Moreover, let $K^* = \sum_{h=1}^{k} \mathbbm{1}(\rho_h > h)$ that counts the number of active elements in $\boldsymbol{\theta}$, which is referred to as the active dimension afterwards.

The indicators $\rho_h$ are updated by sampling from a categorical distribution whose probabilities are derived by marginalizing $\theta_h$ from the joint density of the latent positions $\mathbf{Z}_{\cdot h}$ (the $h$th column of $\mathbf{Z}$). The conditional posterior probabilities are:
\begin{equation}\label{eq:prho}
    \Pr(\rho_h=l \mid \text{rest}) \propto
    \begin{cases}
        \omega_l \mathcal{N}(\mathbf{Z}_{\cdot h}; \boldsymbol{0}, \theta_0 \mathbf{I}_n), & \text{if } l \le h, \\
        \omega_l t_{2a_{\theta}}\left(\mathbf{Z}_{\cdot h}; \boldsymbol{0}, (b_{\theta}/a_{\theta})\mathbf{I}_n\right), & \text{if } l > h.
    \end{cases}
\end{equation}
This mixture representation, involving a narrow Gaussian (spike) and a heavy-tailed Student-t distribution (slab, resulting from integrating the Inverse-Gamma prior), allows for efficient updates while enabling robust shrinkage.

The Gibbs sampler proceeds by iteratively updating the network parameters $(\boldsymbol{\alpha}, \mathbf{Z})$, the node attribute parameters (e.g., $\boldsymbol{\gamma}, \mathbf{B}, \boldsymbol{\sigma}^2$ for the linear factor model), and the parameters of the shrinkage prior $(\boldsymbol{\theta}, \boldsymbol{\rho}, \boldsymbol{v}, \boldsymbol{\omega})$. 
Algorithm~\ref{alg:gibbs1} summarizes the steps for the joint model with Gaussian node attributes. 
Detailed derivations and the algorithm for Bernoulli node variables (MIRT model), which requires an additional layer of Pólya-Gamma augmentation, are provided in the Supplementary Material.

\begin{algorithm}[!ht]
    \caption{Sampling procedure for the joint latent space model (Gaussian)}
    \label{alg:gibbs1}
    \begin{algorithmic}
    \REQUIRE Observed data $\mathbf{A}$ and $\mathbf{Y}$
    \vspace{1mm}\hrule\vspace{1mm}
    \STATE \textbf{Part 1: Update network parameters $(\boldsymbol{\alpha},\mathbf{Z})$ and augmentation data}
    \FOR{$i=1$ to $n$}
        \STATE Update augmentation data: For $i^\prime \neq i$, sample $d^{A}_{ii^\prime} \sim \text{PG}(1,\Theta_{ii^\prime}^{A})$.
        \STATE Update $\alpha_i$: Sample from its Gaussian full conditional. (Details in Supp. Mat.).
        \STATE Update $\boldsymbol{z}_i$: Sample from its Gaussian full conditional. (Details in Supp. Mat.).
    \ENDFOR
    \vspace{1mm}\hrule\vspace{1mm}
    \STATE \textbf{Part 2: Update node parameters $(\boldsymbol{\gamma},\mathbf{B}, \boldsymbol{\sigma}^2)$}
    \STATE Update $\boldsymbol{\gamma}$: Sample from its Gaussian full conditional. (Details in Supp. Mat.).
    \FOR{$j=1$ to $q$}
        \STATE Update $\boldsymbol{\beta}_j$: Sample from $\mathcal{N}(M_{\beta_j}, V_{\beta_j})$, where $V_{\beta_j} = (\sigma_{B}^{-2}\mathbf{I}_k+\sigma^{-2}_j \mathbf{Z}^\top  \mathbf{Z})^{-1}$ and $M_{\beta_j} = V_{\beta_j} \mathbf{Z}^\top \sigma^{-2}_j(\mathbf{Y}_{\cdot j}-\gamma_j\boldsymbol{1}_n)$.
        \STATE Update $\sigma_j^2$: Sample from $\text{IG}\left(a_{\sigma} + n/2, b_{\sigma} + 0.5\|\mathbf{Y}_{\cdot j} - \gamma_j\boldsymbol{1}_n - \mathbf{Z}\boldsymbol{\beta}_{j}\|^2_2\right)$.
    \ENDFOR
    \vspace{1mm}\hrule\vspace{1mm}
    \STATE \textbf{Part 3: Update shrinkage hyperparameters $(\boldsymbol{\rho}, \boldsymbol{\theta},\boldsymbol{v}, \boldsymbol{\omega})$}
    \FOR{$h=1$ to $k$}
        \STATE Sample $\rho_h$ from the categorical distribution defined by Eq.~(\ref{eq:prho}).
        \IF{$\rho_h \leq h$ (Spike)}
             \STATE Set $\theta_h = \theta_0$.
        \ELSE
             \STATE Sample $\theta_h \sim \text{IG}\left(a_{\theta} + n/2, b_{\theta} + 0.5 \|\mathbf{Z}_{\cdot h}\|^2_2\right)$.
        \ENDIF
    \ENDFOR
    \STATE Update stick-breaking variables $v_1, \dots, v_{k-1}$ from their respective Beta conditionals.
    \STATE Compute $\boldsymbol{\omega}$ using Eq.~(\ref{eq:stickbreaking}).
    \end{algorithmic}
\end{algorithm}

\subsection{Adaptive Truncation for Computational Efficiency}

The theoretical guarantees in Section~\ref{sec:theory} require the truncation level $k$ to be large, potentially growing with $n$. However, in practice, the true dimension $k_0$ is usually much smaller than $n$. Running the sampler with a large, fixed $k$ leads to significant computational inefficiency, as most of the computational effort is dedicated to updating parameters for inactive (spike) dimensions.

To maintain computational efficiency without sacrificing theoretical rigor, we adopt an adaptive Gibbs sampling strategy \citep{bhattacharya2011sparse, legramanti2020bayesian} that dynamically adjusts the truncation level $k$ during the MCMC iterations. 
At each iteration $t$ (after the burn-in period $\bar{t}$), the sampler proposes to either expand or contract the latent space with a small and diminishing probability $p(t) = \exp(\eta_0 + \eta_1 t) (\eta_0\leq 0, \eta_1<0).$ The adaptation procedure works as follows:

\begin{enumerate}
    \item \textbf{Calculate active dimension:} $K^* = \sum_{h=1}^{k}\mathbbm{1}(\rho_h > h)$.
    \item \textbf{Contraction:} If more than one columns are currently inactive (i.e., $K^* < k-1$), reduce it to $k_{new} = K^* + 1$ and discard the parameters associated with columns $h > k_{new}$.
    \item \textbf{Expansion:} If all current columns are active (i.e., $K^* = k$), add it to $k_{new} = k+1$, sampling the parameters for the new dimension from their respective priors.
\end{enumerate}
This adaptation, formalized in Algorithm~\ref{alg:gibbs_adaptive}, ensures that the sampler explores dimensions as needed without the high computational cost of a large, fixed $k$. The diminishing probability of adaptation satisfies the conditions of \cite{roberts2007coupling}, ensuring the ergodicity of the Markov chain. 

This fully Bayesian framework offers several advantages. First, the incorporation of the COSS prior allows for automatic and adaptive inference of the latent dimension $k$, integrating dimension selection within the estimation procedure rather than relying on computationally intensive post-hoc criteria as in frequentist approaches { \citep{zhang2022joint,wang2022modeling,wang2023joint}}. Second, the model structure and prior specification facilitate the construction of an efficient Gibbs sampler using data augmentation techniques, avoiding reliance on numerical approximations such as variational inference or Laplace approximations. Finally, we establish strong theoretical guarantees, demonstrating that the posterior distribution concentrates consistently on low-dimensional structures near the true model, as detailed in the next section.

\begin{algorithm}[!ht] 
    \caption{Adaptive Gibbs Sampler with Dynamic Truncation}
    \label{alg:gibbs_adaptive}
    \begin{algorithmic}[1]
    \STATE \textbf{Input:} Last truncation index $k^{(t-1)}$, total iteration $T$, burn-in $\bar{t}$, adaptation schedule parameters $\eta_0, \eta_1$.
    \vspace{2mm}
    \FOR{$t=1$ to $T$}
    \STATE Execute one cycle of Algorithm~\ref{alg:gibbs1} with current truncation $k^{(t-1)}$.
    \IF{$t > \bar{t}$ and Uniform(0,1) $< \exp(\eta_0 + \eta_1 t)$}
        \STATE Determine the number of active columns, $K^{*(t)} = \sum_{h=1}^{k^{(t-1)}} \mathbbm{1}(\rho_h^{(t)} > h)$.
        \IF{$K^{*(t)} < k^{(t-1)}$}
            \STATE Set the new truncation level $k^{(t)} = K^{*(t)} + 1$.
            \STATE Drop all inactive columns from $\mathbf{Z}$ and $\mathbf{B}$ and their associated prior parameters.
        \ELSE
            \STATE Set $k^{(t)} = k^{(t-1)} + 1$.
            \STATE Add a new column to $\mathbf{Z}$ and $\mathbf{B}$, initialized from the {slab} distribution.
        \ENDIF
    \ELSE
        \STATE Set $k^{(t)} = k^{(t-1)}$ (no adaptation).
    \ENDIF
    \ENDFOR
    \vspace{2mm}
    \STATE \textbf{Output:} One posterior sample and the current truncation level $k^{(t)}$.
    \end{algorithmic}
\end{algorithm}

\section{Theoretical Guarantees}\label{sec:theory}

This section provides theoretical justification for the proposed Bayesian joint latent space model, focusing on two key aspects: the ability of the posterior distribution to recover the true latent dimension and the accuracy of parameter estimation. We investigate the asymptotic consistency of the posterior as the number of nodes $n$ increases.

We assume the data $(\mathbf{A}, \mathbf{Y})$ are generated from the models defined in (\ref{eq:network_model}) and (\ref{eq:pym}) under a true set of parameters $\{\boldsymbol{\alpha}_0, \boldsymbol{\gamma}_0,\mathbf{B}_0,\boldsymbol{\phi}_0\}$ and a true latent dimension $k_0$. Denote the expectation under this true process by $\mathbb{E}_0$.
To establish our theoretical results, we rely on several standard regularity conditions concerning the data generating process and prior specification. These conditions align with the literature on latent space models and high-dimensional Bayesian statistics \citep[e.g., ][]{ma2020universal,zhang2022joint}. Specifically, the prior truncation level $k$ grows polynomially with $n$ (i.e., $k \asymp n^\lambda$ for $\lambda \in (0,1]$), while the true dimension $k_0$ grows much slower than $k$. This condition is easily met if $k_0$ is assumed fixed. Furthermore, we allow $q$ to be high-dimensional, growing at most linearly with $n$. Finally, we impose standard boundedness conditions on the true parameters and mild conditions on the link function $f(Y_{ij}\mid\cdot)$.
A formal statement of these assumptions and the required prior specifications is provided in the Supplementary Material.

\subsection{Consistency of Latent Dimension Selection}

Our first main result concerns the posterior distribution of the active latent dimension $K^*$.
We demonstrate that the COSS prior effectively controls the complexity of the latent space, ensuring the estimated dimension does not significantly overshoot the true dimension $k_0$.

\begin{theorem}[Posterior Concentration for the Latent Dimension]\label{th:dimension_bound}%
    Consider the joint latent space model defined in Eqs.~(\ref{eq:network_model}) and (\ref{eq:pym}) with the cumulative shrinkage prior (\ref{eq:priorZ}) placed on $\mathbf{Z}$, and standard Gaussian priors for $\boldsymbol{\alpha}, \boldsymbol{\gamma}, \mathbf{B}$. Under the regularity conditions detailed in the Supplementary Material, 
    there exists a constant $C_1>1$ such that
    \begin{equation*}
        \lim_{n\rightarrow\infty}\mathbb{E}_0^{(n)}\left\{\Pr\left(K^*>C_1k_0\mid \mathbf A, \mathbf{Y}\right)\right\} = 0.
    \end{equation*}
\end{theorem}

    \begin{remark}\label{rmk:exact}
    Establishing exact model selection consistency (i.e.,
    $\Pr(K^* = k_0 \mid \mathbf{A}, \mathbf{Y}) \to 1$) would
    require showing that the constant $C_1=1$ in Theorem~1. This is generally unattainable in Bayesian shrinkage frameworks because the exponential rates governing the prior
    penalty and the marginal likelihood involve unresolved leading
    constants whose relative magnitude cannot be determined in
    closed form, which is a difficulty shared by related results in the
    spike-and-slab literature \citep{rovckova2016fast,li2025sparse,loyal2025spike}.
    To complement the overestimation control established above, we show in Section~B.5 of the Supplementary Material that, under additional minimum signal strength conditions, the
    posterior probability of underestimation ($\Pr(K^* < k_0 \mid
    \mathbf{A}, \mathbf{Y})$) also vanishes asymptotically.
    Together, these results imply that the posterior concentrates around $k_0$.
    \end{remark}

{In practice, Theorem~\ref{th:dimension_bound} and Remark~\ref{rmk:exact} together show that the COSS prior automatically penalizes superfluous dimensions, ensuring the posterior concentrates on parsimonious models whose complexity is bounded relative to the true latent dimensionality. While the posterior mode need not equal $k_0$ exactly in finite samples, the asymptotic guarantees indicate that practitioners can safely set the initial truncation level $k$ to a conservatively large value and rely on the COSS prior to shrink away unnecessary dimensions during posterior computation.}

\subsection{Posterior Contraction Rate}

Next, we establish the consistency of the posterior distribution for the model parameters and the rate at which it contracts around the truth, based on the following definition.

{
Let $\boldsymbol{\Theta} = \{\boldsymbol{\Theta}^A, \boldsymbol{\Theta}^Y\}$ denote the collection of model parameters, where $\boldsymbol{\Theta}^{A} = \{\Theta^{A}_{ii'}\}_{1 \le i < i'\le n}$ and $\boldsymbol{\Theta}^Y = \{\Theta^Y_{ij}\}_{i \le n, j \le q}$ are the natural parameters for the network and node attributes, respectively.
Each $\Theta^{A}_{ii'}$ determines a marginal density $p_{\Theta_{ii'}^A}$ for the edge variable $A_{ii'}$, and each $\Theta^Y_{ij}$ determines a marginal density $p_{\Theta_{ij}^Y}$ for the node attribute $Y_{ij}$. To quantify the discrepancy between two parameter configurations $\boldsymbol{\Theta}_1$ and $\boldsymbol{\Theta}_2$, we define the root-mean-squared Hellinger metric:
\begin{equation}
    H_n(\boldsymbol{\Theta}_1, \boldsymbol{\Theta}_2) = \left[ \frac{2}{n(n + 2q - 1)} \left( \sum_{i=1}^{n}\sum_{i^\prime = 1}^{i-1} h^2(p_{\Theta_{1,ii'}^A}, p_{\Theta_{2,ii'}^A}) + \sum_{i=1}^n \sum_{j=1}^q h^2(p_{\Theta_{1,ij}^Y}, p_{\Theta_{2,ij}^Y}) \right) \right]^{1/2},
\end{equation} 
where $h^2(p_1, p_2) = \int (\sqrt{p_1(x)} - \sqrt{p_2(x)})^2 dx$ denotes the squared Hellinger distance between two densities. This metric averages the Hellinger distances over all $\binom{n}{2}$ edge pairs and $nq$ attribute entries, providing a single summary of how well the estimated parameters approximate the truth across both data sources.
}

\begin{theorem}[Adaptive Posterior Contraction Rate]\label{th:hellinger_consistency}
    Under the same conditions as Theorem~\ref{th:dimension_bound}, the posterior distribution contracts around the true parameters $\boldsymbol{\Theta}_0=\{\boldsymbol{\Theta}_0^A,\boldsymbol{\Theta}_0^Y\}$ at the rate $\epsilon_n = \sqrt{k_0 \log k / n}$, that is:
    \begin{equation*}
        \lim_{n\rightarrow\infty}\mathbb{E}_0^{(n)}\left[\Pr\left\{H_n(\boldsymbol\Theta,\boldsymbol\Theta_0)>C_2\epsilon_n \mid \mathbf A, \mathbf{Y}\right\}\right] = 0,
    \end{equation*}
    where $k_0,k$ are the true and pre-specified latent dimensions and $C_2 > 0$ is a constant. 
\end{theorem}

{Theorem~\ref{th:hellinger_consistency} establishes that the posterior distribution contracts around the true parameters at the rate $\epsilon_n$. This rate is nearly minimax optimal: when the true dimension $k_0$ is known in advance, the optimal rate for latent space models is $\sqrt{k_0/n}$ \citep{ma2020universal,xie2020optimal,zhang2022joint}.
The extra logarithmic factor $\sqrt{\log k}$ reflects the statistical price paid for treating the dimension as unknown and learning it from the data \citep{loyal2025spike}. 
Furthermore, the rate adapts to the unknown true dimension $k_0$ without requiring its specification, so the model automatically achieves high estimation accuracy commensurate with the true latent complexity. Since the truncation level $k$ enters only through the logarithmic factor, setting $k$ conservatively large has a negligible effect on the contraction rate, further confirming the robustness of the COSS prior in practice.}

This general guarantee of distributional convergence can be translated into more interpretable metrics for specific models. For instance, for binary node attributes (the MIRT model), we can show that the estimated probabilities converge to the true probabilities in terms of the Frobenius norm, see Corollary S1 in the Supplementary Material.

\section{Simulations}\label{sec:simulation}

We conducted simulation studies to evaluate the performance of the proposed joint latent space model equipped with the COSS prior. The simulations aimed to (1) assess the improvement in parameter recovery achieved by the JLSM compared to modeling the network and node attributes separately, and (2) benchmark the accuracy of the proposed dimension selection approach against standard model selection techniques. We investigated these aspects under varying sample sizes and network densities.

\subsection{Study 1}\label{sec:simulation1}

\subsubsection{Design and evaluation criteria}
In this study, we evaluated the proposed method for estimating accuracy of parameter estimation, latent dimension recovery, and model selection under different sample sizes and model specifications.
Data were generated from the joint latent space model under conditions reflecting different sample sizes and numbers of node attributes, with $(n, q)$ pairs set to $(50, 10)$, $(100, 20)$, $(150, 30)$, and $(300, 60)$. The true latent dimension was fixed at $k_0=3$. 
For each of the 100 replications per condition, model parameters were drawn as follows: { node heterogeneity parameters were drawn from $\alpha_i \sim U[-0.5,0.5]$}
; intercept parameters were drawn from $\gamma_j \sim \mathcal{N}(0,1)$; latent positions were generated independently from $z_{ih} \sim \mathcal{N}(0,1)$; and a sparse loading matrix $\mathbf{B}_0 \in \mathbb{R}^{q \times k_0}$ of simple structure, where each row contained exactly one non-zero element drawn from $U[0.25, 1.25]$. The network adjacency matrix $\mathbf{A}$ was generated from Eq.~(\ref{eq:network_model}), and the node attribute matrix $\mathbf{Y}$ was generated for two scenarios: continuous (Gaussian) and binary (Bernoulli) variables.

We compared the performance of our proposed Joint LSM to two specialized variants: a Network-Only LSM using only $\mathbf{A}$, and a Node-Only model (factor analysis or MIRT) using only $\mathbf{Y}$. The COSS prior was implemented in all three models. 

For implementation, the truncation level was initialized at $k=8$. Hyperparameters were set as $a_\theta = b_\theta = 3$ and $a=8$. The spike variance $\theta_0$ was $0.1$ for the Network LSM and JLSM, and $0.05$ for the Node LVM. Weakly informative priors were used for other parameters ($\sigma_\alpha = \sigma_\gamma = 100$, $\sigma_B = 1$). The MCMC algorithms were run for 15,000 total iterations, with 10,000 discarded as burn-in, and the remainder thinned every 5 iterations, yielding 1,000 posterior samples. Adaptive Gibbs sampling was initiated after $\bar{t}=500$ iterations, with adaptation parameters $(\eta_0, \eta_1) = (-1, -5 \times 10^{-4})$.

We evaluated performance using several normalized metrics calculated using posterior means. To assess the recovery of the latent structure, we used the normalized Frobenius error of the inner product matrix, $\Delta_Z = \|\hat{\mathbf Z}\hat{\mathbf Z}^\top - \mathbf Z_0\mathbf Z^{\top}_0\|_F/n$. Similarly, loading matrix estimation was evaluated by $\Delta_B = \|\hat{\mathbf B} \hat{\mathbf B}^\top - \mathbf B_0\mathbf B_0^\top\|_F/q$. The use of inner products accounts for the rotational invariance of the latent space. Parameter estimation accuracy for node heterogeneity and intercepts were measured by their normalized $L_2$ errors, $\Delta_{{\alpha}} = \|\hat{\boldsymbol{\alpha}} - \boldsymbol{\alpha}_0\|_2/\sqrt{n}$ and $\Delta_{{\gamma}} = \|\hat{\boldsymbol{\gamma}} - \boldsymbol{\gamma}_0\|_2/\sqrt{q}$, respectively.
For dimension recovery, the latent dimension was estimated using the posterior mode, 
$$\hat{K} = \argmax_{l}\hat{\Pr}(K^* = l\mid \mathbf{A},\mathbf{Y}).$$ 
We report the accuracy (Acc; proportion of replications where $\hat{K}=k_0$) and the mean absolute bias (MAB), calculated only over replications where $\hat{K} \neq k_0$.

\subsubsection{Results}
The results of parameter recovery are summarized in Table~\ref{tab:S1}, demonstrating a clear advantage for the joint modeling approach as well as the validity of the proposed dimension selection approach. 
The JLSM consistently outperforms both the Network LSM and the Node LVM across all configurations and metrics. This demonstrates the statistical efficiency gained by integrating both data sources, allowing the model to borrow strength and yield more accurate estimates of the shared latent structure and associated parameters. As expected, estimation accuracy improves as the sample size increases, and performance is generally better with Gaussian attributes than Bernoulli attributes, reflecting the higher information content of continuous data.

\begin{figure}[ht]
    \centering
    \includegraphics[width=0.99\linewidth]{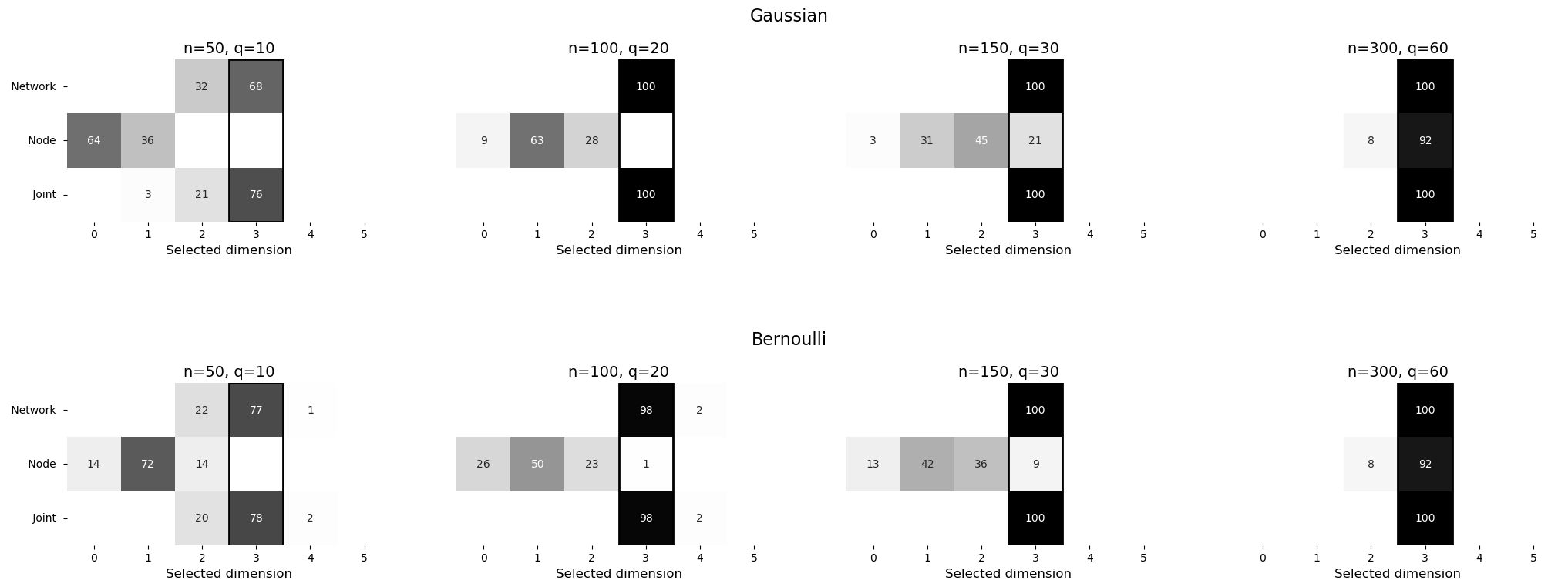}
    \caption{Frequency of dimension selection across 100 simulations for the three models using the COSS prior. The column outlined in black indicates the true value $k_0 = 3$. Darker cells indicate higher percentages.}
    \label{fig:D1}
\end{figure}

Figure~\ref{fig:D1} illustrates the frequency of dimension selection across the three models. As the sample size increases, the accuracy of all models improves significantly, showing the consistency of the latent dimension selection. Moreover, the Joint LSM identified the true latent dimension with higher accuracy than the specialized models, especially in smaller samples.

We further compared the dimension selection performance of the COSS prior against standard post-hoc criteria. These benchmarks involved fitting a standard JLSM with independent $\mathcal{N}(0,1)$ priors on $z_{ih}$ across a range of fixed dimensions and selecting the optimal dimension using information criteria (AIC, BIC, DIC, WAIC) and K-fold cross-validation (K-fold CV) with $K=5$ folds, where we selected the dimension maximizing the average held-out log-likelihood. We also considered the ``K-fold CV 1SE'' criterion to favor parsimony by selecting the smallest model within one standard deviation of the best model. See the Supplementary Material for more details on the criteria calculation.

\begin{figure}[ht]
    \centering
    \includegraphics[width=0.99\linewidth]{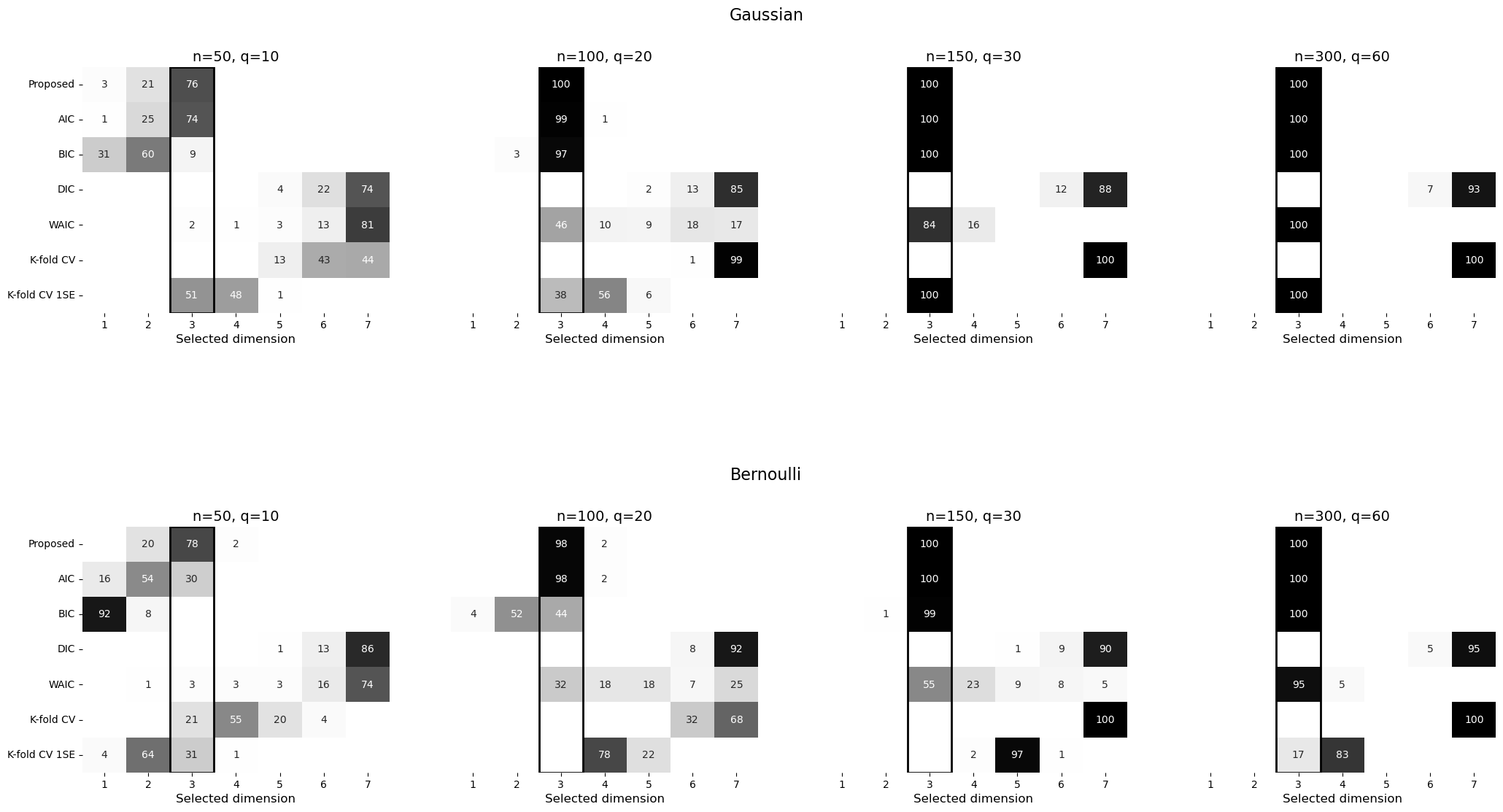}
    \caption{Comparison of dimension selection methods applied to the JLSM across 100 simulations. The column outlined in black indicates the true value $k_0 = 3$. Darker cells indicate higher percentages.}
    \label{fig:D2}
\end{figure}

Figure~\ref{fig:D2} compares the dimension selection performance of the COSS prior (labeled ``Proposed'') against competing methods. The COSS prior was highly accurate and robust across all conditions. 
In contrast, traditional criteria exhibit notable biases: DIC and K-fold CV consistently overestimate the latent dimension. WAIC also tends to overestimate in small samples. Conversely, BIC often underestimates the dimension when the sample size is small ($n=50$) but improves as $n$ increases. AIC performs reasonably well, though it is slightly outperformed by the proposed method in small sample scenarios. Overall, the cumulative shrinkage prior provides robust and accurate dimension selection. Crucially, it achieves this within a single MCMC run, making it computationally more efficient than methods requiring fitting multiple models, and it naturally incorporates uncertainty about the latent dimension into the posterior inference.

\begin{table}[ht]
    \centering
    \small
    \caption{Parameter recovery simulation results for true $k_0=3$. Values are means over 100 replications, with standard errors in parentheses. The values in the parentheses of the `Acc' column are the mean absolute bias (MAB) calculated only when $\hat{K} \neq k_0$.}
    \begin{tabular}{cclccccc}
    \toprule
    $\mathbf{Y}$ & $(n,q)$ & Model & $\Delta_\alpha$ & $\Delta_\gamma$ & $\Delta_B$ & $\Delta_Z$ & Acc (MAB) \\
    \midrule
    \multirow{15}{*}{Gaussian}
    & \multirow{3}{*}{(50,10)}
    & Network   & $0.470 (0.089)$ & --               & --               & $1.067 (0.116)$ & $0.680 (1.000)$ \\
    &                        & Node LVM  & --              & $0.188 (0.047)$ & $0.335 (0.060)$ & $1.676 (0.190)$ & $0.000 (2.640)$ \\
    &                        & JLSM & $0.457 (0.053)$ & $ 0.193 (0.056)$ & $0.213 (0.050)$ & $0.954 (0.113)$ & $0.760 (1.125)$ \\
    \cline{2-8}
    & \multirow{3}{*}{(100,20)}
    & Network   & $0.318 (0.050)$ & --               & --               & $0.758 (0.072)$ & $1.000 (0.000)$ \\
    &                        & Node LVM  & --              & $0.123 (0.027)$ & $0.266 (0.043)$ & $1.443 (0.184)$ & $0.000 (2.000)$ \\
    &                        & JLSM & $0.310 (0.046)$ & $0.128 (0.029)$ & $0.135 (0.021)$ & $0.655 (0.067)$ & $1.000 (0.000)$ \\
    \cline{2-8}
    & \multirow{3}{*}{(150,30)}
    & Network   & $0.261 (0.044)$ & --               & --               & $0.619 (0.051)$ & $1.000 (0.000)$ \\
    &                        & Node LVM  & --              & $0.101 (0.019)$ & $0.312 (0.039)$ & $1.223 (0.227)$ & $0.210 (1.468)$ \\
    &                        & JLSM & $0.256 (0.043)$ & $0.107 (0.024)$ & $0.102 (0.011)$ & $0.535 (0.045)$ & $1.000 (0.000)$ \\
    \cline{2-8}
    & \multirow{3}{*}{(300,60)}
    & Network   & $0.187 (0.031)$ & --               & --               & $0.445 (0.032)$ & $1.000 (0.000)$ \\
    &                        & Node LVM  & --              & $0.075 (0.013)$ & $0.086 (0.022)$ & $0.715 (0.100)$ & $0.920 (1.000)$ \\
    &                        & JLSM & $0.183 (0.030)$ & $0.078 (0.017)$ & $0.068 (0.005)$ & $0.379 (0.032)$ & $1.000 (0.000)$ \\
    \midrule
    \multirow{12}{*}{Bernoulli}
    & \multirow{3}{*}{(50,10)}
    & Network   & $0.453 (0.081)$ & --               & --               & $1.042 (0.107)$ & $0.770 (1.000)$ \\
    &                        & Node LVM  & --              & $0.456 (0.377)$ & $0.514 (0.220)$ & $1.718 (0.158)$ & $0.000 (2.080)$ \\
    &                        & JLSM & $0.451 (0.081)$ & $ 0.428 (0.349)$ & $0.319 (0.064)$ & $1.025 (0.112)$ & $0.780 (1.000)$ \\
    \cline{2-8}
    & \multirow{3}{*}{(100,20)}
    & Network   & $0.322 (0.053)$ & --               & --               & $0.759 (0.062)$ & $0.980 (1.000)$ \\
    &                        & Node LVM  & --              & $ 0.286 (0.056)$ & $0.375 (0.076)$ & $1.689 (0.136)$ & $0.001 (2.030)$ \\
    &                        & JLSM & $0.322 (0.052)$ & $0.280 (0.057)$ & $0.249 (0.033)$ & $0.741 (0.064)$ & $0.980 (1.000)$ \\
    \cline{2-8}
    & \multirow{3}{*}{(150,30)}
    & Network   & $0.265 (0.041)$ & --               & --               & $0.628 (0.046)$ & $1.000 (0.000)$ \\
    &                        & Node LVM  & --              & $0.225 (0.041)$ & $0.312 (0.039)$ & $1.600 (0.114)$ & $0.090 (1.747)$ \\
    &                        & JLSM & $0.264 (0.040)$ & $0.216 (0.035)$ & $0.215 (0.023)$ & $0.612 (0.046)$ & $1.000 (0.000)$ \\
    \cline{2-8}
    & \multirow{3}{*}{(300,60)}
    & Network   & $0.180 (0.025)$ & --               & --               & $0.440 (0.027)$ & $1.000 (0.000)$ \\
    &                        & Node LVM  & --              & $0.168 (0.020)$ & $0.198 (0.016)$ & $1.301 (0.071)$ & $0.920 (1.000)$ \\
    &                        & JLSM & $0.178 (0.025)$ & $0.151 (0.018)$ & $0.156 (0.015)$ & $0.426 (0.027)$ & $1.000 (0.000)$ \\
    \bottomrule
    \end{tabular}
    \label{tab:S1}
\end{table}

\subsection{Study 2}%
In the second study, we investigated the impact of network density on estimation accuracy, focusing on the ability of the joint model to leverage node attributes when network information is scarce. We fixed the sample size to $(n,q) = (100,20)$ and $k_0=3$. We varied the network density by adjusting the distribution of the node heterogeneity parameters $\boldsymbol{\alpha}$. By shifting the distribution of $\alpha_i$ from $U[-3, -1]$ (sparse) towards $U[-0.375, -0.125]$ (dense), we generated networks with densities ranging from 0.07 to 0.42. All other data-generating parameters and MCMC settings were identical to those in Study 1. We compared the performance of the JLSM and the Network LSM.

\begin{figure}[ht]
    \centering
    \includegraphics[width=0.8\linewidth]{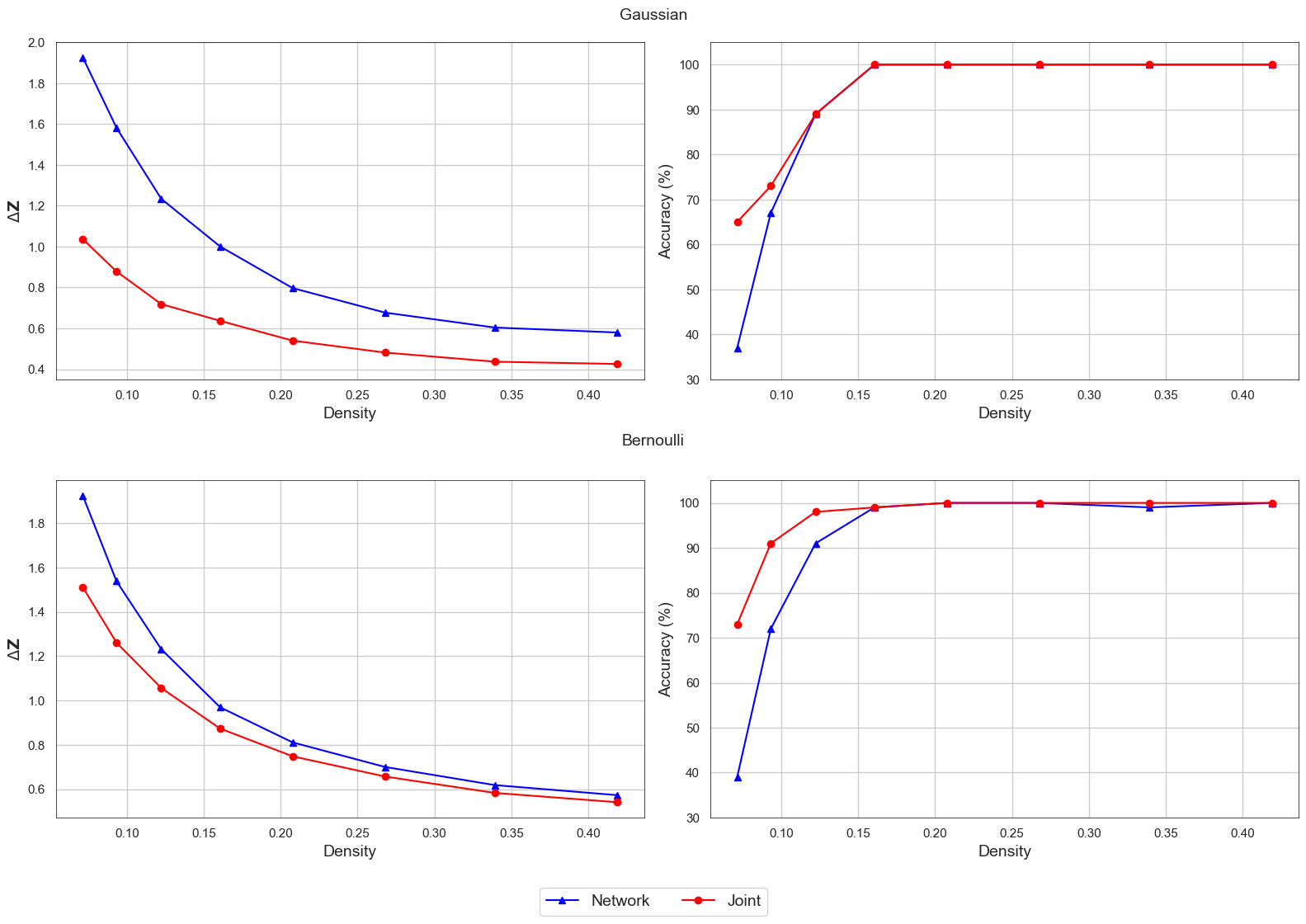}
    \caption{Latent structure estimation error ($\Delta_Z$, left panels) and dimension recovery accuracy (right panels) versus network density for Gaussian (top) and Bernoulli (bottom) attributes.}
    \label{fig:density}
\end{figure}

Figure~\ref{fig:density} illustrates the latent structure recovery error ($\Delta_Z$) and the dimension recovery accuracy (Acc) as a function of network density. 
The performance of the Network-Only LSM was highly dependent on network density. As the network became sparser (moving from right to left on the x-axis), the performance of the Network LSM deteriorates significantly across both metrics. 
In particular, in highly sparse regimes (e.g., density below 0.10), the Network LSM exhibits substantial estimation error, and its dimension recovery accuracy drops sharply to below 40\%.
This is expected, as sparse networks offer limited information for identifying the underlying latent space structure and its dimensionality.

In contrast, the Joint LSM performed consistently well across all levels of network density. 
Notably, the JLSM maintains significantly higher accuracy in dimension selection (over 65\% for Gaussian and 70\% for Bernoulli) even at the lowest densities tested. 
This stability arises because the model effectively borrows information from the node attributes when the network signal is weak. 
The attributes provide the necessary data to identify the true latent dimension and estimate the model parameters accurately. 
This finding highlights a key practical advantage of our approach: it ensures reliable adaptive dimension selection even when one data source is uninformative, a common challenge in applied research.

\section{Real Data Analysis}\label{section:7}
{We demonstrate the proposed COSS prior framework through three empirical applications, each highlighting a different aspect of the methodology.
The first analyzes the French Financial Elites network \citep{kadushin1995friendship}, previously studied by \cite{wang2023joint}, to provide a direct comparison with existing JLSM approaches on a common benchmark; here the COSS prior automatically recovers the latent dimensionality while revealing interpretable social structure.
The second uses Facebook ego-networks from the Stanford SNAP repository \citep{leskovec2012learning} to examines whether the predictive benefit of joint modeling is preserved when the latent dimension is selected adaptively rather than fixed a priori; we evaluate this through out-of-sample imputation of missing node attributes.
The third applies the model to the Teenage Friends and Lifestyle Study \citep{pearson2006homophily}, a friendship network among secondary school students in Glasgow, providing independent empirical evidence on a dataset not previously analyzed in the JLSM literature.
} 

\subsection{French Financial Elites Network}

Our first application analyzes the French Financial Elites dataset, originally collected by \cite{kadushin1995friendship} to study social connections within top financial circles during the final years of France's Socialist government.
We obtained the processed version of this dataset from the \texttt{jlsm} R package \citep{Wang2021jlsm}, which has also served as the empirical basis in \cite{wang2022modeling} and \cite{wang2023joint}. We use it here to provide a direct, equitable comparison with these existing JLSM approaches on a common benchmark.
The dataset comprises a friendship network among 28 prominent financial elites (network density 0.168) together with 13 binary node attributes covering education, career, social class, and political alignment. Notable among these attributes are affiliation with the \'Ecole Nationale d'Administration (ENA) and inclusion in the Social Register, which are primary markers of elite status in the French context. A full description of the variables is provided in \cite{kadushin1995friendship} and \cite{wang2023joint}.

To determine the optimal number of latent dimensions, we compared the performance of our proposed COSS prior against standard model selection techniques by fitting the JLSM 100 times under various specifications. First, we employed our proposed approach, using the COSS prior with a truncation level of $k=5$, allowing the effective dimension to be automatically inferred from $\{1, 2, 3, 4\}$. Second, we fitted the JLSM using a standard normal prior for the latent positions and selected the dimension based on AIC, BIC, 5-fold CV, and CV 1-SE rule (5-fold CV 1-SE). For the CV, the dimension was estimated by maximizing the average held-out conditional likelihood.

\begin{figure}[ht]
\centering
\includegraphics[width=0.45\linewidth]{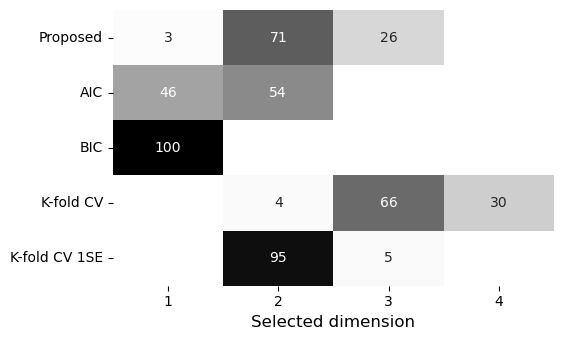}
\caption{Estimated dimension for the French financial elites dataset. Darker cells indicate higher selection percentages.}
\label{fig:D28}
\end{figure}

Figure~\ref{fig:D28} summarizes the results. BIC consistently selects a one-dimensional space, likely due to its tendency to favor overly simple models in small samples. While CV frequently overestimates the dimension, favoring three dimensions. The proposed COSS prior, AIC, and K-fold CV 1-SE predominantly select two dimensions. Given the consistency across these methods and the known biases of BIC and standard CV in this context, we conclude that a two-dimensional latent space is most appropriate for this dataset.

\begin{figure}[ht]
    \centering
    \includegraphics[width=0.9\linewidth]{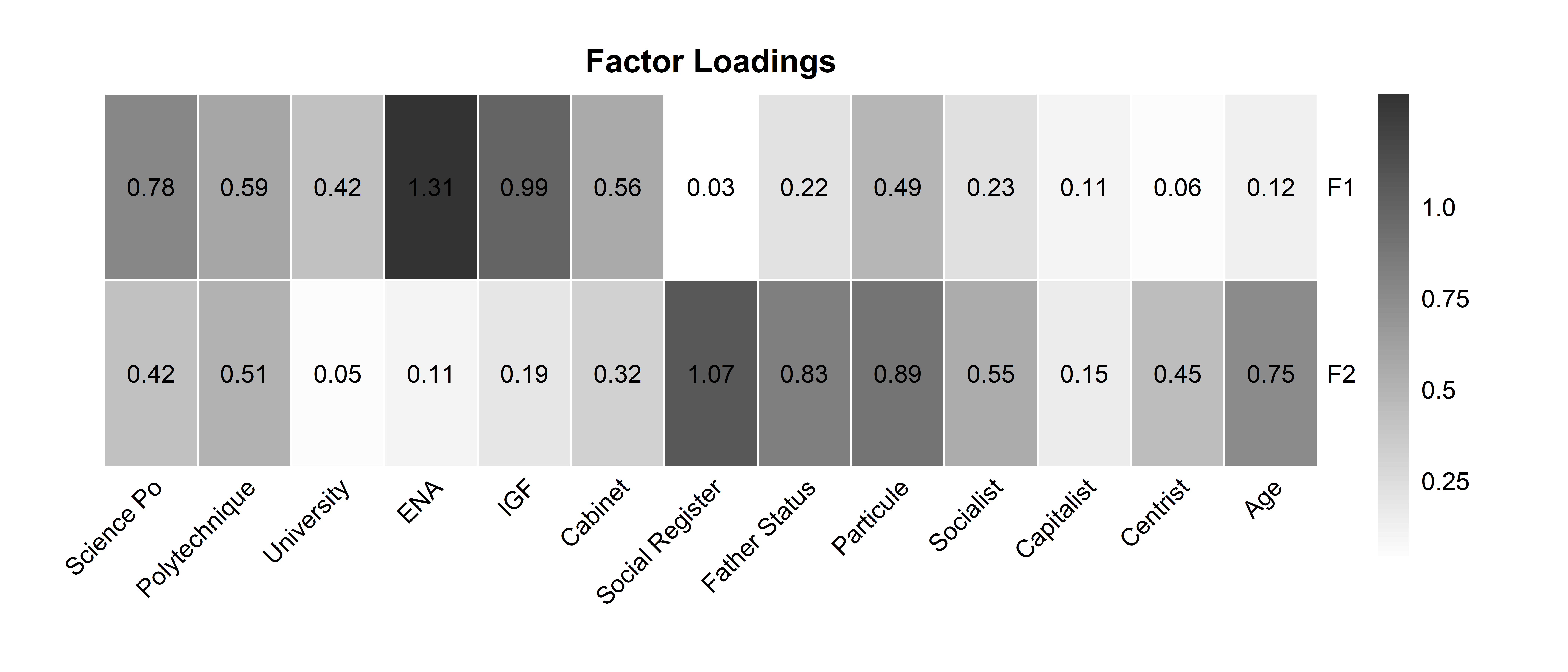}
    \caption{Heatmap of the estimated loading matrix $\mathbf{B}$ in absolute value under Geomin rotation for the French financial elites dataset.}
    \label{fig:loading}
\end{figure}

Next, we interpret the two-dimensional latent space by examining the factor loadings and the visualization of the latent positions. Figure~\ref{fig:loading} presents a heatmap of the absolute values of the estimated loading matrix after a Geomin rotation \citep{yates1987multivariate}. The loadings reveal a clear structure. The first dimension (F1) is strongly associated with attributes related to elite education and high-ranking career positions (ENA, IGF, Science Po, Polytechnique, Cabinet). The second dimension (F2) is characterized by indicators of established social class (Social Register, Father Status, particule) and older age. 
This aligns with the social divisions within the French elite described by \cite{kadushin1995friendship}.

\begin{figure}[ht]
    \centering
    \includegraphics[width=0.9\linewidth]{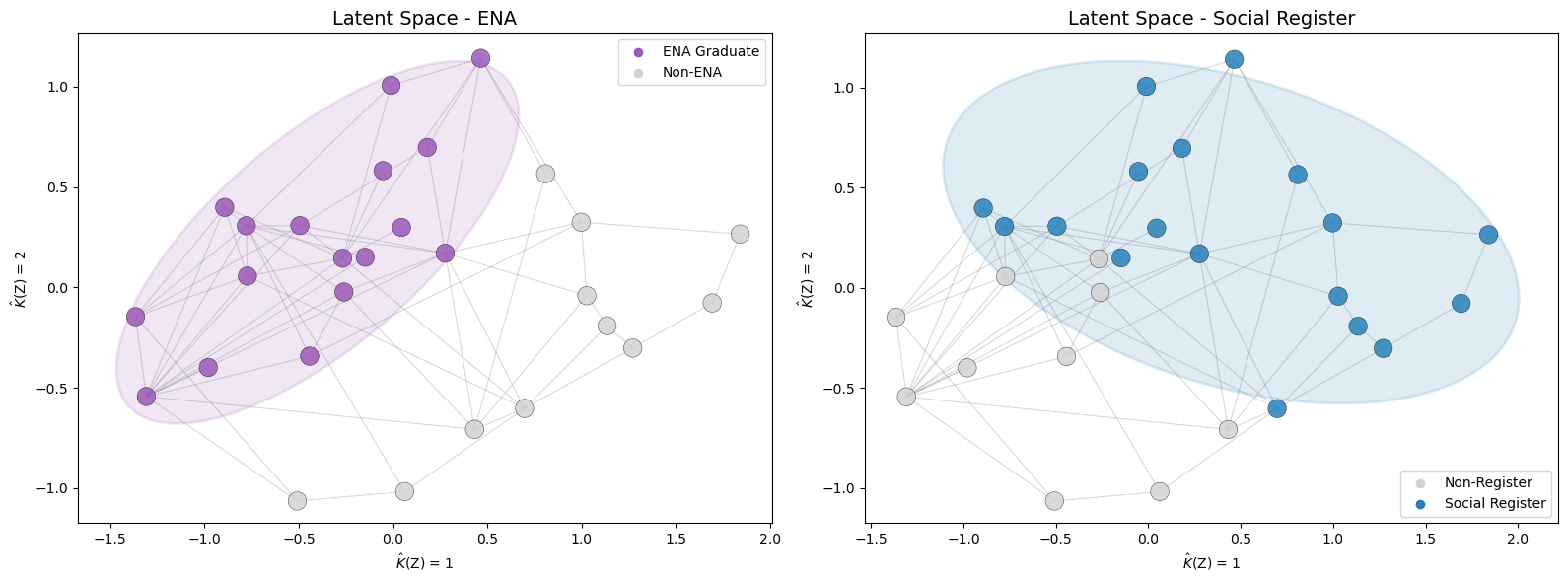}
    \caption{Latent Space Positioning of French Elites by ENA (Left Panel) and Social Register (Right Panel). Ellipses highlight the clustering of the respective groups.}
    \label{fig:f28}
\end{figure}

To further explore the estimated latent space, we visualize it in Figure~\ref{fig:f28}. In the inner product latent space model, nodes with similar latent positions (i.e., those clustered together in the visualization) are more likely to form connections. We highlight ENA and Social Register, the attributes with the strongest loadings on the first and second dimensions, respectively. 
The left panel demonstrates a clear separation based on ENA affiliation. ENA graduates (purple nodes) form a cohesive, densely connected cluster (purple ellipse), distinct from the more dispersed non-ENA individuals (gray nodes). This reinforces the interpretation of the first dimension as representing educational attainment and the associated strong social ties formed in these institutions. 
The right panel highlights social class. Individuals listed in the Social Register (blue nodes) also cluster tightly (blue ellipse), indicating a strong tendency for connection among the established upper class. 
Our joint modeling approach provides quantitative support for the qualitative divisions previously identified, illustrating a social structure segmented by education and class.

\subsection{Facebook Social Network Data}\label{sec:data2}

In this section, we evaluate our model's performance on a missing data imputation task using data from Facebook ego network \citep{zhang2022joint}\footnote{The dataset is publicly available at \url{https://snap.stanford.edu/data/}}. 
The dataset comprises 10 ego-networks, where each network represents the connections among the friends of a specific user. The data include anonymized binary attributes for each user (e.g., an original attribute like \enquote{political = Democratic Party} is transformed into \enquote{political = anonymized feature 1}). We analyzed 8 of the 10 networks, excluding two that had too few nodes or highly sparse attribute vectors. We also removed attributes with extremely low or high prevalence within each network to ensure stable estimation as in \cite{zhang2022joint}. The goal is to impute missing attributes for a user based on their available profile information and their social connections. 

In our experiment, we randomly masked 5\% of the entries in the node attribute matrix $\mathbf{Y}$. We then imputed these missing values using both the network-only latent space model and our proposed joint latent space model. For the joint model, missing values were re-imputed as unknown parameters within each iteration of the Gibbs sampler. For the network-only model, we first estimated the latent positions $\mathbf{Z}_{\text{net}}$ and then used these to predict the missing attributes by regressing $\mathbf{Y}$ on $\mathbf{Z}_{\text{net}}$. As a benchmark, we also applied the Multiple Imputation by Chained Equations (MICE) method \citep{azur2011multiple}, a standard technique that utilizes only the node attributes.

Table~\ref{tab:real_res} summarizes the prediction performance in terms of mean AUROC values and estimated latent dimensions $\hat{K}$ over 30 replications. The results show that both network-based models outperform MICE across all networks, demonstrating that network structure contains predictive information about node attributes that traditional attribute-only methods fail to capture. 
Furthermore, the joint latent space model consistently achieves the highest AUROC scores, indicating that there are meaningful associations between network features and node attributes, and that integrating them within a unified framework enhances predictive accuracy.
As our simulation studies indicated, this ability to flexibly model the dimensionality does not detract from, but rather enhances, the model's performance.

Finally, we find that the estimated latent dimension $\hat{K}$ is higher in the joint model than in the network-only model. This suggests that the latent structures underlying the network and node attributes are not fully aligned. The joint model accommodates this misalignment by adding additional dimensions that primarily capture the node attribute information. As discussed in Section 2 and \cite{zhang2022joint}, this expanded latent space enables the model to represent both data sources effectively without sacrificing performance.

\begin{table}[htbp]
    \centering
    \caption{Node variable imputation results for the Facebook dataset. Values reported are the mean AUROC and estimated dimensions $\hat{K}$ across 30 replications.}
    \begin{tabular}{lcccccccc}
    \toprule
    \multirow{2}{*}{Dataset} &\multirow{2}{*}{$n$} & \multirow{2}{*}{$q$} & \multirow{2}{*}{Density} & \multicolumn{2}{c}{$\hat{K}$} & \multicolumn{3}{c}{AUROC} \\
    \cmidrule(lr){5-6} \cmidrule(lr){7-9}
    ~ & ~ & ~ & ~ & {Network}& {Joint}& {Network}& {Joint} & {MICE}\\
    \midrule
    Circle 1 & 347 & 56 & 0.042 & 4.68 & 5.43 & 0.845 & 0.904 & 0.801\\
    Circle 2 & 547 & 58 & 0.030 & 5.22 & 6.93 & 0.857 & 0.921 & 0.853\\
    Circle 3 & 227 & 87 & 0.124 & 6.00 & 7.53 & 0.820 & 0.875 & 0.716\\
    Circle 4 & 159 & 55 & 0.134 & 3.00 & 6.20 & 0.804 & 0.866 & 0.740\\
    Circle 5 & 170 & 37 & 0.155 & 5.13 & 6.70 & 0.791 & 0.869 & 0.779\\
    Circle 6 & 792 & 100& 0.049 & 8.53 & 9.57 & 0.872 & 0.931 & 0.853 \\
    Circle 7 & 755 & 66 & 0.105 & 8.01 & 9.80 & 0.868 & 0.918 & 0.874 \\
    Circle 8 & 1045& 153& 0.049 & 11.69& 12.08& 0.877 & 0.914 & 0.844 \\
    \bottomrule
    \end{tabular}
    \label{tab:real_res}
\end{table}

\subsection{Teenage Friends and Lifestyle Study}\label{section:Glasgow_data}
Our third application analyzes data from the Teenage Friends and
Lifestyle Study \citep{bush1997role,michell1996peer,michell2000smoke,pearson2003drifting} \footnote{{The dataset is publicly available  at \url{https://www.stats.ox.ac.uk/~snijders/siena/Glasgow_data.htm}.}}, a longitudinal study that tracked a cohort of Glasgow secondary school students over two years (1995--1997) to examine social influences on adolescent health behaviors, particularly smoking.
We focus on the older cohort at the first measurement point and restrict the sample to the $n = 129$ students present at all three waves, consistent with prior analyses of this dataset \citep{pearson2006homophily}.

The friendship network $\mathbf{A}\in\{0,1\}^{n\times n}$ is constructed from nominations of up to six friends per student. The original data distinguished ``best friends'' from ``just friends''; we collapse both categories into a single binary indicator resulting in an undirected graph, so that $A_{ij} = 1$ if either student nominated the other. The network density is 0.039. The node attribute matrix $\mathbf{Y} \in \mathbb{R}^{n \times q}$ comprises $q = 34$ mixed-type variables capturing three aspects of adolescent lifestyle: substance use (alcohol, tobacco, and cannabis, measured on ordinal scales of 1--5, 1--3, and 1--4, respectively), leisure activities (15 items rated on a 4-point frequency scale), and musical tastes (16 binary genre preferences). Details on the treatment of ordinal and mixed-type node variables within our framework are provided in Section~C.2 and Section~C.3 of the Supplementary Material.

We applied the COSS prior with truncation level $k = 11$ and compared its dimension selection against a standard JLSM fitted with a normal prior, where the dimensionality was chosen by AIC, BIC, DIC, WAIC, 5-fold CV, and the 5-fold CV one-standard-error (1-SE) rule. For the cross-validation methods, the optimal dimension was selected by maximizing the average held-out conditional likelihood. Missing values in $\mathbf{Y}$ were handled according to the procedure described in Section~\ref{sec:data2}.

\begin{figure}[ht]
\centering
\includegraphics[width=0.6\linewidth]{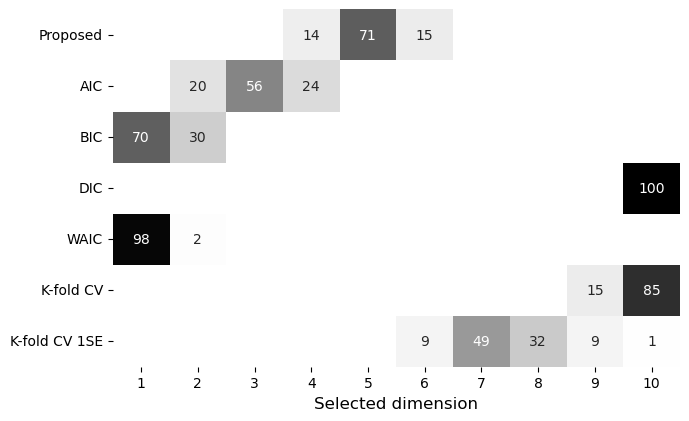}
\caption{Estimated dimension for the Teenage Friends and Lifestyle Study dataset. Darker cells indicate higher selection percentages.}
\label{fig:data3SD}
\end{figure}

Figure~\ref{fig:data3SD} summarizes the results of the dimension selection. BIC and WAIC both select a one-dimensional solution, likely reflecting their tendency to favor parsimonious models in small samples. At the other extreme, CV and DIC consistently select ten dimensions, suggesting overfitting. The COSS prior predominantly selects five dimensions, while AIC and the CV 1-SE rule select three and seven, respectively.

To assess these competing selections, we compare goodness of fit following the predictive approach of \cite{hunter2008goodness}, evaluating the adaptive selection pipeline as a whole. The COSS prior infers the dimension automatically, whereas the competing criteria require it to be fixed in advance. Because the standard criteria diverge sharply here (from $\hat{K}=1$ to $10$), we add transitivity as a structural measure to identify the dimension that best reflects the observed network, since the tendency toward triadic closure is a hallmark of friendship networks shaped by homophily. Table~\ref{tab:goodness} reports the AUROC for edge prediction and the transitivity error $\Delta_{\text{transitivity}}$, the mean squared error between the observed transitivity and that of networks simulated from the fitted model over 100 replications. A lower value indicates more accurate recovery of the observed structure. To separate the effect of the prior from that of the dimension, the table also includes a normal-prior baseline fitted at the dimension selected by COSS.

\begin{table}[htbp]
\centering
\caption{
    Goodness-of-fit comparison for the adjacency matrix $\mathbf{A}$: COSS prior versus normal-prior JLSM at $\hat{K}=3$ (selected by AIC) and $\hat{K}=7$ (selected by 5-fold CV 1-SE), as well as at the COSS-selected dimension ($\hat{K}_{\text{COSS}}$).
    $\Delta_{\text{transitivity}}$ denotes the mean squared error of the transitivity.
    Values are means over 100 replications, with standard errors in parentheses. 
    }
\label{tab:goodness}
\begin{tabular}{lcccc}
\toprule
 & \textbf{COSS} & \textbf{Normal ($\hat{K}=3$)} & \textbf{Normal ($\hat{K}=7$)} & \textbf{Normal ($\hat{K}_{\text{COSS}}$)} \\
\midrule
AUROC            & 0.9929 (0.0021) & 0.9243 (0.0033) & 0.9932 (0.0002) & 0.9747 (0.0097) \\
$\Delta_{\text{transitivity}}$   & 0.0150 (0.0040) & 0.0516 (0.0012) & 0.0290 (0.0008) & 0.0366 (0.0071) \\
\bottomrule
\end{tabular}
\end{table}

The COSS prior outperforms the AIC-selected model ($\hat{K} = 3$) in both AUC and transitivity recovery. Compared with the CV 1-SE model ($\hat{K} = 7$), the COSS prior achieves comparable AUROC while attaining a markedly lower transitivity error ($\Delta_{\text{transitivity}}$), indicating better recovery of the network's structural properties with a more parsimonious representation.
Even at the matched dimension, the COSS prior still outperforms the normal-prior baseline, indicating that its advantage stems not only from adaptive dimension selection but also from its more flexible prior form (Section~\ref{section:coss_prior}).

Applying Geomin rotation to the estimated loading matrix $\hat{\mathbf{B}}$ reveals five interpretable factors: (1) Risk and Street Culture, (2) Cultural Appreciation, (3) Sports Activity, (4) Social Engagement, and (5) Alternative Subculture. A detailed interpretation of each factor and the corresponding heatmap are provided in Section E.1 of the Supplementary Material.

Figure~\ref{fig:latent_spacedata3} displays the estimated student
positions $\hat{\mathbf{Z}}$ projected onto selected pairs of
factor dimensions, with nodes colored by observed covariates and
overlaid with friendship edges.
The left panel projects Factor~1 (Risk and Street Culture) against
Factor~5 (Alternative Subculture). A clear horizontal gradient
separates smokers, who cluster at high values of Factor~1, from
the larger group of non-smokers. The density of friendship ties
among high-risk students confirms a homophilous selection process,
whereby shared risk behaviors drive the formation of tightly
connected peer groups.
The right panel projects Factor~2 (Cultural Appreciation) against
Factor~5. Here, students with a preference for folk music form a
compact, well-separated cluster, suggesting that this musical
taste acts as a strong social marker defining a distinct cultural
niche within the school network.

\begin{figure}[ht]
    \centering
    \includegraphics[width=0.9\linewidth]{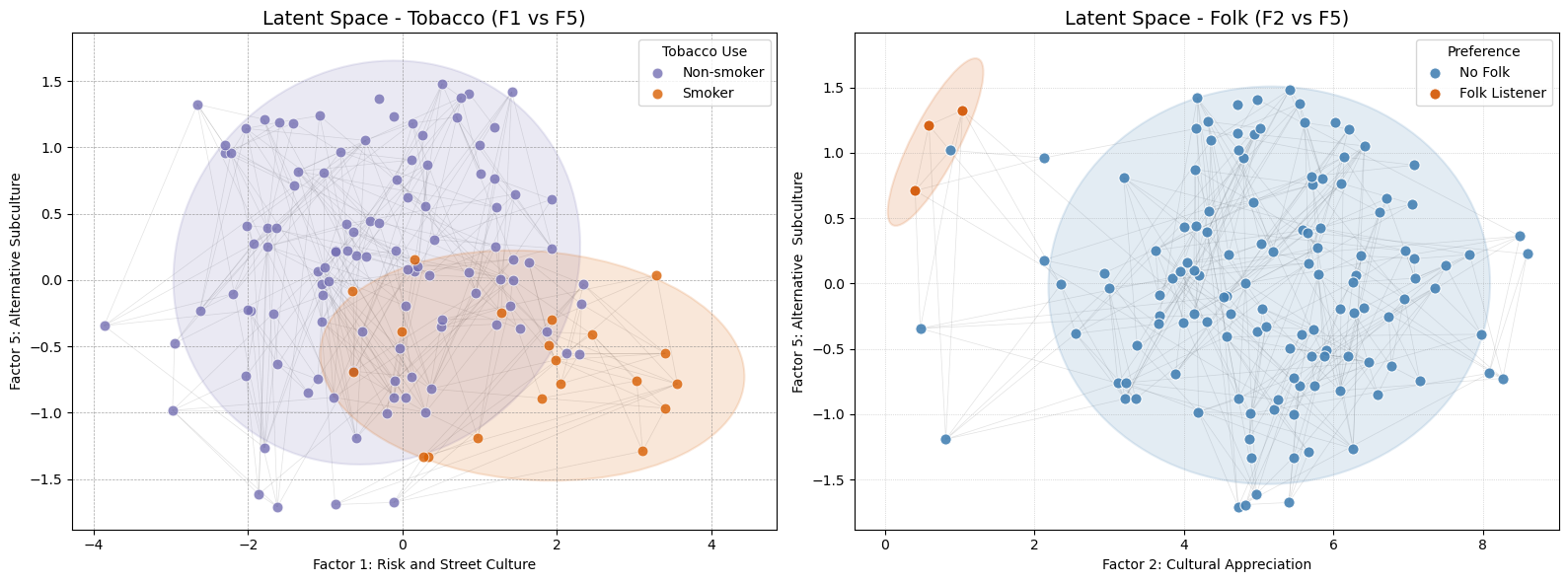}
    \caption{
        Estimated student positions projected onto selected latent
        factor pairs, overlaid with friendship ties (gray lines).
        Left: colored by tobacco use. Right: colored by folk music preference.    
    }
    \label{fig:latent_spacedata3}
\end{figure}

These projections confirm that the five-dimensional latent space
recovered by the COSS prior captures meaningful social geometry:
students' latent positions correspond closely to both their
observed lifestyle profiles and their friendship ties,
reinforcing the interpretation of these dimensions as key drivers
of adolescent social organization. Further details are discussed
in Section E.2 of the Supplementary Material.

\section{Conclusion}\label{section:8}

Network models in psychometrics increasingly integrate relational data with high-dimensional node attributes. While joint latent space models offer a powerful framework for this integration, their utility has been limited by the challenge of selecting the latent dimension, which typically relies on computationally intensive and sometimes unreliable post-hoc procedures. 
{We address this challenge by introducing a cumulative ordered spike-and-slab (COSS) prior into a Bayesian JLSM. This prior induces stochastic ordering on the latent dimensions and progressively shrinks superfluous ones, enabling adaptive dimension selection and parameter estimation within a single posterior computation.}

The proposed methodology offers advantages across several aspects.
First, we developed an efficient and scalable MCMC algorithm by leveraging a hierarchical representation of the prior and Pólya-Gamma data augmentation. Second, we established asymptotic guarantees for the proposed method, proving that the posterior concentrates on parsimonious models near the true dimension and achieves a near-optimal, adaptive Hellinger contraction rate for parameter estimation. Third, simulations confirmed that joint modeling improves estimation accuracy, especially with sparse networks. Furthermore, our adaptive approach outperforms standard dimension selection criteria. Applications to real data demonstrated the model's utility in revealing latent social structures and enhancing predictive accuracy. The code implementing the proposed method with scripts to reproduce the simulation and real-data analyses is available at \url{https://github.com/binlv0211-prog/COSS_JLSM}. 

This adaptive framework suggests several directions for future research.
First, our model assumes that the network and node attributes arise from a single, fully shared latent space. While we have shown this is a reasonable and powerful approximation, future work could explore more flexible architectures that decompose the latent space into components that are shared versus specific to each data modality. 
Second, enhancing interpretability by addressing rotational invariance remains a promising direction. Although the shrinkage prior resolves column permutation invariance, further constraints such as structured sparsity priors on the loading matrix or the inclusion of relevant covariates could be incorporated to achieve an identifiable and interpretable latent structure.
{Third, the methodology can be extended to accommodate richer data structures common in psychometrics, such as dynamic networks and censored responses.}

\section*{Financial Support}
The research was supported by the National Key R\&D Program of China (Grant Nos. 2023YFA1008700 and 2023YFA1008703) and the National Natural Science Foundation of China (Grant Nos. 12301373, 12271168, 12531013, 12271166, and 12371289).

\section*{Competing Interests}
The authors declare none.

\bibliographystyle{apalike}
\bibliography{refs}
\end{document}

\maketitle

\tableofcontents 

\setcounter{equation}{0}
\section{Computational Details}\label{SM:al}
\setcounter{equation}{0}
\subsection{Gibbs Sampler for JLSM with Gaussian Node Attributes}
\textit{Algorithm} 1. One cycle of the Gibbs sampler for joint latent space model with the cumulative ordered spike-and-slab prior.

\textit{Step}.~1: For $i=1$ to $n$ and $i^\prime=1$ to $i-1$, sample  augmented data $d^{A}_{ii^\prime}$ from the full conditional Pólya-Gamma distribution:
\begin{equation*}
d_{ii^\prime}^{A} \mid \text{rest} = d_{i^\prime i}^{A} \mid \text{rest} \sim \text{PG}(1, \Theta_{ii^\prime}^{A}).
\end{equation*}

\textit{Step}.~2: For $i=1$ to $n$, sample $\alpha_i$ from its conditional normal distribution:
\begin{equation}\label{eq:eq:postalpha}
\alpha_i \mid \text{rest} \sim \mathcal{N}\left(\sigma_{\alpha_i}^2\sum_{i^\prime\neq i}\left\{A_{ii^\prime}-1/2-d_{ii^\prime}^{A}(\alpha_{i^\prime}+\boldsymbol z_i^\top \boldsymbol z_{i^\prime})\right\}, \sigma_{\alpha_i}^2\right),
\end{equation}
where $\sigma_{\alpha_i}^2 = 1/\left(\sum_{i^\prime\neq i}d_{ii^\prime}^{A} + \sigma_{\alpha}^{-2}\right)$.

\textit{Step}.~3: For $i=1$ to $n$, sample $\boldsymbol{z}_i$ from its conditional normal distribution given $\mathbf Z^{(-i)}$ and $\boldsymbol{\alpha}^{(-i^\prime)}$, where
$
\mathbf Z^{(-i)} = [\boldsymbol z_1,..., \boldsymbol z_{i-1},\boldsymbol z_{i+1},...,\boldsymbol z_n]^\top $, $ 
\boldsymbol{\alpha}^{(-i^\prime)} = (\alpha_1+\alpha_i,...,\alpha_{i-1}+\alpha_i,\alpha_{i+1}+\alpha_i,...,\alpha_n+\alpha_i)^\top
$,
\begin{equation}\label{eq:postZ}
\boldsymbol z_i \mid \text{rest} \sim \mathcal{N}(\mu_{\boldsymbol z_i},\Sigma_{\boldsymbol z_i}),
\end{equation}
where
\begin{align*}
\Sigma_{\boldsymbol z_i} &= \left(\mathbf Z ^{(-i)\top}\mathbf D^{(-i)}_{A}\mathbf Z^{(-i)} + \mathbf B^\top\Sigma_{Y}^{-1}\mathbf B  + \Sigma^{-1}_{Z}\right)^{-1}, \\
\mu_{\boldsymbol z_i} &= \Sigma_{\boldsymbol z_i}\left(\mathbf Z^{(-i)\top}(\boldsymbol\kappa_{A^{(-i)}} - \mathbf D^{(-i)}_{A}\boldsymbol{\alpha}^{(-i^\prime)}) + \mathbf B^\top\Sigma^{-1}_{Y}(\boldsymbol{y}_i-\boldsymbol{\gamma})\right),
\end{align*}
with
$
\mathbf D^{(-i)}_{A} = \text{diag}(d_{i1}^{A},...,d_{i,i-1}^{A},d_{i,i+1}^{A},...,d_{in}^{A}), 
\Sigma_{Z} = \text{diag}(\theta_1,...,\theta_k)$,
$
\Sigma_{Y} = \text{diag}(\sigma_1^2,...,\sigma_q^2)$,
$
\boldsymbol \kappa_{A^{(-i)}} = \left( A_{i1}-\tfrac{1}{2},...,A_{i,i-1}-\tfrac{1}{2},A_{i,i+1}-\tfrac{1}{2},...,A_{in}-\tfrac{1}{2}\right)^\top.
$

\textit{Step}.~4: Sample $\boldsymbol{\gamma}$ from its conditional normal distribution:
\begin{equation}\label{eq:postgamma}
\boldsymbol\gamma|\text{rest} \sim \mathcal{N} \left(\Sigma_{\gamma}
    \begin{bmatrix}
    \sigma_1^{-2}\sum_{i=1}^{n}\{Y_{i1}-\boldsymbol z^\top _i\boldsymbol\beta_{1}\}\\
    \vdots\\
    \sigma_q^{-2}\sum_{i=1}^{n}\{Y_{iq}-\boldsymbol z^\top _i\boldsymbol\beta_{q}\}
\end{bmatrix}
,\Sigma_{\gamma} \right),    
\end{equation}
where $\Sigma_{\gamma}=\text{diag}\left(\frac{\sigma_{\gamma}^2\sigma_1^2}{n\sigma_{\gamma}^2 + \sigma_1^{2}},\cdots,\frac{\sigma_{\gamma}^2\sigma_q^2}{n\sigma_{\gamma}^2 + \sigma_q^{2}}\right)$.

\textit{Step}.~5: For each $j=1$ to $q$, sample the $j$-th column $\boldsymbol\beta_j$ of $\mathbf B$ from its conditional normal distribution:
$$
\boldsymbol\beta_j|-\sim \mathcal{N}((\sigma_B^{-2}\mathbf{I}_k+\sigma^{-2}_j\mathbf Z^\top \mathbf Z)^{-1}\mathbf Z^\top \sigma^{-2}_j(\mathbf Y_{\cdot j}-\gamma_j\boldsymbol{1}_n),(\sigma_B^{-2}\mathbf{I}_k+\sigma^{-2}_j\mathbf Z^\top \mathbf Z)^{-1}).
$$

\textit{Step}.~6: For each $j=1$ to $q$, sample the variance parameter $\sigma_j^2$ from its conditional inverse gamma distribution:
\begin{equation*}
\sigma_j^2 \mid \text{rest} \sim \text{IG}\left(a_{\sigma} + \frac{n}{2}, b_{\sigma} + \frac{1}{2}\|\mathbf{Y}_{\cdot j} - \gamma_j\boldsymbol{1}_n - \mathbf{Z}\boldsymbol{\beta}_{j}\|^2_2\right).
\end{equation*}

\textit{Step}.~7: For $h=1$ to $k$, sample the indicator variables $\rho_h$ from their conditional distribution:
\begin{equation*}
    \Pr(\rho_h=l \mid \text{rest}) \propto
    \begin{cases}
        \omega_l \mathcal{N}(\mathbf Z_{\cdot h}; 0, \theta_0 \mathbf{I}_n), & l=1,\dots,h,\\
        \omega_l t_{2a_{\theta}}\left(\mathbf Z_{\cdot h}; 0, (a_{\theta}/b_{\theta})\mathbf{I}_n\right), & l=h+1,\dots,k.
    \end{cases}
\end{equation*}

\textit{Step}.~8: Sample $v_1$ from its conditional Beta distribution:
\begin{equation*}
v_1 \mid \text{rest} \sim \text{Beta}\left(\kappa + \sum_{j=1}^{k}\mathbbm{1}(\rho_j = 1), 1 + \sum_{j=1}^{k}\mathbbm{1}(\rho_j > 1)\right).
\end{equation*}

\textit{Step}.~9: For $h=2$ to $k-1$, sample $v_h$ from its conditional Beta distribution:
\begin{equation*}
v_h \mid \text{rest} \sim \text{Beta}\left(a + \sum_{j=1}^{k}\mathbbm{1}(\rho_j = h), 1 + \sum_{j=1}^{k}\mathbbm{1}(\rho_j > h)\right).
\end{equation*}
Set $v_k = 1$ and compute $\omega_1,\dots,\omega_k$ according to Eq.~(5) in the main text.

\textit{Step}.~10: For $h=1$ to $k$, if $\rho_h \leq h$, set $\theta_h = \theta_0$. Otherwise, sample $\theta_h$ from its conditional inverse gamma distribution:
\begin{equation*}
    \theta_h \mid \text{rest} \sim \text{IG}\left(a_{\theta} + \frac{n}{2}, b_{\theta} + \frac{1}{2}\|\mathbf{Z}_{\cdot h}\|^2_2\right).
\end{equation*}

\subsection{Gibbs Sampler for JLSM with Bernoulli Node Attributes}
We can easily obtain the Gibbs sampler when node variables $\mathbf Y$ satisfy the MIRT model by further introducing augmentation data $d^{Y}_{ij}$.

\textit{Algorithm} 2. One cycle of the Gibbs sampler for joint latent space model with the cumulative ordered spike-and-slab prior.

\textit{Step}.~1: For $i=1$ to $n$ and $i^\prime=1$ to $i-1$, sample  augmented data $d^{A}_{ii^\prime}$ from the full conditional Pólya-Gamma distribution:
\begin{equation*}
d_{ii^\prime}^{A} \mid \text{rest} = d_{i^\prime i}^{A} \mid \text{rest} \sim \text{PG}(1,\Theta_{ii^\prime}^{A}).
\end{equation*}

\textit{Step}.~2: For $i=1$ to $n$ and $j=1$ to $q$, sample augmented data $d^{Y}_{ij}$ from the full conditional Pólya-Gamma distribution:
$$
d^{Y}_{ij}|\text{rest} \sim \text{PG}(1,\Theta_{ij}^{Y}).
$$

\textit{Step}.~3: For $i=1$ to $n$, sample $\alpha_i$ from its conditional normal distribution:
\begin{equation*}
\alpha_i \mid \text{rest} \sim \mathcal{N}\left(\sigma_{\alpha_i}^2\sum_{i^\prime\neq i}\left\{A_{ii^\prime}-1/2-d_{ii^\prime}^{A}(\alpha_{i^\prime}+\boldsymbol z_i^\top \boldsymbol z_{i^\prime})\right\}, \sigma_{\alpha_i}^2\right),
\end{equation*}
where $\sigma_{\alpha_i}^2 = 1/\left(\sum_{i^\prime\neq i}d_{ii^\prime}^{A} + \sigma_{\alpha}^{-2}\right)$.

\textit{Step}.~4: For $i=1$ to $n$, sample $\boldsymbol z_i$ from its conditional normal distribution given $\mathbf Z^{(-i)}$ and $\boldsymbol{\alpha}^{(-i^\prime)}$, where
$
\mathbf Z^{(-i)} = [\boldsymbol z_1,...,\boldsymbol z_{i-1},\boldsymbol z_{i+1},...,\boldsymbol z_n]^\top $, $ 
\boldsymbol{\alpha}^{(-i^\prime)} = (\alpha_1+\alpha_i,...,\alpha_{i-1}+\alpha_i,\alpha_{i+1}+\alpha_i,...,\alpha_n+\alpha_i)^\top
$,
\begin{equation*}
\boldsymbol z_i \mid \text{rest} \sim \mathcal{N}(\mu_{\boldsymbol z_i},\Sigma_{\boldsymbol z_i}),
\end{equation*}
where
\begin{align*}
    \Sigma_{\boldsymbol z_i}&=(\mathbf Z^{(-i)^\top}\mathbf D^{(-i)}_{A}\mathbf Z^{(-i)}+\mathbf B^\top\mathbf D_{Y}^{(i)}\mathbf B +\Sigma^{-1}_{Z})^{-1},\\
    \mu_{\boldsymbol z_i} &=\Sigma_{\boldsymbol z_i}(\mathbf Z^{(-i)\top}(\boldsymbol \kappa_{A^{(-i)}}-\mathbf D^{(-i)}_{A}\boldsymbol{\alpha}^{(-i^\prime)}) + \mathbf B^\top(\boldsymbol \kappa_{Y^{(i)}}-\mathbf D_{Y}^{(i)}\boldsymbol{\gamma})),
\end{align*}
with
$\mathbf D^{(-i)}_{A}=\text{diag}(d_{i1}^{A},..,d_{i,i-1}^{A},d_{i,i+1}^{A},...,d_{in}^{A})$,
$D_{Y}^{(i)} = \text{diag}(d_{i1}^{Y},\dots d_{iq}^{Y})$,
$\Sigma_{Z}=\text{diag}(\theta_1,...,\theta_k)$,
$\boldsymbol\kappa_{A^{(-i)}}=(A_{i1}-\frac{1}{2},...,A_{i,i-1}-\frac{1}{2},A_{i,i+1}-\frac{1}{2},...,A_{in}-\frac{1}{2})^\top $ and $\boldsymbol\kappa_{Y^{(i)}} = (Y_{i1} - \frac{1}{2},\dots,Y_{iq}-\frac{1}{2})^\top $.

\textit{Step}.~5: Sample $\boldsymbol{\gamma}$ from its conditional normal distribution:
$$
\boldsymbol\gamma|\text{rest} \sim \mathcal{N} \left(\Sigma_{\gamma}
    \begin{bmatrix}
    \sum_{i=1}^{n}\{Y_{i1}-\frac{1}{2}-d_{i1}^{Y}\boldsymbol z^\top _i\boldsymbol\beta_1\}\\
    \vdots\\
    \sum_{i=1}^{n}\{Y_{iq}-\frac{1}{2}-d_{iq}^{Y}\boldsymbol z^\top _i\boldsymbol\beta_q\}
\end{bmatrix}
,\Sigma_{\gamma} \right)
$$
where $\Sigma_{\boldsymbol\gamma}=\left(\text{diag}\left(\sum_{i=1}^nd_{i1}^{Y},\cdots,\sum_{i=1}^nd_{iq}^{Y}\right) + \sigma_{\gamma}^{-2}\mathbf{I}_q\right)^{-1}$.

\textit{Step}.~6: For each $j=1$ to $q$, sample the $j$-th column $\boldsymbol{\beta}_{j}$ of $\mathbf{B}$ from its conditional normal distribution:
$$
\boldsymbol\beta_j|\text{rest}\sim \mathcal{N}((\sigma_B^{-2}\mathbf{I}_k+ \mathbf Z^\top \mathbf D_{Y}^{(j)^\prime}\mathbf Z)^{-1}\mathbf Z^\top (\boldsymbol\kappa_{Y^{(j)^\prime}}-\mathbf D_{Y}^{(j)^\prime} \gamma_j\boldsymbol{1}_n),(\sigma_B^{-2}\mathbf{I}_k+\mathbf Z^\top \mathbf D_{Y}^{(j)^\prime}\mathbf Z)^{-1}),
$$
where $\mathbf D_{Y}^{(j)^\prime} = \text{diag}(d_{1j}^{Y},\dots,d_{nj}^{Y})$, $\boldsymbol\kappa_{Y^{(j)^\prime}} = (Y_{1j}-\frac{1}{2},\dots,Y_{nj}-\frac{1}{2})^\top$.

\textit{Step}.~7: For $h=1$ to $k$, sample the indicator variables $\rho_h$ from their conditional distribution:
\begin{equation*}
    \Pr(\rho_h=l \mid \text{rest}) \propto
    \begin{cases}
        \omega_l \mathcal{N}(\mathbf Z_{\cdot h}; 0, \theta_0 \mathbf{I}_n), & l=1,\dots,h,\\
        \omega_l t_{2a_{\theta}}\left(\mathbf Z_{\cdot h}; 0, (a_{\theta}/b_{\theta})\mathbf{I}_n\right), & l=h+1,\dots,k.
    \end{cases}
\end{equation*}

\textit{Step}.~8: Sample $v_1$ from its conditional Beta distribution:
\begin{equation*}
v_1 \mid \text{rest} \sim \text{Beta}\left(\kappa + \sum_{j=1}^{k}\mathbbm{1}(\rho_j = 1), 1 + \sum_{j=1}^{k}\mathbbm{1}(\rho_j > 1)\right).
\end{equation*}

\textit{Step}.~9: For $h=2$ to $k-1$, sample $v_h$ from its conditional Beta distribution:
\begin{equation*}
v_h \mid \text{rest} \sim \text{Beta}\left(a + \sum_{j=1}^{k}\mathbbm{1}(\rho_j = h), 1 + \sum_{j=1}^{k}\mathbbm{1}(\rho_j > h)\right).
\end{equation*}
Set $v_k = 1$ and compute $\omega_1,\dots,\omega_k$ according to Eq.~(5) in the main text.

\textit{Step}.~10: For $h=1$ to $k$, if $\rho_h \leq h$, set $\theta_h = \theta_0$. Otherwise, sample $\theta_h$ from its conditional inverse gamma distribution:
\begin{equation*}
    \theta_h \mid \text{rest} \sim \text{IG}\left(a_{\theta} + \frac{n}{2}, b_{\theta} + \frac{1}{2}\|\mathbf{Z}_{\cdot h}\|^2_2\right).
\end{equation*}

\subsection{MCMC algorithm for JLSM with Multinomial Node Attributes}\label{sec:Multinomial_MHgibbs}
The MCMC algorithm for the Nominal Response Model (NRM) -based nodal attributes $\mathbf{Y}$ is obtained by embedding a Metropolis–Hastings step within the overall Gibbs sampling cycle to update the non-conjugate latent parameters.

\textit{Algorithm} 3. One cycle of the MCMC algorithm for joint latent space model with the cumulative ordered spike-and-slab prior.

\textit{Step}.~1: For $i=1$ to $n$ and $i^\prime=1$ to $i-1$, sample  augmented data $d^{A}_{ii^\prime}$ from the full conditional Pólya-Gamma distribution:
\begin{equation*}
d_{ii^\prime}^{A} \mid \text{rest} = d_{i^\prime i}^{A} \mid \text{rest} \sim \text{PG}(1,\Theta_{ii^\prime}^{A}).
\end{equation*}

\textit{Step}.~2: For $i=1$ to $n$, $j=1$ to $q$ and $l = 1$ to $L$. sample  augmented data $d^{Y}_{ijl}$ from the full conditional Pólya-Gamma distribution:
$$
d^{Y}_{ijl}|\text{rest} \sim \text{PG}(1,\psi_{ijl}).
$$
where $\psi_{ijl} = \boldsymbol{z}_i^T \boldsymbol{\beta}_{jl} + \gamma_{jl} - \log\sum_{m\ne l}\exp(\boldsymbol{z}_i^T \boldsymbol{\beta}_{jm} + \gamma_{jm})$.
 
\textit{Step}.~3: For $i=1$ to $n$, sample $\alpha_i$ from its conditional normal distribution:
\begin{equation*}
\alpha_i \mid \text{rest} \sim \mathcal{N}\left(\sigma_{\alpha_i}^2\sum_{i^\prime\neq i}\left\{A_{ii^\prime}-1/2-d_{ii^\prime}^{A}(\alpha_{i^\prime}+\boldsymbol z_i^\top \boldsymbol z_{i^\prime})\right\}, \sigma_{\alpha_i}^2\right),
\end{equation*}
where $\sigma_{\alpha_i}^2 = 1/\left(\sum_{i^\prime\neq i}d_{ii^\prime}^{A} + \sigma_{\alpha}^{-2}\right)$.

\textit{Step}. 4: For $i = 1$ to $n$, propose a candidate latent position $\boldsymbol{z}_i^* \sim \mathcal{N}(\boldsymbol{z}_i^{(t-1)}, \tau^2 \mathbf{I}_k)$. Accept the proposal and set $\boldsymbol{z}_i^{(t)} = \boldsymbol{z}_i^*$ with probability $r_i$; otherwise, set $\boldsymbol{z}_i^{(t)} = \boldsymbol{z}_i^{(t-1)}$. The acceptance probability is given by:
\begin{equation*}
    r_i = \min \left\{ 1, \frac{\mathcal{L}(\boldsymbol{z}_i^* \mid \text{rest}) \, \Pr(\boldsymbol{z}_i^* \mid \Sigma_Z)}{\mathcal{L}(\boldsymbol{z}_i^{(t-1)} \mid \text{rest}) \, \Pr(\boldsymbol{z}_i^{(t-1)} \mid \Sigma_Z)} \right\}
\end{equation*}
More explicitly, the ratio is calculated as:
\begin{equation*}
    r_i = \min \left\{ 1, \frac{\prod_{i^\prime \ne i} \Pr(A_{ii^\prime} \mid \boldsymbol{z}_i^*, \dots) \prod_{j=1}^{q} \Pr(Y_{ij} \mid \boldsymbol{z}_i^*, \dots) \exp \left( -\frac{1}{2} \boldsymbol{z}_i^{*\top} \Sigma_Z^{-1} \boldsymbol{z}_i^* \right)}{\prod_{i^\prime \ne i} \Pr(A_{ii^\prime} \mid \boldsymbol{z}_i^{(t-1)}, \dots) \prod_{j=1}^{q} \Pr(Y_{ij} \mid \boldsymbol{z}_i^{(t-1)}, \dots) \exp \left( -\frac{1}{2} \boldsymbol{z}_i^{(t-1)\top} \Sigma_Z^{-1} \boldsymbol{z}_i^{(t-1)} \right)} \right\}
\end{equation*}
where $\Sigma_{Z}=\text{diag}(\theta_1,...,\theta_k)$, $\boldsymbol{z}_i^{(t)}$ is the current simulated value of $\boldsymbol{z}_i$ and  $\boldsymbol{z}_i^{(t-1)}$ is the simulated value of
the former iteration. 

\textit{Step}. 5: For $j=1$ to $q$ and $l=2$ to $L$, sample $\boldsymbol\gamma_{l}$ from its  conditional  distribution
\begin{equation*}
\boldsymbol\gamma_{l}|\text{rest} \sim \mathcal{N} \left(\Sigma_{\gamma_l}
    \begin{bmatrix}
    \sum_{i=1}^{n}\{y_{i1l}-\frac{1}{2}-d_{i1}^{Y}(\boldsymbol z^\top _i\boldsymbol\beta_{1l} - c_{i1l})\}\\
    \vdots\\
    \sum_{i=1}^{n}\{y_{iql}-\frac{1}{2}-d_{iq}^{Y}(\boldsymbol z^\top _i\boldsymbol\beta_{ql} - c_{iql})\}
\end{bmatrix}
,\Sigma_{\gamma_l} \right)    
\end{equation*}
where $\Sigma_{\boldsymbol\gamma_l}=\left(\text{diag}\left(\sum_{i=1}^nd_{i1l}^{Y},\cdots,\sum_{i-1}^nd_{iql}^{Y}\right) + \sigma_{\gamma}^{-2}\mathbf{I}_q\right)^{-1}$, $c_{ijl} = \log\sum_{m\ne l}\exp(\boldsymbol{z}_i^T \boldsymbol{\beta}_{jm} + \gamma_{jm})$ and $y_{ijl} = \mathbbm{1}(Y_{ij}=l)$.

\textit{Step}. 6: For $j=1$ to $q$ and $l=2$ to $L$, sample $\boldsymbol{\beta}_{jl}$ from its conditional normal distribution:
$$
\boldsymbol\beta_{jl}|\text{rest}\sim \mathcal{N}(({\sigma_B^{-2}}\mathbf{I}_k+ \mathbf Z^\top \mathbf D_{Y}^{(jl)}\mathbf Z)^{-1}\mathbf Z^\top (\boldsymbol\kappa_{Y^{(jl)}}-\mathbf D_{Y}^{(j)^\prime} (\gamma_{jl}\boldsymbol{1}_n - \boldsymbol{c}_{jl})),({\sigma_B^{-2}}\mathbf{I}_k+\mathbf Z^\top \mathbf D_{Y}^{(jl)}\mathbf Z)^{-1}),
$$
where $\mathbf D_{Y}^{(jl)} = \text{diag}(d_{1jl}^{Y},\dots,d_{njl}^{Y})$, $\boldsymbol\kappa_{Y^{(jl)}} = (y_{1jl}-\frac{1}{2},\dots,y_{njl}-\frac{1}{2})^\top$ and $\boldsymbol{c}_{jl} = (c_{1jl},\dots,c_{njl})^\top$.

\textit{Step}.~7: For $h=1$ to $k$, sample the indicator variables $\rho_h$ from their conditional distribution:
\begin{equation*}
    \Pr(\rho_h=l \mid \text{rest}) \propto
    \begin{cases}
        \omega_l \mathcal{N}(\mathbf Z_{\cdot h}; 0, \theta_0 \mathbf{I}_n), & l=1,\dots,h,\\
        \omega_l t_{2a_{\theta}}\left(\mathbf Z_{\cdot h}; 0, (a_{\theta}/b_{\theta})\mathbf{I}_n\right), & l=h+1,\dots,k.
    \end{cases}
\end{equation*}

\textit{Step}.~8: Sample $v_1$ from its conditional Beta distribution:
\begin{equation*}
v_1 \mid \text{rest} \sim \text{Beta}\left(\kappa + \sum_{j=1}^{k}\mathbbm{1}(\rho_j = 1), 1 + \sum_{j=1}^{k}\mathbbm{1}(\rho_j > 1)\right).
\end{equation*}

\textit{Step}.~9: For $h=2$ to $k-1$, sample $v_h$ from its conditional Beta distribution:
\begin{equation*}
v_h \mid \text{rest} \sim \text{Beta}\left(a + \sum_{j=1}^{k}\mathbbm{1}(\rho_j = h), 1 + \sum_{j=1}^{k}\mathbbm{1}(\rho_j > h)\right).
\end{equation*}
Set $v_k = 1$ and compute $\omega_1,\dots,\omega_k$ according to Eq.~(6).

\textit{Step}.~10: For $h=1$ to $k$, if $\rho_h \leq h$, set $\theta_h = \theta_0$. Otherwise, sample $\theta_h$ from its conditional inverse gamma distribution:
\begin{equation*}
    \theta_h \mid \text{rest} \sim \text{IG}\left(a_{\theta} + \frac{n}{2}, b_{\theta} + \frac{1}{2}\|\mathbf{Z}_{\cdot h}\|^2_2\right).
\end{equation*}

\subsection{Gibbs Sampler for JLSM with Ordinal Node Attributes}\label{sec:ordinal_gibbs}
We can easily obtain the Gibbs sampler when node variables $\mathbf Y$ satisfy the Grade Response Model (GRM) by further introducing augmentation data $d^{Y}_{ij}$ and $\psi_{ij}$.

\textit{Algorithm} 4. One cycle of the Gibbs sampler for joint latent space model with the cumulative ordered spike-and-slab prior.

\textit{Step}.~1: For $i=1$ to $n$ and $i^\prime=1$ to $i-1$, sample  augmented data $d^{A}_{ii^\prime}$ from the full conditional Pólya-Gamma distribution:
\begin{equation*}
d_{ii^\prime}^{A} \mid \text{rest} = d_{i^\prime i}^{A} \mid \text{rest} \sim \text{PG}(1,\Theta_{ii^\prime}^{A}).
\end{equation*}

\textit{Step}.~2: For $i=1$ to $n$ and $j=1$ to $q$, sample  augmented data $\psi_{ij}$ from its full conditional truncated normal distribution:
\begin{equation*}
    \psi_{ij}\mid d_{ij}^Y, \boldsymbol{z}_i,\boldsymbol{\beta}_j,\gamma_{j,l-1},\gamma_{j,l},Y_{ij}=l\sim \mathcal{N}(\boldsymbol{z}_i\boldsymbol{\beta}_j^\top, \frac{1}{d_{ij}^Y})\mathbbm{1}(\gamma_{j,l-1},\gamma_{j,l}).
\end{equation*}

\textit{Step}.~3:  For $i=1$ to $n$ and $j=1$ to $q$, sample augmented data $d^{Y}_{ij}$ from the full conditional Pólya-Gamma distribution:
$$
d^{Y}_{ij}|\text{rest} \sim \text{PG}(2,-\psi_{ij} + \boldsymbol{z}_i\boldsymbol{\beta}_j^\top).
$$

\textit{Step}.~4: For $i=1$ to $n$, sample $\alpha_i$ from its conditional normal distribution:
\begin{equation*}
\alpha_i \mid \text{rest} \sim \mathcal{N}\left(\sigma_{\alpha_i}^2\sum_{i^\prime\neq i}\left\{A_{ii^\prime}-1/2-d_{ii^\prime}^{A}(\alpha_{i^\prime}+\boldsymbol z_i^\top \boldsymbol z_{i^\prime})\right\}, \sigma_{\alpha_i}^2\right),
\end{equation*}
where $\sigma_{\alpha_i}^2 = 1/\left(\sum_{i^\prime\neq i}d_{ii^\prime}^{A} + \sigma_{\alpha}^{-2}\right)$.

\textit{Step}.~5: For $i=1$ to $n$, sample $\boldsymbol z_i$ from its conditional normal distribution given $\mathbf Z^{(-i)}$ and $\boldsymbol{\alpha}^{(-i^\prime)}$, where
$
\mathbf Z^{(-i)} = [\boldsymbol z_1,...,\boldsymbol z_{i-1},\boldsymbol z_{i+1},...,\boldsymbol z_n]^\top $, $ 
\boldsymbol{\alpha}^{(-i^\prime)} = (\alpha_1+\alpha_i,...,\alpha_{i-1}+\alpha_i,\alpha_{i+1}+\alpha_i,...,\alpha_n+\alpha_i)^\top
$,
\begin{equation*}
\boldsymbol z_i \mid \text{rest} \sim \mathcal{N}(\mu_{\boldsymbol z_i},\Sigma_{\boldsymbol z_i}),
\end{equation*}
where
\begin{align*}
    \Sigma_{\boldsymbol z_i}&=(\mathbf Z^{(-i)^\top}\mathbf D^{(-i)}_{A}\mathbf Z^{(-i)} +\mathbf B^\top\mathbf D_{Y}^{(i)}\mathbf B +\Sigma^{-1}_{Z})^{-1},\\
    \mu_{\boldsymbol z_i} &=\Sigma_{\boldsymbol z_i}(\mathbf Z^{(-i)\top}(\boldsymbol \kappa_{A^{(-i)}}-\mathbf D^{(-i)}_{A}\boldsymbol{\alpha}^{(-i^\prime)}) + \mathbf B^\top\mathbf D_{Y}^{(i)}\boldsymbol{\psi_i}),
\end{align*}
with
$\mathbf D^{(-i)}_{A}=\text{diag}(d_{i1}^{A},..,d_{i,i-1}^{A},d_{i,i+1}^{A},...,d_{in}^{A})$,
$D_{Y}^{(i)} = \text{diag}(d_{i1}^{Y},\dots d_{iq}^{Y})$,
$\Sigma_{Z}=\text{diag}(\theta_1,...,\theta_k)$,
$\boldsymbol\kappa_{A^{(-i)}}=(A_{i1}-\frac{1}{2},...,A_{i,i-1}-\frac{1}{2},A_{i,i+1}-\frac{1}{2},...,A_{in}-\frac{1}{2})^\top $, and $\boldsymbol\psi_i = (\psi_{i1},\dots,\psi_{iq})^\top$.

\textit{Step}.~6: For $j=1$ to $q$ and $l=1$ to $L-1$, sample $\gamma_{il}$ from its  conditional  distribution
\begin{equation*}
    \gamma_{il}\mid \text{rest} \sim f_{\text{prior}}(\gamma_{il})\mathbbm{1}\left(\Omega_U,\Omega_L\right), 
\end{equation*}
where \( f_{\text{prior}}(\gamma_{jl}) \) denotes the prior distribution of \(\gamma_{jl}\), \(\Omega_U = \max \left( \max_{i: Y_{ij}=l} \psi_{ij}, \gamma_{j,l-1} \right) \), and \(\Omega_L = \min \left( \min_{i: Y_{ij}=l+1} \psi_{ij}, \gamma_{j,l+1} \right) \) represent the upper and lower bounds of the truncated prior distribution, respectively.

\textit{Step}.~7: For each $j=1$ to $q$, sample the $j$-th column $\boldsymbol{\beta}_{j}$ of $\mathbf{B}$ from its conditional normal distribution:
$$
\boldsymbol\beta_j|\text{rest}\sim \mathcal{N}((\sigma_B^{-2}\mathbf{I}_k+ \mathbf Z^\top \mathbf D_{Y}^{(j)^\prime}\mathbf Z)^{-1}\mathbf Z^\top \mathbf D_{Y}^{(j)^\prime}\boldsymbol{\psi}^\prime_j,(\sigma_B^{-2}\mathbf{I}_k+\mathbf Z^\top \mathbf D_{Y}^{(j)^\prime}\mathbf Z)^{-1}),
$$
where $\mathbf D_{Y}^{(j)^\prime} = \text{diag}(d_{1j}^{Y},\dots,d_{nj}^{Y})$ and $\boldsymbol{\psi}^\prime_j = (\psi_{1j},\dots,\psi_{nj})^\top$.

\textit{Step}.~8: For $h=1$ to $k$, sample the indicator variables $\rho_h$ from their conditional distribution:
\begin{equation*}
    \Pr(\rho_h=l \mid \text{rest}) \propto
    \begin{cases}
        \omega_l \mathcal{N}(\mathbf Z_{\cdot h}; 0, \theta_0 \mathbf{I}_n), & l=1,\dots,h,\\
        \omega_l t_{2a_{\theta}}\left(\mathbf Z_{\cdot h}; 0, (a_{\theta}/b_{\theta})\mathbf{I}_n\right), & l=h+1,\dots,k.
    \end{cases}
\end{equation*}

\textit{Step}.~9: Sample $v_1$ from its conditional Beta distribution:
\begin{equation*}
v_1 \mid \text{rest} \sim \text{Beta}\left(\kappa + \sum_{j=1}^{k}\mathbbm{1}(\rho_j = 1), 1 + \sum_{j=1}^{k}\mathbbm{1}(\rho_j > 1)\right).
\end{equation*}

\textit{Step}.~10: For $h=2$ to $k-1$, sample $v_h$ from its conditional Beta distribution:
\begin{equation*}
v_h \mid \text{rest} \sim \text{Beta}\left(a + \sum_{j=1}^{k}\mathbbm{1}(\rho_j = h), 1 + \sum_{j=1}^{k}\mathbbm{1}(\rho_j > h)\right).
\end{equation*}
Set $v_k = 1$ and compute $\omega_1,\dots,\omega_k$ according to Eq.~(5) in the main text.

\textit{Step}.~11: For $h=1$ to $k$, if $\rho_h \leq h$, set $\theta_h = \theta_0$. Otherwise, sample $\theta_h$ from its conditional inverse gamma distribution:
\begin{equation*}
    \theta_h \mid \text{rest} \sim \text{IG}\left(a_{\theta} + \frac{n}{2}, b_{\theta} + \frac{1}{2}\|\mathbf{Z}_{\cdot h}\|^2_2\right).
\end{equation*}

\subsection{Gibbs Sampler for JLSM with Mix-type Node Attributes}\label{sec:mix_gibbs}
\textit{Algorithm} 5 . One cycle of the Gibbs sampler for joint latent space model with the cumulative ordered spike-and-slab prior.

\textit{Step}.~1: For $i=1$ to $n$ and $i^\prime=1$ to $i-1$, sample  augmented data $d^{A}_{ii^\prime}$ from the full conditional Pólya-Gamma distribution:
\begin{equation*}
d_{ii^\prime}^{A} \mid \text{rest} = d_{i^\prime i}^{A} \mid \text{rest} \sim \text{PG}(1,\Theta_{ii^\prime}^{A}).
\end{equation*}

\textit{Step}.~2: For $i=1$ to $n$ and $j=1$ to $q_b$, sample augmented data $d^{Y_b}_{ij}$ from the full conditional Pólya-Gamma distribution:
$$
d^{Y_b}_{ij}|\text{rest} \sim \text{PG}(1,\Theta_{ij}^{Y_b}).
$$

\textit{Step}.~3: For $i=1$ to $n$ and $j=1$ to $q_o$, sample  augmented data $\psi_{o,ij}$ from its full conditional truncated normal distribution:
\begin{equation*}
    \psi_{o,ij}\mid d_{ij}^{Y_o}, \boldsymbol{z}_i,\boldsymbol{\beta}_{o,j},\gamma_{o,j,l-1},\gamma_{o,j,l},Y_{o,ij}=l\sim \mathcal{N}(\boldsymbol{z}_i\boldsymbol{\beta}_{o,j}^\top, \frac{1}{d_{ij}^{Y_o}})\mathbbm{1}(\gamma_{o,j,l-1},\gamma_{o,j,l}).
\end{equation*}

\textit{Step}.~4:  For $i=1$ to $n$ and $j=1$ to $q_o$, sample augmented data $d^{Y_o}_{ij}$ from the full conditional Pólya-Gamma distribution:
$$
d^{Y_o}_{ij}|\text{rest} \sim \text{PG}(2,-\psi_{o,ij} + \boldsymbol{z}_i\boldsymbol{\beta}_{o,j}^\top).
$$

\textit{Step}.~5: For $i=1$ to $n$, sample $\alpha_i$ from its conditional normal distribution:
\begin{equation*}
\alpha_i \mid \text{rest} \sim \mathcal{N}\left(\sigma_{\alpha_i}^2\sum_{i^\prime\neq i}\left\{A_{ii^\prime}-1/2-d_{ii^\prime}^{A}(\alpha_{i^\prime}+\boldsymbol z_i^\top \boldsymbol z_{i^\prime})\right\}, \sigma_{\alpha_i}^2\right),
\end{equation*}
where $\sigma_{\alpha_i}^2 = 1/\left(\sum_{i^\prime\neq i}d_{ii^\prime}^{A} + \sigma_{\alpha}^{-2}\right)$.

\textit{Step}.~6: For $i=1$ to $n$, sample $\boldsymbol z_i$ from its conditional normal distribution given $\mathbf Z^{(-i)}$ and $\boldsymbol{\alpha}^{(-i^\prime)}$, where
$
\mathbf Z^{(-i)} = [\boldsymbol z_1,...,\boldsymbol z_{i-1},\boldsymbol z_{i+1},...,\boldsymbol z_n]^\top $, $ 
\boldsymbol{\alpha}^{(-i^\prime)} = (\alpha_1+\alpha_i,...,\alpha_{i-1}+\alpha_i,\alpha_{i+1}+\alpha_i,...,\alpha_n+\alpha_i)^\top
$,
\begin{equation*}
\boldsymbol z_i \mid \text{rest} \sim \mathcal{N}(\mu_{\boldsymbol z_i},\Sigma_{\boldsymbol z_i}),
\end{equation*}
where
\begin{align*}
    \Sigma_{\boldsymbol z_i}&=(\mathbf Z^{(-i)^\top}\mathbf D^{(-i)}_{A}\mathbf Z^{(-i)} + \mathbf B_g^\top\Sigma_{Y_g}^{-1}\mathbf B_g + \mathbf B_b^\top\mathbf D_{Y_b}^{(i)}\mathbf B_b +\mathbf B_o^\top\mathbf D_{Y_o}^{(i)}\mathbf B_o +\Sigma^{-1}_{Z})^{-1},\\
    \mu_{\boldsymbol z_i} &=\Sigma_{\boldsymbol z_i}(\mathbf Z^{(-i)\top}(\boldsymbol \kappa_{A^{(-i)}}-\mathbf D^{(-i)}_{A}\boldsymbol{\alpha}^{(-i^\prime)}) + \mathbf B_g^\top\Sigma^{-1}_{Y_g}(\boldsymbol{y}_{g,i}-\boldsymbol{\gamma}_g) + \mathbf B_b^\top(\boldsymbol \kappa_{Y_b^{(i)}}-\mathbf D_{Y_b}^{(i)}\boldsymbol{\gamma}_b) + \mathbf B_o^\top\mathbf D_{Y_o}^{(i)}\boldsymbol{\psi}_{o,i}),
\end{align*}
with
$\mathbf D^{(-i)}_{A}=\text{diag}(d_{i1}^{A},..,d_{i,i-1}^{A},d_{i,i+1}^{A},...,d_{in}^{A})$,
$\Sigma_{Y_g} = \text{diag}(\sigma_1^2,...,\sigma_{q_g}^2)$,
$D_{Y_b}^{(i)} = \text{diag}(d_{i1}^{Y_b},\dots d_{iq_b}^{Y_b})$,
$D_{Y_o}^{(i)} = \text{diag}(d_{i1}^{Y_o},\dots d_{iq_o}^{Y_o})$,
$\Sigma_{Z}=\text{diag}(\theta_1,...,\theta_k)$,
$\boldsymbol\kappa_{A^{(-i)}}=(A_{i1}-\frac{1}{2},...,A_{i,i-1}-\frac{1}{2},A_{i,i+1}-\frac{1}{2},...,A_{in}-\frac{1}{2})^\top $, 
$\boldsymbol\kappa_{Y_b^{(i)}} = (Y_{b,i1} - \frac{1}{2},\dots,Y_{b,iq}-\frac{1}{2})^\top $
and $\boldsymbol\psi_{o,i} = (\psi_{o,i1},\dots,\psi_{o,iq})^\top$.

\textit{Step}.~7: Sample $\boldsymbol{\gamma}_g$ from its conditional normal distribution:
\begin{equation*}
\boldsymbol\gamma_g|\text{rest} \sim \mathcal{N} \left(\Sigma_{\gamma_g}
    \begin{bmatrix}
    \sigma_1^{-2}\sum_{i=1}^{n}\{Y_{g,i1}-\boldsymbol z^\top _i\boldsymbol\beta_{g,1}\}\\
    \vdots\\
    \sigma_{q_g}^{-2}\sum_{i=1}^{n}\{Y_{g,iq}-\boldsymbol z^\top _i\boldsymbol\beta_{g,q_g}\}
\end{bmatrix}
,\Sigma_{\gamma_g} \right),    
\end{equation*}
where $\Sigma_{\gamma_g}=\text{diag}\left(\frac{\sigma_{\gamma_g}^2\sigma_1^2}{n\sigma_{\gamma_g}^2 + \sigma_1^{2}},\cdots,\frac{\sigma_{\gamma_g}^2\sigma_{q_g}^2}{n\sigma_{\gamma_g}^2 + \sigma_{q_g}^{2}}\right)$.

\textit{Step}.~8: Sample $\boldsymbol{\gamma}_b$ from its conditional normal distribution:
$$
\boldsymbol\gamma_b|\text{rest} \sim \mathcal{N} \left(\Sigma_{\gamma_b}
    \begin{bmatrix}
    \sum_{i=1}^{n}\{Y_{b,i1}-\frac{1}{2}-d_{i1}^{Y_b}\boldsymbol z^\top _i\boldsymbol\beta_{b,1}\}\\
    \vdots\\
    \sum_{i=1}^{n}\{Y_{b,iq}-\frac{1}{2}-d_{iq_b}^{Y_b}\boldsymbol z^\top _i\boldsymbol\beta_{b,q}\}
\end{bmatrix}
,\Sigma_{\gamma_b} \right)
$$
where $\Sigma_{\boldsymbol\gamma_b}=\left(\text{diag}\left(\sum_{i-1}^nd_{i1}^{Y_b},\cdots,\sum_{i-1}^nd_{iq_b}^{Y_b}\right) + \sigma_{\gamma_b}^{-2}\mathbf{I}_{q_b}\right)^{-1}$.

\textit{Step}.~9: For $j=1$ to $q_o$ and $l=1$ to $L-1$, sample $\gamma_{il}$ from its  conditional  distribution
\begin{equation*}
    \gamma_{o,il}\mid \text{rest} \sim f_{\text{prior}}(\gamma_{o,il})\mathbbm{1}\left(\Omega_U,\Omega_L\right), 
\end{equation*}
where \( f_{\text{prior}}(\gamma_{o,jl}) \) denotes the prior distribution of \(\gamma_{o,jl}\), \(\Omega_U = \max \left( \max_{i: Y_{o,ij}=l} \psi_{o,ij}, \gamma_{o,j,l-1} \right) \), and \(\Omega_L = \min \left( \min_{i: Y_{o,ij}=l+1} \psi_{o,ij}, \gamma_{o,j,l+1} \right) \) represent the upper and lower bounds of the truncated prior distribution, respectively.

\textit{Step}.~10: For each $j=1$ to $q_g$, sample the $j$-th column $\boldsymbol\beta_{g,j}$ of $\mathbf B_g$ from its conditional normal distribution:
$$
\boldsymbol\beta_{g,j}|-\sim \mathcal{N}(({\sigma_{B_g}^{-2}}\mathbf{I}_k+\sigma^{-2}_j\mathbf Z^\top \mathbf Z)^{-1}\mathbf Z^\top \sigma^{-2}_j(\mathbf Y_{g,\cdot j}-\gamma_{g,j}\boldsymbol{1}_n),({\sigma_{B_g}^{-2}}\mathbf{I}_k+\sigma^{-2}_j\mathbf Z^\top \mathbf Z)^{-1}).
$$

\textit{Step}.~11: For each $j=1$ to $q_b$, sample the $j$-th column $\boldsymbol\beta_{b,j}$ of $\mathbf B_b$ from its conditional normal distribution:
$$
\boldsymbol\beta_{b,j}|\text{rest}\sim \mathcal{N}(({\sigma_{B_b}^{-2}}\mathbf{I}_k+ \mathbf Z^\top \mathbf D_{Y_b}^{(j)^\prime}\mathbf Z)^{-1}\mathbf Z^\top (\boldsymbol\kappa_{Y_b^{(j)^\prime}}-\mathbf D_{Y_b}^{(j)^\prime} \gamma_{b,j}\boldsymbol{1}_n),({\sigma_{B_b}^{-2}}\mathbf{I}_k+\mathbf Z^\top \mathbf D_{Y_b}^{(j)^\prime}\mathbf Z)^{-1}),
$$
where $\mathbf D_{Y_b}^{(j)^\prime} = \text{diag}(d_{1j}^{Y_b},\dots,d_{nj}^{Y_b})$, $\boldsymbol\kappa_{Y_b^{(j)^\prime}} = (Y_{b,1j}-\frac{1}{2},\dots,Y_{b,nj}-\frac{1}{2})^\top$.

\textit{Step}.~12: For each $j=1$ to $q_o$, sample the $j$-th column $\boldsymbol{\beta}_{o,j}$ of $\mathbf{B_o}$ from its conditional normal distribution:
$$
\boldsymbol\beta_{o,j}|\text{rest}\sim \mathcal{N}((\sigma_{B_o}^{-2}\mathbf{I}_k+ \mathbf Z^\top \mathbf D_{Y_o}^{(j)^\prime}\mathbf Z)^{-1}\mathbf Z^\top \mathbf D_{Y_o}^{(j)^\prime}\boldsymbol{\psi}^\prime_{o,j},(\sigma_{B_o}^{-2}\mathbf{I}_k+\mathbf Z^\top \mathbf D_{Y_o}^{(j)^\prime}\mathbf Z)^{-1}),
$$
where $\mathbf D_{Y_o}^{(j)^\prime} = \text{diag}(d_{1j}^{Y_o},\dots,d_{nj}^{Y_o})$ and $\boldsymbol{\psi}^\prime_{o,j} = (\psi_{o,1j},\dots,\psi_{o,nj})^\top$.

\textit{Step}.~13: For $h=1$ to $k$, sample the indicator variables $\rho_h$ from their conditional distribution:
\begin{equation*}
    \Pr(\rho_h=l \mid \text{rest}) \propto
    \begin{cases}
        \omega_l \mathcal{N}(\mathbf Z_{\cdot h}; 0, \theta_0 \mathbf{I}_n), & l=1,\dots,h,\\
        \omega_l t_{2a_{\theta}}\left(\mathbf Z_{\cdot h}; 0, (a_{\theta}/b_{\theta})\mathbf{I}_n\right), & l=h+1,\dots,k.
    \end{cases}
\end{equation*}

\textit{Step}.~14: Sample $v_1$ from its conditional Beta distribution:
\begin{equation*}
v_1 \mid \text{rest} \sim \text{Beta}\left(\kappa + \sum_{j=1}^{k}\mathbbm{1}(\rho_j = 1), 1 + \sum_{j=1}^{k}\mathbbm{1}(\rho_j > 1)\right).
\end{equation*}

\textit{Step}.~15: For $h=2$ to $k-1$, sample $v_h$ from its conditional Beta distribution:
\begin{equation*}
v_h \mid \text{rest} \sim \text{Beta}\left(a + \sum_{j=1}^{k}\mathbbm{1}(\rho_j = h), 1 + \sum_{j=1}^{k}\mathbbm{1}(\rho_j > h)\right).
\end{equation*}
Set $v_k = 1$ and compute $\omega_1,\dots,\omega_k$ according to Eq.~(5) in the main text.

\textit{Step}.~16: For $h=1$ to $k$, if $\rho_h \leq h$, set $\theta_h = \theta_0$. Otherwise, sample $\theta_h$ from its conditional inverse gamma distribution:
\begin{equation*}
    \theta_h \mid \text{rest} \sim \text{IG}\left(a_{\theta} + \frac{n}{2}, b_{\theta} + \frac{1}{2}\|\mathbf{Z}_{\cdot h}\|^2_2\right).
\end{equation*}

\subsection{Definitions of competing model selection criteria}
Let $\{\xi^{(s)}\}_{s=1}^S$ denote the $S$ posterior samples from the Bayesian model under consideration. In addition, $\hat{\xi}$ denote the posterior mean of the parameters based on these posterior samples. Then the $\text{AIC}$ is defined as
\begin{equation*}
    \text{AIC} = -2 \log \Pr(A,Y \mid \hat{\xi}) + 2d, 
\end{equation*}
where $d$ is the number of the parameters, that is, $nk + qk + n + 2q$ for gaussian node varibales and $nk + qk + n + q$ for bernoulli node variables. Similarly, the $\text{BIC}$ is defined as 
\begin{equation*}
    \text{BIC} = -2\log\Pr(A,Y\mid\hat{\xi}) + 2d\log n,
\end{equation*}
where $d$ is same as in the $\text{AIC}$.  The $\text{DIC}$ is defined as 
\begin{equation*}
    \text{DIC} = -2\log\Pr(A,Y\mid \hat{\xi}) + 2p_{\text{DIC}},
\end{equation*}
where $p_{\text{DIC}} = 2\left(\log \Pr(A,Y\mid \hat{\xi}) - \frac{1}{S}\sum_{s=1}^{S}\log\Pr(A,Y\mid \xi^{(s)})\right)$. The $\text{WAIC}$ is defined as
\begin{equation*}
    \begin{aligned}
        \text{WAIC} &= -2\Bigg(\sum_{i=1}^{n}\sum_{i^\prime=1}^{i-1}\log\left(\frac{1}{S}\sum_{s=1}^S\Pr(A_{ii^\prime}\mid \xi^{(s)})\right) +\sum_{i=1}^{n}\sum_{j=1}^q\log\left(\frac{1}{S}\sum_{s=1}^S\Pr(Y_{ij}\mid\xi^{(s)})\right) \\
        & - \sum_{i=1}^{n}\sum_{i^\prime=1}^{i-1}Var\left(\log\Pr(A_{ii^\prime}\mid\xi)\right) - \sum_{i=1}^{n}\sum_{j=1}^q Var\left(\log\Pr(Y_{ij}\mid\xi)\right)\Bigg),
    \end{aligned}  
\end{equation*}
where $Var\left(\log\Pr(A_{ii^\prime}\mid\xi)\right)$ and $Var\left(\log\Pr(Y_{ij}\mid\xi)\right)$ denote the sample variance of the log-likelihood evaluated over the
 posterior samples $\{\xi^{(s)}\}_{s=1}^S$. Lastly,we
considered the K-fold cross-validation as follows:\
\begin{itemize}
    \item We randomly divide the nodes in the $n\times n$  adjacency matrix $A$ and $n\times q$ node variables $Y$  into $K$ equal sizes, using one set of data points as the test data and other sets of data points as the training data. 
    \item For each $k\in\{1,\dots,K\}$, we fit the training data to JLSM with the number of dimensions equaling to $k$. We obtain the posterior mean of the model parameters and use them as parameters to obtain predictions for the test data. 
    \item Based on the observed test data and the predicted probabilities for the test data, we obtain the held-out log-likelihood.  
\end{itemize}

\section{Theoretical Details and Proofs}\label{SM:proof}
\subsection{Preliminaries}
\subsubsection{Notations and Regularity Conditions}

Let $\{a_n\}$ and $\{b_n\}$ be two sequences of positive numbers.
\begin{itemize}
    \item We write $a_n \lesssim b_n$ or $b_n \gtrsim a_n$ if there exists a constant $C > 0$ such that $a_n \le C b_n$ for all sufficiently large $n$.
    \item We write $a_n \asymp b_n$ if $a_n \lesssim b_n$ and $a_n \gtrsim b_n$.
    \item For a real number $x$, $\lceil x \rceil$ denotes the smallest integer greater than or equal to $x$, and $\lfloor x \rfloor$ denotes the largest integer less than or equal to $x$.
\end{itemize}

Our main theoretical results rely on the following conditions regarding the growth of the model and the properties of the  data-generating process.

\begin{conditionS}[Hyperparameter]\label{con:Hyperparameter}
The Hyperparameter of the COSS prior satisfies:
    \begin{enumerate}[label=(\alph*)]
        \item $a>0$, $\kappa = k^{1 + \delta_n}$ where $\delta_n>0$ and $\delta_n\asymp n$.
        \item $a_\theta>2$ and $b_\theta>0$.
    \end{enumerate}    
\end{conditionS}    

\begin{assumptionS}[Growth Conditions]\label{ass:growth}
    The truncation level $k$, the number of attributes $q$, and the true dimension $k_0$ satisfy:
    \begin{enumerate}[label=(\alph*)]
        \item $k = \lceil n^\lambda \rceil$ for some constant $\lambda \in (0,1]$.
        \item $q \lesssim n$.
        \item $k_0 / \log k \to 0$ as $n \to \infty$.
    \end{enumerate}
\end{assumptionS}

\begin{assumptionS}[Boundedness of True Parameters]\label{ass:boundedness}
    There exist positive constants $C_\alpha, C_\gamma, C_Z, C_Z', C_B, C_B'$ such that the true parameters satisfy:
    \begin{enumerate}[label=(\alph*)]
        \item (Heterogeneity) $\max_{1 \le i \le n} |\alpha_{0,i}| \le C_\alpha$.
        \item (Intercepts) $\max_{1 \le j \le q} |\gamma_{0,j}| \le C_\gamma$.
        \item (Latent Positions) $\max_{i,h} |z_{0,ih}| \le C_Z$ and $\max_{i,i'} |\mathbf{z}_{0,i}^\top \mathbf{z}_{0,i'}| \le C_Z'$.
        \item (Loadings) $\max_{j,h} |\beta_{0,jh}| \le C_B$ and $\max_{i,j} |\mathbf{z}_{0,i}^\top \boldsymbol{\beta}_{0,j}| \le C_B'$.
    \end{enumerate}
\end{assumptionS}

\begin{assumptionS}[Regularity of the GLM Component]\label{ass:glm_reg}
    The function $f_Y(\cdot) = b(\cdot)
    $ from the exponential family formulation of the node attribute model satisfies:
    \begin{enumerate}[label=(\alph*)]
        \item (Bounded Variance) For any compact set $\mathcal{K} \subset \R$, there exists a constant $C_f$ such that $\sup_{\Theta \in \mathcal{K}} f_Y''(\Theta) \le C_f$.
        \item (Polynomial Tail Growth) Define $\psi_Y(L) = 1 + \sup_{|x| \le L(1+2L)} f_Y''(x)$. For any constants $M_1, M_2 > 0$, we have $\log \psi_Y(M_1 n^{M_2}) \lesssim \log n$. This is satisfied by Gaussian and Bernoulli (logit/probit link) models.
    \end{enumerate}
\end{assumptionS}

\subsection{Proof of Prior Properties (Proposition 1 and Corollary 1)}

\begin{proof}[Proof of proposition 1] 
The proof proceeds directly:
\begin{equation*}
\begin{aligned}
    \Pr(|\theta_h|>\epsilon) = \mathbb{E}[\Pr(|\theta_h|>\epsilon\mid \pi_h)] 
    = \mathbb{E}(1-\pi_h)P_{\text{slab}}(|\theta_h|>\epsilon).
\end{aligned}
\end{equation*}
Since $\mathbb{E}(1-\pi_h) = \frac{1}{
\kappa +1}\left(\frac{1}{a+1}\right)^{h-1}, (1\le h\le k-1)$ is decreasing in $h$ and $\pi_k=1$, the result follows.
\end{proof}

\begin{proof}[Proof of corollary 1] 
The proof proceeds directly:
\begin{equation*}
\begin{aligned}
    \Pr(|z_{ih}| \le \epsilon) = \mathbb{E}[\Pr(|z_{ih}| \le \epsilon\mid\pi_h)] 
    = \mathbb{E}(1-\pi_h)P_{\text{slab}}(|z_{ih}| \le \epsilon) + \mathbb{E}(\pi_h)P_{\text{spike}}(|z_{ih}| \le \epsilon).
\end{aligned}
\end{equation*}
Since $P_{\text{slab}}(|z_{ih}| \le \epsilon) > P_{\text{spike}}(|z_{ih}| \le \epsilon)$ if $b_\theta/a_\theta>\theta_0$ where $P_{\text{spike}} = t_{2a_\theta}(0,b_\theta/a_\theta)$ and $P_{\text{spike}} = \mathcal{N}(0,\theta_0)$, the result follows.
\end{proof}

\subsection{Proofs of Main Asymptotic Results}

\subsubsection{Technical Lemmas}

\paragraph{Outline of proofs}
The proofs of Theorems 1 and 2 follow the general framework for posterior concentration developed by \cite{10.1214/009053606000001172} and adapted for sparse generalized linear models  by \cite{10.1093/biomet/asaa074}. The main steps are as follows:
\begin{enumerate}
    \item We first show that the prior distribution assigns sufficient mass to a shrinking Kullback-Leibler (KL) neighborhood around the true parameter value $\boldsymbol{\Theta}_0$. This involves constructing a specific element in the support of the prior that is close to the truth and leveraging the properties of the cumulative shrinkage process.
    \item We define a sequence of subsets of the parameter space, called a ``sieve,'' indexed by the number of active latent dimensions. Lemma~\ref{lemma:priorh_supp} is critical for showing that the prior probability of parameter sets outside these sieves decays exponentially, allowing us to control the model complexity.
    \item For any two parameter sets $\boldsymbol{\Theta}_1$ and $\boldsymbol{\Theta}_2$, we construct uniformly consistent statistical tests to distinguish between them. The power of these tests depends on the Hellinger distance $H_n(\boldsymbol{\Theta}_1, \boldsymbol{\Theta}_2)$.
    \item By combining these three components, the general theory guarantees that the posterior probability of the parameter space outside a Hellinger ball of radius $\epsilon_n$ around the true value $\boldsymbol{\Theta}_0$ converges to zero in probability. Theorem 1 is obtained as an intermediate result in this process, where the posterior mass is shown to concentrate on the sieve corresponding to the true dimension $k_0$.
\end{enumerate}

The remaining results analyze properties of the posterior distribution. When possible,  we will prove the results for a general fractional posterior to simplify the proofs of parameter consistency in Section 5. As in \cite{10.1093/biomet/asaa074}, we write the fractional posterior as
\begin{equation*}
    \Pr(\mathcal{B}\mid \mathbf A, \mathbf Y) = \frac{N_n(\mathcal{\mathcal{B}})}{D_n},
\end{equation*}
where 
\begin{align}
    \label{eq:Nn}N_n(\mathcal{\mathcal{B}})&=\int_\mathcal{B}\prod_{i<i^\prime}\frac{p_{\Theta^{A}_{ii^\prime}}({A}_{ii^\prime})}{p_{\Theta^{A}_{0ii^\prime}}({A}_{ii^\prime})}\prod_{i,j}\frac{p_{\Theta^{Y}_{ij}}(Y_{ij})}{p_{\Theta^{Y}_{0ij}}(Y_{ij})} d\Pi(\boldsymbol{\alpha}) d\Pi(\boldsymbol{\gamma})d\Pi(\mathbf Z)d\Pi(\mathbf B),\\ \label{eq:Dn}
    D_n &= N_n(\mathbb{R}^n\times\mathbb{R}^q\times\mathbb{R}^{n\times k}\times\mathbb{R}^{q\times k}),
\end{align}
for any measurable set $\mathcal{B}\subset{\mathbb{R}^n\times\mathbb{R}^q\times\mathbb{R}^{n\times k}\times\mathbb{R}^{q\times k}}$ and prior distributions $\Pi(\boldsymbol{\alpha})$, $\Pi(\boldsymbol{\gamma})$, $\Pi(\mathbf Z)$ and $\Pi(\mathbf B)$ for $\boldsymbol{\alpha}$, $\boldsymbol{\gamma}$, $\mathbf Z$ and $\mathbf B$ respectively. Before proving Theorem 1 and Theorem 2, we need the following lemmas.

\begin{lemma}{(\cite{10.1093/acprof:oso/9780199535255.001.0001,10.3150/20-BEJ1198})}\label{lemma2}
Let \( X \sim \Gamma(a, b) \) and \( h_1(u) = 1 + u - \sqrt{1 + 2u} \) for \( u > 0 \). For every \( t > 0 \), we have

\[
Pr(X > t) \leq \exp\left\{-ah_1\left(\frac{t}{ab}\right)\right\} \leq \exp\left(-\frac{t}{2b} + a\right).
\]
\end{lemma}

\begin{lemma}{(Lemma S6 of \cite{loyal2025spike}})\label{lemma1}
    Let \( X_1, \ldots, X_n \) be independent Bernoulli random variables with \( \Pr(X_i = 1) = p_i \) for \( i = 1, \ldots, n \), such that \( p_1 > p_2 > \cdots > p_n \). Let \( S_n = \sum_{i=1}^n X_i \) and \( p = \sum_{i=1}^n p_i \), then
    $$
    \Pr(S_n > an) \leq \left\{ \left( \frac{p_1}{a} \right)^{a} e^{a} \right\}^n,
    $$
    for \( p_1 \leq a < 1 \).
\end{lemma}

\begin{lemma}[Prior Concentration on Sparsity]\label{lemma:priorh_supp}
    Let $\boldsymbol{\theta} = (\theta_1, \dots, \theta_k)$ follow the prior defined by (4)--(5) in the main text with $P_{\text{spike}}=\delta_0$. If the hyperparameters are set such that $\kappa = k^{1+\delta}$ with $\delta > 6/\log k$, then for any $t \ge 1$:
    \begin{equation*}
        \Pr(\|\boldsymbol{\theta}\|_0 > t) \le 2e^{-t(\delta/6)\log k}.
    \end{equation*}
    where $\|\boldsymbol{\theta}\|_0 = \sum_{h=1}^k \mathbbm{1}(\theta_h \neq 0)$ is the number of active (non-zero variance) dimensions.
\end{lemma}

\begin{proof}[proof of lemma~\ref{lemma:priorh_supp}] 
Since the inequality follows trivially when $t\ge k$, we assume $t<k$. Defined the random variable $S_k = \sum_{h=1}^k\mathbbm{1}\{\theta_h \ne 0\}$. We bounded $\Pr(S_k > t)$ as follows:
\begin{equation*}
\Pr(S_k > t) \le \mathbb{E}\left\{\Pr(S_k>t\mid \{\pi_h\}_{h=1}^k)\mathbbm{1}\{\pi_1 > 1-\tilde{\pi}\}\right\} + \Pr(\pi_1\le 1-\tilde{\pi}),
\end{equation*}
where $\tilde{\pi} = t\delta\log k/6k^{1+\delta}$. Condition on $\{\pi_h\}_{h=1}^k$, $S_k$ is sum of independent Bernoulli random variables with success probabilities
\begin{equation*}
p_h = \Pr(\theta_h \ne 0) = 1 - \pi_h,
\end{equation*}
with the property $p_1>p_1>\dots,>p_k$ by construction. In addition, on the event $\{\pi_1 > 1-\tilde{\pi}\}\}$, we have $p_1 = 1 - \pi_1 < t\delta\log k/6k^{1+\delta} = (t/6k)\delta\log k \exp(-\delta\log k)<t/k<1$, where we use the fact that  $e^{-x} < 1/x$ for $x>0$. As we apply  Lemma~\ref{lemma1} to bound $\Pr(S_k>t\mid \{\pi_h\}_{h=1}^k)$, we have
\begin{equation*}
    \begin{aligned}
        \Pr(S_k>t\mid \{\pi_h\}_{h=1}^k) &= \Pr(S_k>\frac{t}{k}k\mid \{\pi_h\}_{h=1}^k) \\
        &\le \left\{\left((1-\pi_1)\frac{k}{t}\right)^{t/k}e^{t/k}\right\}^k \\
        &\le \left(\frac{t\delta\log k}{6k^{1+\delta}}\frac{k}{t}\right)^\top  e^\top \\
        &= \left(\frac{\log k^{\delta/2}}{3k^\delta}\right)^\top  e^\top \\
        &\le \left(\frac{1}{3k^{\delta/2}}\right)^\top \\
        &\le e^{-t(\delta/2)\log k},
    \end{aligned}
\end{equation*}
where we use the fact that $\log x < (1/e)x$ for any $x>0$ in fifth line.

Now, we bound $\Pr(\pi_1\le 1-\tilde{\pi}) = \Pr(v_1\le 1-\tilde{\pi})$, where $v_1\sim\text{Beta}(k^{1+\delta},1)$. Then, we have,
\begin{equation*}
    \begin{aligned}
        \Pr(v_1\ge \tilde{\pi}) &=  k^{1+\delta}\int_0^{1-\tilde{\pi}} v^{k^{1+\delta}-1} dv\\  
        &=\left(1-\frac{t\delta\log k}{6k^{1+\delta}}\right)^{k^{1+\delta}}\\ 
        &= \left(1-\frac{t\delta\log k}{6k^{1+\delta}}\right)^{\frac{k^{1+\delta}}{\log k}\log k} \\
        &\le e^{-t(\delta/6)\log k},
    \end{aligned}
\end{equation*}
where we used the fact that $(1-b/x)^x\le e^{-b}$ for $b\in\mathbb{R}$ and $x>0$ in the last line.
\end{proof}

\begin{lemma}\label{lemma:A}
    Suppose that the conditions of Theorem 1 hold. Then $\Pr(\mathcal{A}_n^{C}) = o(1)$ where the event $\mathcal{A}_n$ is 
    \begin{equation*}
        \mathcal{A}_n = \left\{A_{ii^\prime}(1\le i <i^\prime\le n), Y_{ij}(1\le i\le n,1\le j\le q): D_n \ge e^{-C_1n(n+q)\epsilon_n^2}\right\},
    \end{equation*}
    for 
\begin{equation*}
    \epsilon_n = \sqrt{\frac{\delta_nk_0\log k}{n(n+q)}},
\end{equation*}
for some constant $C_1>1$.
\end{lemma}

\begin{proof}[proof of lemma~\ref{lemma:A}]  Denote the KL-neighborhood for the model parameters by
\begin{equation*}
\begin{aligned}
     \mathcal{B}_n = \Bigg\{(\boldsymbol{\alpha},\boldsymbol{\gamma},\mathbf Z,\mathbf B): \frac{2}{n(n+2q-1)}&\left(\sum_{i<i^\prime}K(p^{A}_{ii^\prime,},p^{A}_{0ii^\prime}) + \sum_{i,j}K(p_{ij}^{Y},p_{0ij}^{Y})\right)\le \epsilon_n^2, \\
     &\frac{2}{n(n+2q-1)}\left(\sum_{i<i^\prime}V(p^{A}_{ii^\prime,},p^{A}_{0ii^\prime}) + \sum_{i,j}V(p_{ij}^{Y},p_{0ij}^{Y})\right)\le \epsilon_n^2\Bigg\}.    
\end{aligned}
\end{equation*}
For two densities $p$ and $q$, we let $K(p,q) = \mathbb{E}_p\{\log(p/q)\}$ and $V(p,q) = \mathbb{E}_p\{\log(p/q) - K(p,q)\}^2$ denote the Kullback-Leibler (KL) divergence and second moment of the KL ball, respectively.

By Lemma~10 of \cite{10.1214/009053606000001172}, we have for any $C>0$, 

\begin{equation*}
    \Pr\left\{D_n \ge e^{-(1 + C)n(n+q)\epsilon_n^2}\Pr(\mathcal{B}_n)\right\} \ge 1 - \frac{1}{C^2n(n+q)\epsilon_n^2},
\end{equation*}

Therefore,it suffices to show that the prior probability $\Pr(\mathcal{B}_n)\ge e^{-C_2n(n+q)\epsilon_n^2}$ for some $C_2>0$. By Taylor expanding $K(p^{A}_{ii^\prime,},p^{A}_{0ii^\prime})$, $K(p_{ij}^{Y},p_{0ij}^{Y})$, $V(p^{A}_{ii^\prime,},p^{A}_{0ii^\prime})$, and $V(p_{ij}^{Y},p_{0ij}^{Y})$. Lemma~1 of \cite{10.1093/biomet/asaa074} established that

\begin{align*}
    \max\left\{K(p^{A}_{ii^\prime,},p^{A}_{0ii^\prime}),V(p^{A}_{ii^\prime,},p^{A}_{0ii^\prime})\right\} &= f^{\prime\prime}_{A}(\Theta^{A}_{0ii^\prime})(\Theta^{A}_{ii^\prime}-\Theta_{0ii^\prime}^{A})^2 + o\{(\Theta^{A}_{ii^\prime}-\Theta_{0ii^\prime}^{A})^2\}, \\
    \max\left\{V(p^{Y}_{ij,},p^{A}_{0ij}),V(p^{Y}_{ij,},p^{A}_{0ij})\right\} &= \frac{f^{\prime\prime}_{Y}(\Theta^{Y}_{0ij})}{\phi}(\Theta^{Y}_{ij}-\Theta_{0ij}^{Y})^2 + o\{(\Theta^{Y}_{ij}-\Theta_{0ij}^{Y})^2\}, 
\end{align*}
where $f_{A}(x) = \log(1 + \exp(x))$. 

$f^{\prime\prime}_{A}(\Theta^{A}_{0ii^\prime})$ is bounded by Assumptions~\ref{ass:boundedness}. And  $f^{\prime\prime}_{Y}(\Theta^{Y}_{0ij}) \le C_{f}$ by Assumptions~\ref{ass:boundedness} and \ref{ass:glm_reg}. Therefore, $K(p^{A}_{ii^\prime,},p^{A}_{0ii^\prime})$ and $V(p^{A}_{ii^\prime,},p^{A}_{0ii^\prime})$  can be bounded above by a constant multiple of $(\Theta^{A}_{ii^\prime}-\Theta_{0ii^\prime}^{A})^2$ for sufficiently large $n$. 
Thus, we have for some constant $b_1>0$ which depends on $C_{{Z}}^\prime$, $C_{B}^\prime$, $C_{{\alpha}}$, $C_{\gamma}$ and $C_{f}/\phi$, and sufficiently large $n$,

\begin{equation}\label{eq:pB}
\begin{aligned}
    &\Pr(\mathcal{B}_n) \ge \Pr\left(\sum_{i<i^\prime}(\Theta^{A}_{ii^\prime}-\Theta_{0ii^\prime}^{A})^2 + \sum_{i,j}(\Theta^{Y}_{ij}-\Theta_{0ij}^{Y})^2 \le b_1n(n+q)\epsilon_n^2 \right) \\
    \ge & \Pr\left(\sum_{i,i^\prime}(\Theta^{A}_{ii^\prime}-\Theta_{0ii^\prime}^{A})^2 + \sum_{i,j}(\Theta^{Y}_{ij}-\Theta_{0ij}^{Y})^2 \le b_1n(n+q)\epsilon_n^2 \right) \\
    = &\Pr\Big(\left\|\boldsymbol{\alpha}\boldsymbol{1}_n^\top  + \boldsymbol{1}_n\boldsymbol{\alpha}^\top  + \mathbf Z\mathbf Z^\top  - \boldsymbol{\alpha}_0\boldsymbol{1}_n^\top  - \boldsymbol{1}_n\boldsymbol{\alpha}_0^\top  - \mathbf{Z}_0\mathbf{Z}_0^\top \right\|_F^2 + 
    \left\|\boldsymbol{1}_n\boldsymbol{\gamma}^\top  + \mathbf Z\mathbf B^\top  - \boldsymbol{1}_n\boldsymbol{\gamma}_0^\top  - \mathbf{Z}_0\mathbf{B}_0^\top \right\|_F^2 \le b_1n(n+q)\epsilon_n^2 \Big) \\
    \ge &\Pr\Big(\left\|\boldsymbol{\alpha}\boldsymbol{1}_n^\top  + \boldsymbol{1}_n\boldsymbol{\alpha}^\top  + \mathbf Z\mathbf Z^\top  - \boldsymbol{\alpha}_0\boldsymbol{1}_n^\top  - \boldsymbol{1}_n\boldsymbol{\alpha}_0^\top  - \mathbf{Z}_0\mathbf{Z}_0^\top \right\|_F + 
    \left\|\boldsymbol{1}_n\boldsymbol{\gamma}^\top  + \mathbf Z\mathbf B^\top  - \boldsymbol{1}_n\boldsymbol{\gamma}_0^\top  - \mathbf{Z}_0\mathbf{B}_0^\top \right\|_F \le \sqrt{b_1n(n+q)}\epsilon_n \Big)\\
    \ge & \Pr\left(\|\boldsymbol{\alpha}\boldsymbol{1}_n^\top  + \boldsymbol{1}_n\boldsymbol{\alpha}^\top  - \boldsymbol{\alpha}_0\boldsymbol{1}_n^\top  - \boldsymbol{1}_n\boldsymbol{\alpha}_0^\top \|_F\le \sqrt{\frac{b_1n(n+q)}{16}}\epsilon_n\right) \times \Pr\left(\|\boldsymbol{1}_n\boldsymbol{\gamma}^n - \boldsymbol{1}_n\boldsymbol{\gamma}_0^n\|_F\le\sqrt{\frac{b_1n(n+q)}{16}}\epsilon_n\right) \times  \\
    &\qquad \qquad \qquad \qquad 
    \Pr\left(\|\mathbf Z\mathbf Z^\top -\mathbf{Z}_0\mathbf Z^\top _0\|_F + \|\mathbf Z\mathbf B^\top  - \mathbf{Z}_0\mathbf{B}_0^\top \|_F \le \sqrt{\frac{b_1n(n+q)}{4}}\epsilon_n\right),
\end{aligned}
\end{equation}
where the last line follows from the triangle inequality and the fact that $\boldsymbol{\alpha}$, $\boldsymbol{\gamma}$, $\mathbf{Z}$ and $\mathbf{B}$ is independent of the remaining variables under our prior specification.

Let $\tilde{\mathbf Z}_0 = [\mathbf{Z}_0,\boldsymbol{0}_{n,k-k_0}]\in \mathbb{R}^{n\times k}$ and $\tilde{\mathbf B}_0 = [\mathbf{B}_0,\boldsymbol{0}_{q,k-k_0}]\in\mathbb{R}^{q\times k}$ where $\boldsymbol{0}_{n,k-k_0}$ and $\boldsymbol{0}_{q,k-k_0}$ are $n\times (k-k_0)$ and $q\times(k-k_0)$ matrix of zero. Then, we have 
\begin{align*}
\|\mathbf Z\mathbf Z^\top  - \mathbf{Z}_0\mathbf{Z}_0^\top \|_F &\le \|\mathbf Z-\tilde{ \mathbf Z}_0\|_F^2 + 2\|\mathbf{Z}_0\|_2\|\mathbf Z-\tilde{\mathbf Z}_0\|,\\    
\|\mathbf Z\mathbf B^\top  - \mathbf{Z}_0\mathbf{B}_0^\top \|_F &\le \|\mathbf Z-\tilde{\mathbf Z}_0\|_F\|\mathbf B-\tilde{\mathbf B}_0\|_F + \|\mathbf Z_0\|_2\|\mathbf B-\tilde{\mathbf B}_0\|_F + \|\mathbf{B}_0\|_2\|\mathbf Z-\tilde{\mathbf Z}_0\|_F.
\end{align*}
Combining this with the previous inequality and using Assumptions~\ref{ass:boundedness}. There exists a constant $b_2>0$ such that $2\|\mathbf{Z}_0\|_2 + \|\mathbf{B}_0\|_2 \le b_2\sqrt{nk_0}$ for sufficiently large $n$. Let $\mathbf{Z}_{h}$ and $\mathbf{B}_{h}$ denote the matrix containing the first $h$ columns of $\mathbf{Z}$ and $\mathbf{B}$. Then, We have
\begin{equation*}
\begin{aligned}
    &\Pr\Bigg(\|\mathbf Z \mathbf Z^\top  -\mathbf{Z}_0 \mathbf Z^\top _0\|_F +\| \mathbf Z \mathbf B^\top  - \mathbf{Z}_0\mathbf{B}_0^\top \|_F \le \sqrt{\frac{b_1n(n+q)}{4}}\epsilon_n\Bigg) \\ 
    \ge &\Pr\left(\|\mathbf Z-\tilde{\mathbf Z}_0\|_F + \|\mathbf B-\tilde{\mathbf B}_0\|_F\le \sqrt{\frac{b_1(n+q)}{16b_2k_0}}\epsilon_n\right)\\
    = &\mathbb{E}\left\{\Pr\left(\|\mathbf Z-\tilde{\mathbf Z}_0\|_F + \|\mathbf B-\tilde{\mathbf B}_0\|_F\le \sqrt{\frac{b_1(n+q)}{16b_2k_0}}\epsilon_n\mid \{\pi_h\}_{h=1}^{k}\right)\right\} \\
    \ge & \mathbb{E}\left\{\Pr\left(\|\mathbf Z_{k_0}-\mathbf Z_{0}\|_F\le\sqrt{\frac{b_1(n+q)}{64b_2k_0}}\epsilon_n\right)\Pr\left(\|\mathbf B_{k_0} - \mathbf B_{0}\|_F\le\sqrt{\frac{b_1(n+q)}{64b_2k_0}}\epsilon_n\right)\prod_{h = 1}^{k_0}(1-\pi_h)\prod_{h=k_0+1}^{k}\pi_h\right\}\\
    \ge &\mathbb{E}\left\{\prod_{i=1}^n\prod_{h=1}^{k_0}\Pr\left(|z_{ih}-z_{0ih}|\le\sqrt{\frac{b_1(n+q)}{64b_2k^2_0n}}\epsilon_n\right)\prod_{j=1}^q\prod_{h=1}^{k_0}\Pr\left(|\beta_{jh}-\beta_{0jh}|\le\sqrt{\frac{b_1(n+q)}{64b_2k^2_0q}}\epsilon_n\right)(1-\pi_{k_0})^{k_0}\pi_1^{k-k_0}\right\}.
\end{aligned}
\end{equation*}

We have that $\sqrt{b_1(n+q)/64k_0^2n} > b_3/k_0$. some constant $b_3>0$ and $n$ sufficiently large. Thus,
\begin{equation*}
\begin{aligned}
    &\Pr\left(|z_{ih}-z_{0ih}|\le\sqrt{\frac{b_1(n+q)}{64b_2k^2_0n}}\epsilon_n\right)\ge \Pr\left(|z_{ij}-z_{0ih}|\le\frac{b_3\epsilon_n}{k_0}\right) \\
    = & \frac{a_\theta\Gamma\left(({2a_{\theta}-1})/{2}\right)}{b_\theta\Gamma(a_\theta)\sqrt{2\pi a_\theta}}\int_{|z_{ij}-z_{0ih}|\le{b_3\epsilon_n}/{k_0}}\left(1 + \frac{1}{2a_\theta}\left(\frac{a_\theta z_{ih}}{b_\theta}\right)^2\right)^{(2a_\theta+1)/2}dz_{ih} \\
    \ge & \frac{a_\theta\Gamma\left(({2a_{\theta}-1})/{2}\right)}{b_\theta\Gamma(a_\theta)\sqrt{2\pi a_\theta}}\left(1 + \frac{1}{2a_\theta}\left(\frac{3a_\theta C_Z}{2b_\theta}\right)^2\right)^{(2a_\theta+1)/2}\int_{|z_{ij}-z_{0ih}|\le{b_3\epsilon_n}/{k_0}}dz_{ih}\\
    \ge & \frac{b_4\epsilon_n}{k_0},
\end{aligned}
\end{equation*}
for some constant $b_4>0$. Similarly, we have  $\sqrt{b_1(n+q)/64k_0^2q} > b_5/k_0$ for some constant $b_5>0$ and $n$ sufficiently large. Thus, 
\begin{equation*}
\begin{aligned}
    &\Pr\left(|\beta_{jh}-\beta_{0jh}|\le\sqrt{\frac{b_1(n+q)}{64b_2k^2_0q}}\epsilon_n\right) \ge \Pr\left(|\beta_{jh}-\beta_{0jh}|\le\frac{b_5\epsilon_n}{k_0}\right) \\
    =&(2\pi\sigma_B)^{-1/2}\int_{|\beta_{jh}-\beta_{0jh}|\le b_5\epsilon_n/k_0}\exp\left(\frac{-\beta_{jh}^2}{2\sigma_B^2}\right)d \beta_{jh}\\
    \ge &(2\pi\sigma_B)^{-1/2}\exp\left(\frac{-9C_B^2}{8\sigma_B^2}\right) \int_{|\beta_{jh}-\beta_{0jh}|\le b_5\epsilon_n/k_0} d \beta_{jh}\\
    \ge &\frac{b_6\epsilon_n}{k_0},
\end{aligned}
\end{equation*}
for some constant $b_4>0$. Furthermore, by jensen inequality we have
\begin{equation*}
\begin{aligned}
    \mathbb{E}\{\pi_1^{k-k_0}(1-\pi_{k_0})^{k_0}\} \ge \left\{\mathbb{E}\pi_1^{(k-k_0)/k_00}(1-\pi_{k_0})\right\}^{k_0}.
\end{aligned}
\end{equation*}
Then for some constant $b_7 > 0$ and $n$ sufficiently large, we have
\begin{equation*}
\begin{aligned}
    \mathbb{E}\left\{\pi_1^{(k-k_0)/k_0}(1-\pi_{k_0})\right\} &= \mathbb{E}\left\{v_1^{(k-k_0)/k_0}\left(1-\sum_{h=1}^{k_0}v_h\prod_{l=1}^{h-1}(1-v_l)\right)\right\} \\
    &=\mathbb{E}\left\{v_1^{(k-k_0)/k_0}(1-v_1)\left[1-v_2 - v_3(1-v2)-,\dots,-v_{k_0}\prod_{l=2}^{k_0-1}(1-v_l)\right]\right\}\\
    &= \frac{k^{1+\delta^n}}{\left(k^{1+\delta_n}+(k-k_0)/k_0 + 1\right)\left(k^{1+\delta_n}+(k-k_0)/k_0\right)}\left(\frac{1}{a+1}\right)^{k_0-1}\\
    &\ge \frac{b_7}{k^{1+\delta_n}}\left(\frac{1}{a+1}\right)^{k_0-1}. 
\end{aligned}
\end{equation*}
Hence, for sufficiently large $n$ we have 
\begin{equation*}
\begin{aligned}
    &\Pr\Bigg(\|\mathbf Z\mathbf Z^\top  -\mathbf{Z}_0\mathbf Z^\top _0\|_F +\|\mathbf Z\mathbf B^\top  - \mathbf{Z}_0\mathbf{B}_0^\top \|_F \le \sqrt{\frac{b_1n(n+q)}{4}}\epsilon_n\Bigg) \\ 
    \ge & \exp\left(-\frac{1}{2}(nk_0 + qk_0)\log (k_0n(n+q) - \delta_n\log k - 2(b_4+b_6))-(1+\delta_n)k_0\log k - k_0(k_0-1)\log (a+1) \right) \\
    \ge & \exp (-\tilde{C}\delta_nk_0\log n -(1+\delta_n)k_0\log k)
    \ge \exp(-(\tilde{C}/\lambda)\delta_nk_0\log k -(1+\delta_n)k_0\log k)\\
    \ge&\exp(-C_3\delta_nk_0\log k)\\
    \ge&\exp(-C_3n(n+q)\epsilon_n^2).
\end{aligned}
\end{equation*}
for some universal constant $\tilde{C}$ and $C_3 = \tilde{C}/ \lambda + 2$.  Then
\begin{equation*}
\begin{aligned}
&\Pr\left(\|\boldsymbol{\alpha}\boldsymbol{1}_n^\top  + \boldsymbol{1}_n\boldsymbol{\alpha}^\top  - \boldsymbol{\alpha}_0\boldsymbol{1}_n^\top  + \boldsymbol{1}_n\boldsymbol{\alpha}_0^\top \|_F \le \sqrt{\frac{b_1n(n+q)}{16}}\epsilon_n\right) \\ 
\ge & \Pr\left(\|(\boldsymbol{\alpha - \alpha_0})\|_2\le\sqrt{\frac{b_1(n+q)}{64}}\epsilon_n\right)\\
\ge & \prod_{i=1}^n \Pr(|\alpha_i - \alpha_{0i}|\le \sqrt{\frac{b_1(n+q)}{64n}}\epsilon_n)\\
\ge &\left(b_8\epsilon_n\right)^n 
= \exp(-(n/2)\log n(n+q) + (n/2)\log(\delta_nk_0\log k + 2b_8))\\
\ge &\exp(-b_9\delta_n\log n)
\ge \exp(-(b_9/\lambda)\delta_n\log k)\\
\ge &\exp(-C_4n(n+q)\epsilon_n^2).
\end{aligned}
\end{equation*}

Similarly, we have
\begin{equation*}
\begin{aligned}
&\Pr\left(\|\boldsymbol{1}_n\boldsymbol{\gamma}^n - \boldsymbol{1}_n\boldsymbol{\gamma}_0^n\|_F\le\sqrt{\frac{b_1n(n+q)}{16}}\epsilon_n\right)\\
\ge&\prod_{j=1}^q \Pr\left(|\gamma_j-\gamma_{0j}|\le\sqrt{\frac{b_1(n+q)}{16q}}\epsilon_n\right)\\
\ge &\left(b_{10}\epsilon_n\right)^q
 = \exp(-q/2\log n(n+q) + q/2\log(\delta_nk_0\log k + 2b_{10}))\\
\ge &\exp(-b_{11}\delta_n\log n)
\ge \exp(-(b_{11}/\lambda)\delta_n\log k)\\
\ge &\exp(-C_5n(n+q)\epsilon_n^2).
\end{aligned}
\end{equation*}

Combining the previous three bounds for the probabilities in Eq.~(\ref{eq:pB}) with $C_1 = \max(C_3,C_4,C_5)$ gives the desired bound on $\Pr(\mathcal{B}_n)$. The result follows $C_1 = 1+C + C_2$. 
\end{proof}

\subsubsection{Proof of Theorem 1 (Posterior Concentration for the Latent Dimension)}
\begin{proof}[Proof of Theorem 1]
We prove this result for a general fractional posterior. We write the fractional posterior probability as 
\begin{equation}\label{eq:post}
    \Pr(K^* > Ck_0\mid \mathbf A , \mathbf Y) = \frac{N_n(\{\mathbf{Z}:K^* > Ck_0\})}{D_n},
\end{equation}
where $N_n(\cdot)$ and $D_n$ are defined in Eq.~(\ref{eq:Nn}) and (\ref{eq:Dn}). Now, we introduce an event $\mathcal{A}_n$  with large probability under the true data generating process. In particular, $\mathcal{A}_n = \{\mathbf A, \mathbf Y:D_n\ge e^{-C_1n(n+q)\epsilon_n^2}\}$, where $\epsilon_n$ is the rate in Lemma~\ref{lemma:A}. If we decompose the probability in Eq.~(\ref{eq:post}) into the sum of two complementary conditional probabilities (conditioning on $\mathcal{A}_n$ and $\mathcal{A}_n^C$), we can write:
\begin{equation*}
    \mathbb{E}_0\left\{\Pr(K^* > Ck_0\mid \mathbf A , \mathbf Y)\right\} \le \mathbb{E}_0\left\{\Pr(K^* > Ck_0\mid \mathbf  A ,  \mathbf Y)\mathbbm{1}(\mathcal{A}_n)\right\} + \Pr(\mathcal{A}_n^C).
\end{equation*}
On the event $\mathcal{A}_n$, $D_n$ is bounded below by $e^{-C_1\delta_nk_0\log k}$. On the other hand, the expected value of the numerator can by bounded above by $\Pr(K^* > Ck_0)$ using Fubini’s theorem
\begin{equation*}
\begin{aligned}
    &\mathbb{E}_0\left\{\int\int\int_{K^* > Ck_0}\prod_{i<i^\prime}\frac{p_{\Theta^{A}_{ii^\prime}}(A_{ii^\prime})}{p_{\Theta^{A}_{0ii^\prime}}(A_{ii^\prime})}\prod_{i,j}\frac{p_{\Theta^{Y}_{ij}}(Y_{ij})}{p_{\Theta^{Y}_{0ij}}(Y_{ij})} d\Pi(\boldsymbol{\alpha}) d\Pi(\boldsymbol{\gamma})d\Pi(\mathbf Z)d\Pi(\mathbf B)\right\} \\
    =& \int\int\int_{K^* > Ck_0}\prod_{i<i^\prime}\prod_{i,j}\mathbb{E}_0\left\{\frac{p_{\Theta^{A}_{ii^\prime}}(A_{ii^\prime})}{p_{\Theta^{A}_{0ii^\prime}}(A_{ii^\prime})}\frac{p_{\Theta^{Y}_{ij}}(Y_{ij})}{p_{\Theta^{Y}_{0ij}}(Y_{ij})}\right\} d\Pi(\boldsymbol{\alpha}) d\Pi(\boldsymbol{\gamma})d\Pi(\mathbf Z)d\Pi(\mathbf B) \le  \Pr(K^* > Ck_0)%
\end{aligned}
\end{equation*}
Therefore, we can use Lemma~\ref{lemma:A} to conclude that
\begin{equation*}
    \mathbb{E}_0\left\{\Pr(K^* > Ck_0\mid \mathbf A , \mathbf Y)\right\} \le \Pr(K^* > Ck_0)e^{C_1\delta_nk_0\log k} + o(1), 
\end{equation*}
In addition, for $n$ sufficiently large we have that $\delta_n>6/\log k$, since $k=\lceil n^\lambda \rceil$ by Assumption~\ref{ass:growth}, so we can apply Lemma~\ref{lemma:priorh_supp} to bound the previous expression from above as 
\begin{equation*}
    \mathbb{E}_0\left\{\Pr(K^* > Ck_0\mid A , Y)\right\} \le 3e^{-(C/6-C_1)\delta_nk_0\log k} + o(1),
\end{equation*}
for $n$ large enough,which goes to zero as $n\rightarrow\infty$ and $C>\max\{6C_1,1\}$. Lastly, $C_1$ increases with the constants $C_3$, $C_4$ and $C_5$ defined in the proof of Lemma~\ref{lemma:A} which depend on the type of node variable and increase with $\lambda^{-1}$.
\end{proof}

\subsubsection{Proof of Theorem 2 (Posterior Contraction Rate)}
\begin{proof}[Proof of theorem 2]
Define the contraction rate $\epsilon_n = \left(k_0\log k / n\right)^{1/2}$ and the event $\mathcal{E}_n = \{(\boldsymbol{\alpha},\boldsymbol{\gamma},\mathbf Z,\mathbf B): K^* \le C_1k_0\}$  for $C_1$ in Theorem 1. For a large enough constant $C_2>0$, we have that
\begin{equation*}
\begin{aligned}
\mathbb{E}_0&\left[\Pr\left\{(\boldsymbol{\alpha},\boldsymbol{\gamma},\mathbf Z,\mathbf B): H_n(\boldsymbol\Theta,\boldsymbol\Theta_0)>C_2\epsilon_n\mid \mathbf A,\mathbf Y\right\}\right]\\
&\le\mathbb{E}_0\left[\Pr\left\{(\boldsymbol{\alpha},\boldsymbol{\gamma},\mathbf Z,\mathbf B)\in\mathcal{E}_n: H_n(\boldsymbol\Theta,\boldsymbol\Theta_0)>C_2\epsilon_n\mid \mathbf A,\mathbf Y\right\}\mathbbm{1}(\mathcal{A}_n)\right]  + \mathbb{E}_0\{\Pr(\mathcal{E}_n^C\mid \mathbf A, \mathbf Y)\} + \Pr(\mathcal{A}_n^C),  
\end{aligned}
\end{equation*}
where $\mathcal{A}_n = \{\mathbf A,\mathbf Y:D_n\ge e^{-C_1\delta_n k_0\log k }\}$  is the event in Lemma~\ref{lemma:A}. Also, on the event $\mathcal{A}_n$ the denominator $D_n\ge e^{-C_1^\prime n k_0\log k} > e^{-C_1^\prime n(n+q)\epsilon^2}$ for some constant $C_1^\prime$. By Theorem 1 and Lemma~\ref{lemma:A}, the last two terms tend to zero as $n\rightarrow\infty$. It remains to show that the  first term also goes to zero. To do so,we use the prior-mass-and-testing technique introduced in \cite{10.1214/aos/1016218228}, which involves constructing an appropriate sieve $\mathcal{F}_n\subset{\mathcal{E}_n}$ and a uniformly
 consistent sequence of test functions $\Phi_n$  such that for all $\epsilon>\epsilon_n$, 
\begin{equation}
\begin{aligned}
\Pr(\mathcal{E}_n \setminus \mathcal{F}_n) &\lesssim \exp\{-(1 + C_1') n(n+1) \epsilon_n^2\},\\   
\mathbb{E}_0(\Phi_n) &\lesssim \exp(-Cn(n+1) \epsilon_n^2),\\
\sup_{(\boldsymbol{\alpha}, \boldsymbol{\gamma},\mathbf{Z}, \mathbf{B}) \in \mathcal{F}_{n} : H_{n} (\boldsymbol{\Theta}, \boldsymbol{\Theta}_{0}) > \epsilon} \mathbb{E}_{\Theta} (1 - \Phi_n) &\lesssim \exp(-Cn(n+1) \epsilon_n^2),
\end{aligned}
\end{equation}
where $\mathbb{E}_\Theta$ denotes the expectation under the model with parameters $(\boldsymbol{\alpha}, \boldsymbol{\gamma},\mathbf Z, \mathbf B)$ and $\boldsymbol\Theta  = [\boldsymbol{\alpha}\boldsymbol{1}_n^\top  + \boldsymbol{1}_n\boldsymbol{\alpha}^\top  + \mathbf Z\mathbf Z^\top , \boldsymbol{1}_n\boldsymbol{\gamma}^\top  + \mathbf Z\mathbf B^\top ]$ and the constant $C>C_1^\prime$.

Before constructing such a sieve and test functions $\Phi_n$, we demonstrate that their existence implies the Hellinger consistency of the posterior distribution. We use a standard argument originally \cite{schwartz1965bayes}. Since $\Phi_n$ and the posterior probability are less than or equal to one,
 we have that
\begin{equation}\label{eq:test}
\begin{aligned}
\mathbb{E}_0&\left[\Pr\left\{(\boldsymbol{\alpha},\boldsymbol{\gamma},\mathbf Z,\mathbf B)\in\mathcal{E}_n: H_n(\boldsymbol\Theta,\boldsymbol\Theta_0)>C_2\epsilon_n\mid \mathbf A,\mathbf Y\right\}\mathbbm{1}(\mathcal{A}_n)\right]  \\
&\le\mathbb{E}_0\left[\Pr\left\{(\boldsymbol{\alpha},\boldsymbol{\gamma},\mathbf Z,\mathbf B)\in\mathcal{E}_n: H_n(\boldsymbol\Theta,\boldsymbol\Theta_0)>C_2\epsilon_n\mid \mathbf A,\mathbf Y\right\}\mathbbm{1}(\mathcal{A}_n)(1-\Phi_n)\right] + \mathbb{E}_0(\Phi_n)\\
&\le \left\{\sup_{(\boldsymbol{\alpha}, \boldsymbol{\gamma},\mathbf Z, \mathbf B) \in \mathcal{F}_n : H_n (\boldsymbol\Theta, \boldsymbol\Theta_0) > C_2\epsilon_n} \mathbb{E}_{\Theta} (1 - \Phi_n) +\Pr(\mathcal{E}_n \setminus \mathcal{F}_n)\right\}e^{C_1^\prime n(n+q)\epsilon^2} +\mathbb{E}_0(\Phi_n),
\end{aligned}  
\end{equation}
where the last inequality follows from an application of Fubini’s theorem and the definition of the event \(\mathcal{A}_n\). We have that \({\Pr}(\mathcal{E}_n \setminus \mathcal{F}_n) e^{C_1' n(n+1) \epsilon_n^2} \lesssim e^{-n(n+1) \epsilon_n^2}\), which goes to zero as \(n \to \infty\). Since \(C > C_1'\), the remaining two terms go to zero as long as \(C_2 > 1\). To complete the proof, we proceed to construct the appropriate sieve and test functions satisfying (\ref{eq:test}).

\textit{Sieve construction}. First,we introduce some notation.
for  matrix $\mathbf Z\in\mathbb{R}^{n\times k}$ and $\mathbf B\in\mathbb{R}^{q\times k}$ , and a set $S\subset\{1,\dots,k\}$, we use the notation $\mathbf Z_S = (\mathbf Z_j)_{j\in S}\in\mathbb{R}^{n\times |S|}$ and $\mathbf B_S = (\mathbf B_j)_{j\in S}\in\mathbb{R}^{q\times |S|}$, that the matrix constructed from $\mathbf Z$ and $\mathbf B$  by including only columns in $S$. Define the the sieve 
\begin{equation*}
\begin{aligned}
    \mathcal{F}_n = \{(\boldsymbol{\alpha}, \boldsymbol{\gamma},\mathbf Z,\mathbf B)&\in\mathbb{R}^n\times\mathbb{R}^q\times\mathbb{R}^{n\times k}\times\mathbb{R}^{q\times k}:\\
    &\|\boldsymbol{\alpha}\|_2\le L_n, \|\boldsymbol{\gamma}\|_2\le L_n,\|\mathbf Z\|_F\le L_n,\|\mathbf B\|_F\le L_n,K^*\le C_1k_0\}\subset{\mathcal{E}_n}.
\end{aligned}
\end{equation*}
where $L_n = M_1n^{M_2}$ and $M_1,M_2>0$, are constants chosen subsequently so that the prior probability of falling out side the sieve is bounded above by $e^{-(1+C_1^\prime)n(n+q)\epsilon_n^2}$. Under the prior, we have that
\begin{equation*}
    \Pr(\mathcal{E}_n\setminus\mathcal{F}_n) \le \Pr(\|\boldsymbol{\alpha}\|_2>L_n) + \Pr(\|\boldsymbol{\gamma}\|_2>L_n) + \Pr(\|\mathbf Z\|_F>L_n,K^*\le C_1k_0) + \Pr(\|\mathbf B\|_F>L_n,K^*\le C_1k_0),
\end{equation*}
We bound each probability in turn starting with the second term. We have that 
\begin{equation*}
    \Pr(\|\mathbf Z\|_F>L_n,K^*\le C_1k_0) = \mathbb{E}\{\Pr(\|\mathbf Z\|_F>L_n,K^*\le C_1k_0\mid \{\pi_h\}_{h=1}^k)\},
\end{equation*}
where
\begin{equation}\label{eq:sez}
\begin{aligned}
    \Pr(\|\mathbf Z\|_F>L_n,K^*\le C_1k_0\mid \{\pi_h\}_{h=1}^k) 
    &= \sum_{S:1\le|S|\le C_1k_0} \Pr(S\mid\{\pi_h\}_{h=1}^k)\Pr(\|\mathbf Z_S\|_F>L_n) \\
    &\le  \sum_{S:1\le|S|\le C_1k_0}\Pr(\|\mathbf Z_S\|_F>L_n),
\end{aligned}
\end{equation}
Since $\|\mathbf Z_S\|_F^2 = \sum_{i=n}\sum_{j\in S}z_{ij}^2$ is a sum of independent and identically distributed random variables with expectation $b_{z1} = b_\theta / (a_\theta - 1)$ and variance $b_{z2} = 3b_\theta^2 / ((a_\theta-1)(a_\theta-2)$. Then for $n$ lager enough
\begin{equation*}
    \Pr(\|\mathbf Z_S\|_F>L_n) = \Pr\left(\frac{\sum_{i=n}\sum_{j\in S}z_{ij}^2 - n|S|b_{z1}}{\sqrt{n|S|b_{z2}}} > \frac{L_n^2 - n|S|b_{z1}}{\sqrt{n|S|b_{z2}}}\right) \le \exp\left\{-\frac{(L_n^2 - n|S|b_{z1})^2}{2n|S|b_{z2}}\right\},
\end{equation*}
We can choose $M_2>1/2$, there exist constant $b_1$ for $n$ lager enough such that,
\begin{equation*}
    \Pr(\|\mathbf Z_S\|_F>L_n) \le \exp\{-(b_1/k_0)M_1^4n^{4M_2-1}\},
\end{equation*}
Plugging the previous expression into Eq.~(\ref{eq:sez})  and noting that the bound is independent of $\{\pi_h\}_{h=1}^k$, we have
\begin{equation*}
\begin{aligned}
    \Pr(\|\mathbf Z\|_F>L_n,K^*\le C_1k_0) &\le \sum_{s=1}^{\lfloor C_1k_0\rfloor} \binom{k}{s}\exp\{-(b_1/k_0)M_1^4n^{4M_2-1}\}\\
    &\le \sum_{s=1}^{\lfloor C_1k_0\rfloor} \left(\frac{ek}{s}\right)^s\exp\{-(b_1/k_0)M_1^4n^{4M_2-1}\} \\
    & \le \exp\{-(b_1/k_0)M_1^4n^{4M_2-1} + C_1k_0\log k +\log C_1k_0 + C_1k_0\}\\
    & \le \exp\{-(b_1/k_0)M_1^4n^{4M_2-1} + b_2nk_0\log k\},
\end{aligned}
\end{equation*}
for some constant $b_2>0$ where we use the inequality $\binom{k}{s}\le \left(\frac{ek}{s}\right)^s$ in the second line. Due to Assumption~\ref{ass:growth}, we can choose that $M_2>1$ so that $n^{4M_2-1}/k_0 > n^2\log n>nk_0\log k)$.  With this choice of $M_2$, we have
\begin{equation*}
\begin{aligned}
    \Pr(\|\mathbf Z\|_F>L_n,K^*\le C_1k_0) &\le \exp\{-(b_1M_1^4-b_2 +1)nk_0\log k\} \\
    & \le \exp\{-(1+C_1^\prime)n(n+q)\epsilon_n^2\},
\end{aligned}
\end{equation*}
when we set $M_1 > ((C_1^\prime + b_2)/b_1)^{1/4}$. 

Similarly, we have that 
\begin{equation*}
    \Pr(\|\mathbf B\|_F>L_n,K^*\le C_1k_0) \le \sum_{S:1\le|S|\le C_1k_0}\Pr(\|\mathbf B_S\|_F>L_n),
\end{equation*}
And we have that $(1/\sigma_{B}^2)\|\mathbf B_S\|_F^2\sim \Gamma(n|S|/2,2)$.  Similar as the proof of \cite{loyal2025spike}, we use the sub-gamma tail bound in Lemma ~\ref{lemma2} to conclude that 
\begin{equation*}
    \Pr(\|\mathbf B_S\|_F>L_n) \le \exp\{-(1/4\sigma_B^2)M_1^2n^{2M_2} +nC_1k_0 \},
\end{equation*}
Then we have that 
\begin{equation*}
\begin{aligned}
    \Pr(\|\mathbf B\|_F>L_n,K^*\le C_1k_0) 
    &\le  \sum_{s=1}^{\lfloor C_1k_0\rfloor} \binom{k}{s} \exp\{-(1/4\sigma_B^2)M_1^2n^{2M_2} +nC_1k_0 \}\\ 
    &\le \sum_{s=1}^{\lfloor C_1k_0\rfloor}\left(\frac{ek}{s}\right)^s \exp\{-(1/4\sigma_B^2)M_1^2n^{2M_2} +nC_1k_0 \}\\
    &\le  \exp\{-(1/4\sigma_{B}^2)M_1^2n^{2M_2} +b_3nk_0\log k \},
\end{aligned}
\end{equation*}
for some constant $b_3>0$. We can choose $M_2>3/2$ so that $n^{2M_2}>n^2\log n > nk_0\log k$. We have
\begin{equation*}
\begin{aligned}
    \Pr(\|\mathbf B\|_F>L_n,K^*\le C_1k_0) &\le \exp\{-((1/4\sigma_{B}^2)M_1^2n^{2M_2}-b_3 +1)nk_0\log k\} \\
    & \le \exp\{-(1+C_1^\prime)n(n+q)\epsilon_n^2\},
\end{aligned}
\end{equation*}
when we set $M_1 > \max\{2\sigma_B\sqrt{C_1^\prime + b_3}, ((C_1^\prime + b_2)/b_1)^{1/4}\}$. 

Similarly, we have that $(1/\sigma_{\alpha}^2)\|\boldsymbol{\alpha}\|_2^2\sim\Gamma(n/2,2)$ and $(1/\sigma_{\gamma}^2)\|\boldsymbol{\gamma}\|_2^2\sim\Gamma(n/2,2)$. So we use the sub-gamma tail bound in Lemma~\ref{lemma2} to conclude that
\begin{equation*}
\begin{aligned}
    \Pr(\|\boldsymbol{\alpha}\|_2^2>L_n) &\le \exp\{-(1/4\sigma_{\alpha}^2)M_1^2n^{2M_2} +n\} \le \exp\{-(1+C_1^\prime)n(n+q)\epsilon_n^2\}, \\
    \Pr(\|\boldsymbol{\gamma}\|_2^2>L_n) &\le \exp\{-(1/4\sigma_{\gamma}^2)M_1^2n^{2M_2} +q\} \le \exp\{-(1+C_1^\prime)n(n+q)\epsilon_n^2\},
\end{aligned}
\end{equation*}
due to our previous choice of $M_2$ and when we set $M_1 > \max\{2\sigma_{\alpha}\sqrt{C_1^\prime+1},2\sigma_{\gamma}\sqrt{C_1^\prime+1},2\sigma_B\sqrt{C_1^\prime + b_3}, ((C_1^\prime + b_2)/b_1)^{1/4}\}$

\textit{Test construction}. By Lemma 2 of \cite{10.1214/009053606000001172} there exist local test functions $\phi_n$, such that for any $\boldsymbol{\alpha},\boldsymbol{\gamma},\mathbf Z,\mathbf B$ with $H_n(\boldsymbol\Theta_1,\boldsymbol\Theta_0)>\epsilon$,
\begin{equation*}
    \mathbb{E}_0(\phi_n)\le \exp(-n(n+q)\epsilon^2/4), \quad  \sup_{(\boldsymbol{\alpha}, \boldsymbol{\gamma},\mathbf Z, \mathbf B) \in \mathcal{F}_n : H_n (\boldsymbol\Theta_1, \boldsymbol\Theta_0) > \epsilon/18} \mathbb{E}_{\Theta} (1 - \phi_n) \lesssim \exp(-n(n+q) \epsilon_n^2/4).
\end{equation*}
To construct a uniformly consistent test over the sieve, we aim to apply Lemma 9 of \cite{10.1214/009053606000001172}, which constructs this test by combining these local tests using an $\epsilon$-net argument. As such, we need to control the metric entropy of the sieve. In particular, we must show that
$\log N(\epsilon_n/36, \mathcal{F}_n,H_n)\lesssim n(n+q)\epsilon_n^2$. To do so we re-express $\mathcal{F}_n = \cup_{S:|S|\le C_1k_0}\mathcal{F}_n(S)$ where
\begin{equation*}
\begin{aligned}
    \mathcal{F}_n = \{(\boldsymbol{\alpha}, \boldsymbol{\gamma},\mathbf Z,\mathbf B)&\in\mathbb{R}^n\times\mathbb{R}^q\times\mathbb{R}^{n\times k}\times\mathbb{R}^{q\times k}:\\
    &\|\boldsymbol{\alpha}\|_2\le L_n, \|\boldsymbol{\gamma}\|_2\le L_n,\|\mathbf Z_S\|_F\le L_n,\|\mathbf B_S\|_F\le L_n, \mathbf Z_{S^C} = \boldsymbol{0}\}.
\end{aligned}
\end{equation*}
Since a union bound implies $N(\epsilon_n/36, \mathcal{F}_n,H_n) \le \sum_{S:|S|\le C_1 k_0}N(\epsilon_n/36, \mathcal{F}_n(S),H_n)$, we start by upper bounding $N(\epsilon_n/36, \mathcal{F}_n(S),H_n)$ for given $S$. 

First, we relate the metric entropy under $H_n$ to more natural metrics on the parameter space using anargument developed in \cite{10.1093/biomet/asaa074}. Since the squared Hellinger distance is bounded by the KL divergence, we have for any $(\boldsymbol{\alpha}_1,\boldsymbol{\gamma}_1,\mathbf{Z}_1,\mathbf{B}_1),(\boldsymbol{\alpha}_2,\boldsymbol{\gamma}_2,\mathbf{Z}_2,\mathbf{B}_2)\in\mathcal{F}_n$ with $\boldsymbol\Theta_1^{A} = \boldsymbol{\alpha}_1\boldsymbol{1}_n^\top  + \boldsymbol{1}_n\boldsymbol{\alpha}_1^\top  + \mathbf{Z}_1\mathbf{Z}_1^\top $, $\boldsymbol\Theta^{Y}_1 = \boldsymbol{1}_n\boldsymbol{\gamma}_1^\top  + \mathbf{Z}_1\mathbf{B}_1^\top $, $\boldsymbol\Theta_2^{A} = \boldsymbol{\alpha}_2\boldsymbol{1}_n^\top  + \boldsymbol{1}_n\boldsymbol{\alpha}_2^\top  + \mathbf{Z}_2\mathbf{Z}_2^\top $ and $\boldsymbol\Theta^{Y}_2 = \boldsymbol{1}_n\boldsymbol{\gamma}_2^\top  + \mathbf{Z}_2\mathbf{B}_2^\top $ that 
\begin{align*}
    h^2(p^{A}_{1ii^\prime,},p^{A}_{2ii^\prime}) &\le K(p^{A}_{1ii^\prime,},p^{A}_{2ii^\prime}) = f^{\prime\prime}_{A}(\boldsymbol\Theta^{A}_{1ii^\prime})(\boldsymbol\Theta^{A}_{1ii^\prime}-\boldsymbol\Theta_{2ii^\prime}^{A})^2 + o\{(\boldsymbol\Theta^{A}_{1ii^\prime}-\boldsymbol\Theta_{2ii^\prime}^{A})^2\}, \\
    h^2(p^{Y}_{1ij,},p^{A}_{2ij}) &\le K(p^{Y}_{1ij,},p^{A}_{2ij}) = \frac{f^{\prime\prime}_{Y}(\boldsymbol\Theta^{Y}_{1ij})}{\phi}(\boldsymbol\Theta^{Y}_{1ij}-\boldsymbol\Theta_{2ij}^{Y})^2 + o\{(\boldsymbol\Theta^{Y}_{1ij}-\boldsymbol\Theta_{2ij}^{Y})^2\}, 
\end{align*}
 by a Taylor expansion. Next for any $(\boldsymbol{\alpha}_1,\boldsymbol{\gamma}_1,\mathbf{Z}_1,\mathbf{B}_1)\in 
\mathcal{F}_n(S)$  for a given $S$, we have that $|\boldsymbol\Theta^{A}_{1ii^\prime}|  \le L_n(2 + L_n)$ and $\boldsymbol\Theta^{Y}_{1ij} \le L_n(1 + L_n)$. Let $\psi_{A}(L_n) = 1 + \sup_{x:x\le L_n(2 + L_n)}f_{A}^{\prime\prime}(x)$ where $f_{A}(x) = \log(1 + \exp(x))$. Recall $\psi_{Y}(L_n) = 1 + \sup_{x:x\le L_n(1 + L_n)}f_{Y}^{\prime\prime}(x)$, so that
\begin{align*}
    h^2(p^{A}_{1ii^\prime,},p^{A}_{2ii^\prime}) &\le \psi_{A}(L_n)(\boldsymbol\Theta^{A}_{1ii^\prime}-\boldsymbol\Theta_{2ii^\prime}^{A})^2 + o\{(\boldsymbol\Theta^{A}_{1ii^\prime}-\boldsymbol\Theta_{2ii^\prime}^{A})^2\},\\
    h^2(p^{Y}_{1ij,},p^{A}_{2ij}) &\lesssim \psi_{Y}(L_n)(\boldsymbol\Theta^{Y}_{1ij}-\boldsymbol\Theta_{2ij}^{Y})^2 + o\{(\boldsymbol\Theta^{Y}_{1ij}-\boldsymbol\Theta_{2ij}^{Y})^2\},
\end{align*}
for any $(\boldsymbol{\alpha}_1,\boldsymbol{\gamma}_1,\mathbf{Z}_1,\mathbf{B}_1),(\boldsymbol{\alpha}_2,\boldsymbol{\gamma}_2,\mathbf{Z}_2,\mathbf{B}_2)\in\mathcal{F}_n$. If $|\boldsymbol\Theta^{A}_{1ii^\prime}-\boldsymbol\Theta_{2ii^\prime}^{A}|\le\{\sum_{i<i^\prime}(\boldsymbol\Theta^{A}_{1ii^\prime}-\boldsymbol\Theta_{2ii^\prime}^{A})^2\}^{1/2}\le\psi_{Y}(L_n)^{-1/2}\epsilon_n$ and $|\boldsymbol\Theta^{Y}_{1ij}-\boldsymbol\Theta_{2ij}^{Y}|\le \{\sum_{ij}(\boldsymbol\Theta^{Y}_{1ij}-\boldsymbol\Theta_{2ij}^{Y})^2\}^{-1/2}\le\psi_{Y}(L_n)^{-1/2}\epsilon_n$ which tends to zero since $\psi_{A}(L_n),\psi_{Y}(L_n)\ge 1$, let $\psi(L_n) = \max\{\psi_{A}(L_n), \psi_{Y}(L_n)\}$, the root-averaged-squared Hellinger metric satisfies
\begin{equation*}
    H_n(p_{\Theta_1},p_{\Theta_2}) \lesssim \left\{\frac{2\psi(L_n)}{n(n+q)}\left(\sum_{i<i^\prime}(\boldsymbol\Theta^{A}_{1ii^\prime}-\boldsymbol\Theta_{2ii^\prime}^{A})^2 + \sum_{ij}(\boldsymbol\Theta^{Y}_{1ij}-\boldsymbol\Theta_{2ij}^{Y})^2\right)\right\}^{\frac{1}{2}}\le \epsilon_n, 
\end{equation*}
for any $(\boldsymbol{\alpha}_1,\boldsymbol{\gamma}_1,\mathbf{Z}_1,\mathbf{B}_1),(\boldsymbol{\alpha}_2,\boldsymbol{\gamma}_2,\mathbf{Z}_2,\mathbf{B}_2)\in\mathcal{F}_n$. Furthermore,
\begin{equation*}
\begin{aligned}
&\left\{\sum_{i<i^\prime}(\boldsymbol\Theta^{A}_{1ii^\prime}-\boldsymbol\Theta_{2ii^\prime}^{A})^2 + \sum_{ij}(\boldsymbol\Theta^{Y}_{1ij}-\boldsymbol\Theta_{2ij}^{Y})^2\right\}^{\frac{1}{2}} \\    
\le  &\|\boldsymbol{\alpha}_1\boldsymbol{1}_n^\top  + \boldsymbol{1}_n\boldsymbol{\alpha}_1^\top  + \mathbf Z_1\mathbf Z_1^\top  - \boldsymbol{\alpha}_2\boldsymbol{1}_n^\top  - \boldsymbol{1}_n\boldsymbol{\alpha}_2^\top  - \mathbf Z_2\mathbf Z_2^\top \|_F + 
\left\|\boldsymbol{1}_n\boldsymbol{\gamma}_1^\top  + \mathbf Z_1\mathbf B_1^\top  - \boldsymbol{1}_n\boldsymbol{\gamma}_2^\top  - \mathbf Z_2\mathbf B_2^\top \right\|_F\\
\le& 2\sqrt{n}\|\boldsymbol{\alpha}_1-\boldsymbol{\alpha}_2\|_2 + \sqrt{n}\|\boldsymbol{\gamma}_1-\boldsymbol{\gamma}_2\|_2 + 7L_n\|\mathbf Z_1-\mathbf Z_2\|_F + L_n\|\mathbf B_1 - \mathbf B_2\|_F, 
\end{aligned}
\end{equation*}
for any $(\boldsymbol{\alpha}_1,\boldsymbol{\gamma}_1,\mathbf{Z}_1,\mathbf{B}_1),(\boldsymbol{\alpha}_2,\boldsymbol{\gamma}_2,\mathbf{Z}_2,\mathbf{B}_2)\in\mathcal{F}_n$, using the  the argument presented in the proof of
Lemma~\ref{lemma:A}. Based on the above bound, for some constant $b_4>0$, the metric entropy $\log N(\epsilon_n/36, \mathcal{F}_n(S),H_n)$ is bounded above by
\begin{equation*}
\begin{aligned}
    &\log N\left(\frac{b_4\epsilon_n/288}{(n\psi(L_n))^{1/2}},\{\boldsymbol{\alpha\in\mathbb{R}^n:\|\boldsymbol{\alpha}\|_2}\le L_n\},\|\cdot\|_2\right) +  \log N\left(\frac{b_4\epsilon_n/144}{(n\psi(L_n))^{1/2}},\{\boldsymbol{\gamma}\in\mathbb{R}^q:\|\boldsymbol{\gamma}\|_2\le L_n\},\|\cdot\|_2\right)\\
    +& \log N\left(\frac{b_4\epsilon_n/1008}{\psi(L_n)^{1/2}L_n},\{\mathbf Z\in\mathbb{R}^{n\times |S|}:\|\mathbf Z\|_F\le L_n\},\|\cdot\|_F\right) + \log N\left(\frac{b_4\epsilon_n/144}{\psi(L_n)^{1/2}L_n},\{\mathbf B\in\mathbb{R}^{q\times |S|}:\|\mathbf B\|_F\le L_n\},\|\cdot\|_F\right),
\end{aligned}
\end{equation*}
We bound each term separately. In what follows, recall that the
$\epsilon$-covering number of a $d$-dimensional Euclidean ball of radius $r$  under the $l_2$ metric is upper bounded by $(3r/\epsilon)^d$ for $\epsilon\in(0,1]$. Then we have
\begin{equation*}
\begin{aligned}
    \log N\left(\frac{b_4\epsilon_n/288}{(n\psi(L_n))^{1/2}},\{\boldsymbol{\alpha\in\mathbb{R}^n:\|\boldsymbol{\alpha}\|_2}\le L_n\},\|\cdot\|_2\right) &\le \log\left\{\frac{864(n\psi(L_n))^{1/2}L_n}{b_2\epsilon_n}\right\}^n \\
    &\lesssim n\log n\\
    &\lesssim n(n+q)\epsilon_n^2,
\end{aligned}
\end{equation*}
and 
\begin{equation*}
\begin{aligned}
  \log N\left(\frac{b_4\epsilon_n/144}{(n\psi(L_n))^{1/2}},\{\boldsymbol{\gamma\in\mathbb{R}^q:\|\boldsymbol{\gamma}\|_2}\le L_n\},\|\cdot\|_2\right) &\le \log\left\{\frac{432(n\psi(L_n))^{1/2}L_n}{b_2\epsilon_n}\right\}^q \\
    &\lesssim q\log n\\
    &\lesssim n(n+q)\epsilon_n^2,  
\end{aligned}
\end{equation*}
and 
\begin{equation*}
\begin{aligned}
    \log N\left(\frac{b_4\epsilon_n/1008}{\psi(L_n)^{1/2}L_n},\{\mathbf Z\in\mathbb{R}^{n\times |S|}:\|\mathbf Z\|_F\le L_n\},\|\cdot\|_F\right) &\le \log\left\{\frac{3024\psi(L_n)^{1/2}L^2_n}{b_2\epsilon_n}\right\}^{n\times k_0} \\
    &\lesssim nk_0\log n\\
    &\lesssim n(n+q)\epsilon_n^2,
\end{aligned}
\end{equation*}
and 
\begin{equation*}
\begin{aligned}
    \log N\left(\frac{b_4\epsilon_n/144}{\psi(L_n)^{1/2}L_n},\{\mathbf B\in\mathbb{R}^{q\times |S|}:\|\mathbf B\|_F\le L_n\},\|\cdot\|_F\right) &\le \log\left\{\frac{432\psi(L_n)^{1/2}L^2_n}{b_2\epsilon_n}\right\}^{q\times k_0} \\
    &\lesssim qk_0\log n\\
    &\lesssim n(n+q)\epsilon_n^2,
\end{aligned}
\end{equation*}
where we used Assumption~\ref{ass:glm_reg} to conclude that $\log\psi(L_n) + \log L_n\lesssim\log n$. Combining these bounds, we find that for some constant $b_5>0$, the metric entropy of the sieve $\log N(\epsilon_n/36, \mathcal{F}_n,H_n)$ is bounded by
\begin{equation*}
\begin{aligned}
    \log\left\{\sum_{S:|S|\le C_1k_0}N(\frac{\epsilon_n}{36}, \mathcal{F}_n(S),H_n)\right\}&\le \log\left\{\sum_{s=0}^{\lfloor C_1k_0\rfloor}\tbinom{k}{s}e^{b_5n(n+q)\epsilon_n^2}\right\}\\
    &\lesssim \log(C_1k_0 + 1) C_1k_0\log k + b_5n(n+q)\epsilon_n^2\\
    &\lesssim n(n+q)\epsilon_n^2,
\end{aligned}
\end{equation*}

Now, Lemma 9 of \cite{10.1093/biomet/asaa074} implies that for every $\epsilon>\epsilon_n$, there exists a test $\Phi_n$ such that 
\begin{equation*}
\begin{aligned}
    \mathbb{E}_0(\Phi_n) &\le \frac{1}{2}\exp(b_6n(n+q)\epsilon_n^2 - n(n+q)\epsilon^2/4), \\
    \sup_{(\boldsymbol{\alpha}, \boldsymbol{\gamma},\mathbf Z, \mathbf B) \in \mathcal{F}_n : H_n (\boldsymbol\Theta, \boldsymbol\Theta_0) > \epsilon} \mathbb{E}_\Theta (1 - \Phi_n) &\le \exp(-n(n+q) \epsilon^2/4),
\end{aligned}
\end{equation*}
for some constant $b_6>0$. To complete the proof, we choose $\epsilon = 4C_2\epsilon_n^2$ for $C_2>\max(C_1^\prime,b_6)$.
\end{proof}

\subsection{Corollary on Frobenius Norm Convergence}
The general guarantee of distributional convergence can be translated into more interpretable metrics for specific models. For instance, for binary node attributes (the MIRT model), we can show that the estimated probabilities converge to the true probabilities in terms of the Frobenius norm. 

\begin{corollary}[Consistency for Bernoulli Node Variables]\label{col:bernoulli_consistency}
    Under the conditions of Theorem 2, for a Bernoulli node variable model, the posterior distribution for the matrix of dyadic probabilities converges in Frobenius norm:
    \begin{equation*}
       \lim_{n\to\infty} \mathbb{E}_0 \left[ \Pr\left( \frac{1}{\sqrt{n(n+q)}} \|g^{-1}(\boldsymbol\Theta) - g^{-1}(\boldsymbol\Theta_0)\|_F > C_3 \epsilon_n \mid \mathbf{A}, \mathbf{Y} \right) \right] = 0,
    \end{equation*}
    for some constant $C_3 > 0$, where $g^{-1}$ is the inverse logit function.
\end{corollary}
This corollary confirms that the proposed model consistently recovers the underlying probabilities of interactions and responses at the same near-optimal, adaptive rate, ensuring robust predictive performance.

To obtain the conclusion of corollary~\ref{col:bernoulli_consistency}, we only need to prove the following lemma.

\begin{lemma}{(Lemma S10 of \cite{loyal2025spike})}\label{lemma3}
    Let $\boldsymbol{\Theta}_1$ and $\boldsymbol{\Theta}_2$ be the log-odds matrices that define the probability measures of two Bernoulli joint latent space model such that $\mathbf{P}_1 = g^{-1}(\boldsymbol\Theta_1)$ and $\mathbf{P}_2 = g^{-1}(\boldsymbol\Theta_2)$ where $g^{-1}(\cdot)$ is the inverse logit function and is applied component-wise, then $\frac{1}{\sqrt{n(n+q)}}\|\mathbf{P}_1 - \mathbf{P}_2\|_F\lesssim H_n(\boldsymbol\Theta_1,\boldsymbol\Theta_2)$. 
\end{lemma}
\begin{proof}[proof of lemma~\ref{lemma3}]  
This result does not depend on the joint latent space model structure, only that the entries of the adjacency matrix and node variables are independent Bernoulli random variables. Let $p_{1ij} = [\mathbf{P}_1]_{ij}$ and $p_{2ij} = [\mathbf{P}_2]_{ij}$. The
Hellinger distance between two Bernoulli random variables satisfies
\begin{equation*}
    \begin{aligned}
        h^2(\Theta_{1ij},\Theta_{2ij})  &= \frac{1}{2}\left[\left(\sqrt{p_{1ij}} - \sqrt{p_{2ij}}\right)^2 + \left(\sqrt{1-p_{1ij}} - \sqrt{1- p_{2ij}}\right)^2\right] \\
        & = \frac{1}{2}\left[\left(\frac{1}{2}2 |\sqrt{p_{1ij}} - \sqrt{p_{2ij}}|\right)^2 + \left(\frac{1}{2}2|\sqrt{1-p_{1ij}} - \sqrt{1- p_{2ij}}|\right)^2\right] \\
        &\ge \frac{1}{4} (p_{1ij} - p_{2ij})^2,
    \end{aligned}
\end{equation*}
where we used the fact that $|\sqrt{p_{1ij}}  + \sqrt{p_{2ij}}| < 2$ and $|\sqrt{1-p_{1ij}} + \sqrt{1- p_{2ij}}|$ since $p_{1ij}, p_{2ij}\in [0,1]$. As such
\begin{equation*}
    H^2_n(\boldsymbol{\Theta}_1,\boldsymbol{\Theta}_2) = \frac{2}{n + 2q + 1}\left(\sum_{i<i^\prime}h^2(p_{1ii^{\prime}}^{A},p_{2ii^\prime}^{A}) + \sum_{ij}h^2(p_{1ij}^Y,p_{2ij}^Y)\right) \gtrsim \frac{1}{n(n+q)}\|\mathbf{P}_1 - \mathbf{P}_2\|_F^2.
\end{equation*}
\end{proof}

\subsection{Additional theoretical results under compact support}\label{sec:TruncatedPriors}
This section investigates the posterior consistency of parameters under the assumption of truncated priors for $\boldsymbol{\alpha}$, $\boldsymbol{\gamma}$, $\mathbf{Z}$, and $\mathbf{B}$. To guarantee that the support of the truncated posterior encompasses the true parameter values, we specify the slab distribution for latent positions as a Student’s $t$-distribution, $P_{\text{slab}} = t_{[-A_Z,A_Z],2a_{\theta}}(0,b_{\theta}/a_{\theta})$, with $2a_{\theta}$ degrees of freedom and scale $\sqrt{b_{\theta}/a_{\theta}}$, restricted to the interval $[-A_Z,A_Z]$ where $A_Z > C_Z$. In addition, independent truncated normal priors are placed on the entries of $\boldsymbol{\alpha}$, $\boldsymbol{\gamma}$, and $\mathbf{B}$, such that $\alpha_i \sim \mathcal{N}_{[-A_{\alpha},A_{\alpha}]}(0,\sigma^2_{\alpha})$ for $i = 1,\dots,n$, $\gamma_j \sim \mathcal{N}_{[-A_{\gamma},A_{\gamma}]}(0,\sigma^2_{\gamma})$ for $j=1,\dots,q$, and $\beta_{jh} \sim \mathcal{N}_{[-A_B,A_B]}(0,\sigma^2_B)$ for $j=1,\dots,q$ and $h=1,\dots,k$. For these priors, the truncation intervals are defined by constants $A_\alpha > C_\alpha$, $A_\gamma > C_\gamma$, and $A_B > C_B$, respectively. It should be emphasized that these truncated priors are introduced primarily to facilitate the investigation of the model's theoretical properties. 

The following theorem establishes the parameter consistency and the posterior contraction rate under the aforementioned truncated priors, assuming $k_0$ to be bounded for simplicity. While the proof largely parallels that of Theorem 2, the bounded support inherent in the truncated priors allows us to remove condition (b) of Assumption~\ref{ass:glm_reg} during sieve construction. Furthermore, this bounded nature enables the Hellinger metric to be lower-bounded by the Frobenius norm, thereby facilitating the derivation of the parameter consistency results. Beyond the prior specifications, we further require Assumption~\ref{ass:node_var} to ensure the lower-boundedness of variance for the node variables.

\begin{assumptionS}\label{ass:node_var}
    For any compact set $\mathcal{K} \subset \mathbb{R}$, there exists a constant $C_{f'}$ such that $\inf_{\Theta \in \mathcal{K}} f_Y''(\Theta) \ge C_{f'}$.
\end{assumptionS}

\begin{theoremS}[Adaptive Posterior Contraction Rate, Truncated Priors]\label{th:Truncated_consistency}
    Consider the joint latent space model under Condition~\ref{con:Hyperparameter} and Assumptions \ref{ass:growth}-\ref{ass:node_var}. We further assume that the latent dimension $k_0$ is bounded. Suppose the prior for $\mathbf{Z}$ is specified as in Section 3 with $P_{\text{slap}} = t_{[-A_Z,A_Z],2a_{\theta}}(0,b_{\theta}/a_{\theta})$ for constants $A_Z > C_Z$ and $\theta_0=0$ (in the limit). Priors on other parameters are set as $\alpha_i \sim \mathcal{N}_{[-A_{\alpha},A_{\alpha}]}(0,\sigma^2_{\alpha})$ for $i = 1,\dots,n$, $\gamma_j \sim \mathcal{N}_{[-A_{\gamma},A_{\gamma}]}(0,\sigma^2_{\gamma})$ for $j=1,\dots,q$, and $\beta_{jh} \sim \mathcal{N}_{[-A_B,A_B]}(0,\sigma^2_B)$ for $j=1,\dots,q$ and $h=1,\dots,k$, for constants $A_\alpha > C_\alpha$, $A_\gamma > C_\gamma$, and $A_B > C_B$. Then,
    \begin{equation*}
       \lim_{n\to\infty} \mathbb{E}_0 \left[ \Pr\left( \frac{1}{\sqrt{n(n+q)}} \|\boldsymbol\Theta - \boldsymbol\Theta_0\|_F > C_4 \epsilon_n \mid \mathbf{A}, \mathbf{Y} \right) \right] = 0,
    \end{equation*}
    for some constant $C_4>0$, where $\epsilon_n = \sqrt{k_0 \log k / n}$.
\end{theoremS}

\begin{proof}[Proof of Theorem~\ref{th:Truncated_consistency}] First,we demonstrate that the Hellinger consistency of the posterior with a contraction rate of $\epsilon_n = \sqrt{k_0 \log k / n}$ continues to hold for the posterior under the truncated priors. Note that Theorem~\ref{th:Truncated_consistency} assumptions on the parameter space are stronger than the ones made in Theorem 2. In the statement of Theorem~\ref{th:Truncated_consistency}, $A_Z$, $A_\alpha$, $A_\gamma$ and $A_B$ are chosen large enough so that the true parameters are in the support of the truncated prior. As such, Lemma~\ref{lemma:A}  continues to hold under the conditions of Theorem~\ref{th:Truncated_consistency} since the probability to lie within a metric ball around
the true parameters under the truncated prioris bounded below by that same probability under
the original priors. Since Lemma~\ref{lemma:A} holds, are petition of the proof of Theorem 1 establishes that
\begin{equation*}
    \lim_{n\rightarrow\infty}\mathbb{E}_0\left\{\Pr\left(K^*>C_1k_0\mid \mathbf A, \mathbf{Y}\right)\right\} = 0.
\end{equation*}
under the truncated priors for some $C_1>1$. Next,we can establish that  
\begin{equation*}
    \lim_{n\rightarrow\infty}\mathbb{E}_0\left[\Pr\left\{H_n(\boldsymbol\Theta,\boldsymbol\Theta_0)>C_2\epsilon_n \mid \mathbf A, \mathbf{Y}\right\}\right] = 0,
\end{equation*}    
for some large enough constant $C_2>0$ by repeating the same testing and sieve construction argument in the proof of Theorem 2. However, in this case, the compactness of the parameter space guarantees the existence of an appropriate sieve and uniformly consistent sequence of test functions satisfying Eq~(\ref{eq:test}), see \cite{10.1214/009053606000001172} for details. 

Finally, we establish the parameter consistency of the posterior. To do so,we follow an argument introduced by \cite{10.1093/biomet/asaa074} to demonstrate consistency of $\boldsymbol{\Theta}$ under the Frobenius norm. Consider the event $\mathcal{A}_n = \{(\boldsymbol{\alpha},\boldsymbol{\gamma},\mathbf{Z},\mathbf{B}):K^* \le C_1k_0, H_n(\boldsymbol\Theta,\boldsymbol\Theta_0)\le C_2\epsilon_n\}$, since the probability of its complement goes to zero under the posterior. For $\delta_1$, $\delta_2>0$ in Lemma~\ref{lemma4}, let $\mathcal{D}_n = \{(i,j): \delta_1(\Theta_{ij} - \Theta_{0ij})^2\ge \delta_2, 1\le i\le n, 1\le j \le (n+q)\}$. Then on $\mathcal{A}_n$, we have that 
\begin{equation}\label{eq:H_bound}
    C^2_2\epsilon^2_n \ge H^2_n(\boldsymbol\Theta,\boldsymbol\Theta_0) \ge \frac{\delta_1}{n(n+q)}\sum_{(i,j)\notin\mathcal{D}_n}\tilde{\xi}_{ij}(\Theta_{0ij})(\Theta_{ij}-\Theta_{0ij})^2 + \frac{\delta_2}{n(n+q)}\sum_{(i,j)\in\mathcal{D}_n}\tilde{\xi}_{ij}(\Theta_{0ij}),
\end{equation}
where $\tilde{\xi}_{ij}(\Theta_{0ij})= \min\{f_A^{\prime\prime}(\Theta_{0ij})/4, f_Y^{\prime\prime}(\Theta_{0ij})/4 \} > 0$. Then, Assumption~\ref{ass:boundedness} and Assumption~\ref{ass:node_var} implies that $\min_{ij}\tilde{\xi}_{ij}(\Theta_{0ij})$ is lower bounded by a constant $C_3>0$. As such 
\begin{equation*}
    \begin{aligned}
        \frac{1}{n(n+q)}\sum_{(i,j)\notin\mathcal{D}_n}\tilde{\xi}_{ij}(\Theta_{0ij})(\Theta_{ij}-\Theta_{0ij})^2 \ge &\frac{1}{n(n+q)}\sum_{i,j}\tilde{\xi}_{ij}(\Theta_{0ij})(\Theta_{ij}-\Theta_{0ij})^2 - \\ 
        &\qquad \frac{1}{n(n+q)}\sum_{(i,j)\in\mathcal{D}_n}\tilde{\xi}_{ij}(\Theta_{0ij})\|\Theta_{ij}-\Theta_{0ij}\|_{\infty}^2 \\
        &\ge \frac{C_3}{n(n+1)}\|\Theta-\Theta_0\|_F^2 -\frac{C_2^2\epsilon_n^2}{\delta_2}\|\Theta-\Theta_0\|_{\infty}^2,        
    \end{aligned}
\end{equation*}
where the lower bound for the second term uses Eq~(\ref{eq:H_bound}). Furthermore, on $\mathcal{A}_n$, $\|\Theta-\Theta_0\|_\infty \le |\max\{2A_\alpha+C_1k_0A_Z^2, A_\gamma+C_1k_0A_ZA_B\}|+\|\Theta_0\|_\infty = \mathcal{O}(1)$ due to our assumptions on the truth and the support of the truncated priors. Combining the previous expression with Eq~(\ref{eq:H_bound}) implies that
\begin{equation*}
    C^2_2\epsilon^2_n \ge \frac{1}{n(n+q)}\|\boldsymbol\Theta-\boldsymbol\Theta_0\|_F^2 + {o}(1),
\end{equation*}
which further implies that
\begin{equation*}
   \lim_{n\to\infty} \mathbb{E}_0 \left[ \Pr\left( \frac{1}{\sqrt{n(n+q)}} \|\boldsymbol\Theta - \boldsymbol\Theta_0\|_F > C_4 \epsilon_n \mid \mathbf{A}, \mathbf{Y} \right) \right] = 0,
\end{equation*}
for some constant $C_4$.
\end{proof}

\begin{lemma}{(Lemma A1 of \cite{10.1093/biomet/asaa074})}\label{lemma4}
Let $\xi_{ij}(\Theta_{ij}) = h^2(p_{\Theta_{ij}} - p_{\Theta_{0ij}})$. Then
there exists constants $\delta_1$, $\delta_2>0$ such that $\xi_{ij}(\Theta_{ij})\ge\xi^{\prime\prime}(\Theta_{0ij})\min\{\delta_1(\Theta_{1ij}-\Theta_{0ij})^2,\delta_2\}$. 
\end{lemma}

While Theorem 1 establishes that the posterior probability of overestimating the dimension vanishes asymptotically, we now further demonstrate that the probability of underestimation also becomes negligible. Leveraging the parameter consistency results from Theorem~\ref{th:Truncated_consistency}, we show that the posterior probability of under-specifying the dimension vanishes.  And, we need additional assumption on the true latent position $\mathbf{Z}_0$. 

\begin{assumptionS}[Regularity of the True Latent Position $\mathbf{Z}_0$]\label{ass:regularity_Z}
    The true latent position $\mathbf{Z}_0$ satisfy 
    \begin{enumerate}[label=(\alph*)]
        \item $\mathbf{Z}_0^\top\boldsymbol{1}_n = \boldsymbol0_n$.
        \item $\lambda_{min}(\mathbf{Z}_0^\top\mathbf{Z}_0) \gtrsim \sqrt{nk_0 \log k}$, where $\lambda_{min}(\mathbf{Z}_0^\top\mathbf{Z}_0)$ represents the smallest eigenvalue of $\mathbf{Z}_0^\top\mathbf{Z}_0$. 
    \end{enumerate}
\end{assumptionS}

\begin{theoremS}[Posterior Concentration for the Latent Dimension]\label{th:Truncated_underest}
    Consider the joint latent space model under Condition~\ref{con:Hyperparameter} and Assumptions \ref{ass:growth}-\ref{ass:regularity_Z}. We further assume that the latent dimension $k_0$ is bounded. Suppose the prior for $\mathbf{Z}$ is specified as in Section 3 with $P_{\text{slap}} = t_{[-A_Z,A_Z],2a_{\theta}}(0,b_{\theta}/a_{\theta})$ for constants $A_Z > C_Z$ and $\theta_0=0$ (in the limit). Priors on other parameters are set as $\alpha_i \sim \mathcal{N}_{[-A_{\alpha},A_{\alpha}]}(0,\sigma^2_{\alpha})$ for $i = 1,\dots,n$, $\gamma_j \sim \mathcal{N}_{[-A_{\gamma},A_{\gamma}]}(0,\sigma^2_{\gamma})$ for $j=1,\dots,q$, and $\beta_{jh} \sim \mathcal{N}_{[-A_B,A_B]}(0,\sigma^2_B)$ for $j=1,\dots,q$ and $h=1,\dots,k$, for constants $A_\alpha > C_\alpha$, $A_\gamma > C_\gamma$, and $A_B > C_B$. Then,
    \begin{equation*}
       \lim_{n\to\infty} \mathbb{E}_0 \left[ \Pr\left( K^*<k_0 \mid \mathbf{A}, \mathbf{Y} \right) \right] = 0.
    \end{equation*}
\end{theoremS}

\begin{proof}[Proof of Theorem~\ref{th:Truncated_underest}] Let $\mathbf{J} = \mathbf{I}_n - \frac{1}{n}\boldsymbol{1}_n\boldsymbol{1}_n^\top$. We first consider 
\begin{equation*}
    \mathbf{E}_1 = \boldsymbol\Theta^A - \boldsymbol\Theta_0^A = (\boldsymbol{1}_n\boldsymbol{\delta}_\alpha^\top + \boldsymbol{\delta}_\alpha\boldsymbol{1}_n^\top) + (\mathbf{Z}\mathbf{Z}^\top - \mathbf{Z}_0\mathbf{Z}_0^\top),
\end{equation*}
where $\boldsymbol{\delta}_\alpha = \boldsymbol{\alpha}-\boldsymbol{\alpha}_0$ and $\mathbf{Z}\in 
\mathbb{R}^{n\times K^*}$ only contains the nonzero column. Since $\mathbf{J}\boldsymbol{1}_n = 0$, then, by Lemma~\ref{lemma:5} , we have
\begin{equation*}
    \|\mathbf{E}_1\|_F^2 \ge \|\mathbf{J}\mathbf{E}_1\mathbf{J}\|_F^2 = \|\mathbf{Z}_c\mathbf{Z}_c^\top - \mathbf{Z}_0\mathbf{Z}_0^\top\|^2_F,
\end{equation*}
where $\mathbf{Z}_c = \mathbf{J}\mathbf{Z}\in 
\mathbb{R}^{n\times K^*}$. 

Similarly, we also consider 
\begin{equation*}
    \mathbf{E}_2 = \boldsymbol{\Theta}^Y - \boldsymbol{\Theta}^Y_0 = \boldsymbol{1}_n\boldsymbol{\delta}_\gamma^\top + (\mathbf{Z}\mathbf{B}^\top - \mathbf{Z}_0\mathbf{B}_0^\top),
\end{equation*}
where $\boldsymbol{\delta}_\gamma = \boldsymbol{\gamma}-\boldsymbol{\gamma}_0$ and $\mathbf{B}\in \mathbb{R}^{n\times K^*}$. Then, by Lemma~\ref{lemma:5},  
\begin{equation*}
    \|\mathbf{E}_2\|_F^2\ge \|\mathbf{J}\mathbf{E}_2\|_F^2 = \|\mathbf{Z}_c\mathbf{B}^\top - \mathbf{Z}_2\mathbf{B}_0^\top\|_F^2.
\end{equation*}

Then, we obtain 
\begin{equation*}
    \|\boldsymbol{\Theta}-\boldsymbol{\Theta}_0\|_F^2\ge \|\mathbf{Z}_c\mathbf{Z}_c^\top - \mathbf{Z}_0\mathbf{Z}_0^\top\|^2_F + \|\mathbf{Z}_c\mathbf{B}^\top - \mathbf{Z}_2\mathbf{B}_0^\top\|_F^2 = \|\tilde{\Theta}_c - \tilde{\Theta}_0\|,
\end{equation*}
where $\tilde{\boldsymbol\Theta}_c = \{\mathbf{Z}_c\mathbf{Z}_c^\top, \mathbf{Z}_c \mathbf{B}^\top\}$ and $\tilde{\boldsymbol\Theta}_0 = \{\mathbf{Z}_0\mathbf{Z}_0^\top, \mathbf{Z}_0 \mathbf{B}_0^\top\}$.

Suppose that $K^*<k_0$, by Eckart–Young–Mirsky theorem,  we obtain
\begin{equation*}
    \|\tilde{\boldsymbol\Theta}_c - \tilde{\boldsymbol\Theta}_0\|_F \ge
    \sqrt{\lambda_{min}(\mathbf{Z_0}(\mathbf{Z}^\top_0\mathbf{Z}_0 + \mathbf{B}^\top_0\mathbf{B}_0)\mathbf{Z}_0^\top)}\ge \lambda_{\min}(\mathbf{Z}_0^\top\mathbf{Z}_0)\gtrsim\sqrt{nk_0\log k}.
\end{equation*}
Hence, 
\begin{equation*}
    \lim_{n\to\infty}\mathbb{E}_0 \left[ \Pr\left( K^*<k_0 \mid \mathbf{A}, \mathbf{Y} \right) \right] \le    \lim_{n\to\infty} \mathbb{E}_0 \left[ \Pr\left( \frac{1}{\sqrt{n(n+q)}} \|\boldsymbol\Theta - \boldsymbol\Theta_0\|_F \gtrsim \sqrt{\frac{k_0\log k}{n}}  \mid \mathbf{A}, \mathbf{Y} \right) \right] = 0.
\end{equation*}
\end{proof}

\begin{lemma}\label{lemma:5}
Let $\mathbf{P} \in \mathbb{R}^{n \times n}$ and $\mathbf{Q} \in \mathbb{R}^{m \times m}$ be orthogonal projection matrices (i.e., $\mathbf{P}^2 = \mathbf{P}, \mathbf{P}^\top = \mathbf{P}$ and $\mathbf{Q}^2 = \mathbf{Q}, \mathbf{Q}^\top = \mathbf{Q}$). For any matrix $\mathbf{M} \in \mathbb{R}^{n \times m}$, the following inequalities hold:
\begin{enumerate}
    \item $\|\mathbf{P}\mathbf{M}\|_F \le \|\mathbf{M}\|_F$.
    \item $\|\mathbf{M}\mathbf{Q}\|_F \le \|\mathbf{M}\|_F$.
    \item $\|\mathbf{P}\mathbf{M}\mathbf{Q}\|_F \le \|\mathbf{M}\|_F$.
\end{enumerate}
In particular, the centering matrix $\mathbf{J} = \mathbf{I} - \frac{1}{n}\mathbf{1}\mathbf{1}^\top$ satisfies these properties.    
\end{lemma}

\begin{proof}[Proof of Lemma~\ref{lemma:5}]
The Frobenius norm is induced by the trace inner product, defined as $\langle \mathbf{A}, \mathbf{B} \rangle_F = \operatorname{tr}(\mathbf{A}^\top \mathbf{B})$. We utilize the orthogonal decomposition of $\mathbf{M}$ as follows:

\textit{1. Proof of $\|\mathbf{P}\mathbf{M}\|_F \le \|\mathbf{M}\|_F$:}  
Decompose $\mathbf{M}$ using the identity $\mathbf{I} = \mathbf{P} + (\mathbf{I} - \mathbf{P})$:
$$
\mathbf{M} = \mathbf{P}\mathbf{M} + (\mathbf{I} - \mathbf{P})\mathbf{M}.
$$
The squared Frobenius norm of $\mathbf{M}$ can be expanded as:
$$
\|\mathbf{M}\|_F^2 = \|\mathbf{P}\mathbf{M} + (\mathbf{I} - \mathbf{P})\mathbf{M}\|_F^2 = \|\mathbf{P}\mathbf{M}\|_F^2 + \|(\mathbf{I} - \mathbf{P})\mathbf{M}\|_F^2 + 2\langle \mathbf{P}\mathbf{M}, (\mathbf{I} - \mathbf{P})\mathbf{M} \rangle_F.
$$
The inner product term  satisfies:
$$
\langle \mathbf{P}\mathbf{M}, (\mathbf{I} - \mathbf{P})\mathbf{M} \rangle_F = \operatorname{tr}\left( (\mathbf{P}\mathbf{M})^\top (\mathbf{I} - \mathbf{P})\mathbf{M} \right) = \operatorname{tr}\left( \mathbf{M}^\top \mathbf{P}^\top (\mathbf{I} - \mathbf{P}) \mathbf{M} \right).
$$
Since $\mathbf{P}$ is an orthogonal projection matrix, it is both symmetric ($\mathbf{P}^\top = \mathbf{P}$) and idempotent ($\mathbf{P}^2 = \mathbf{P}$), which implies:
$$
\mathbf{P}^\top(\mathbf{I} - \mathbf{P}) = \mathbf{P}(\mathbf{I} - \mathbf{P}) = \mathbf{P} - \mathbf{P}^2 = \mathbf{P} - \mathbf{P} = \mathbf{0}.
$$
Thus, the inner product is zero, yielding the Pythagorean identity for the Frobenius norm:
$$
\|\mathbf{M}\|_F^2 = \|\mathbf{P}\mathbf{M}\|_F^2 + \|(\mathbf{I} - \mathbf{P})\mathbf{M}\|_F^2.
$$
Since $\|(\mathbf{I} - \mathbf{P})\mathbf{M}\|_F^2 \ge 0$, it follows that $\|\mathbf{M}\|_F^2 \ge \|\mathbf{P}\mathbf{M}\|_F^2$.

\textit{2. Proof of $\|\mathbf{M}\mathbf{Q}\|_F \le \|\mathbf{M}\|_F$:}  
Similarly, decomposing $\mathbf{M}$ from the right as $\mathbf{M} = \mathbf{M}\mathbf{Q} + \mathbf{M}(\mathbf{I} - \mathbf{Q})$ and invoking the cyclic property of the trace, we have:
$$
\langle \mathbf{M}\mathbf{Q}, \mathbf{M}(\mathbf{I} - \mathbf{Q}) \rangle_F = \operatorname{tr}\left( \mathbf{Q}^\top \mathbf{M}^\top \mathbf{M}(\mathbf{I} - \mathbf{Q}) \right) = \operatorname{tr}\left( \mathbf{M}^\top \mathbf{M}(\mathbf{I} - \mathbf{Q})\mathbf{Q}^\top \right).
$$
As $(\mathbf{I} - \mathbf{Q})\mathbf{Q}^\top = \mathbf{Q} - \mathbf{Q}^2 = \mathbf{0}$, the components $\mathbf{M}\mathbf{Q}$ and $\mathbf{M}(\mathbf{I}-\mathbf{Q})$ are orthogonal. Following the same logic as above, we obtain $\|\mathbf{M}\|_F^2 = \|\mathbf{M}\mathbf{Q}\|_F^2 + \|\mathbf{M}(\mathbf{I} - \mathbf{Q})\|_F^2 \ge \|\mathbf{M}\mathbf{Q}\|_F^2$.

\textit{3. Proof of $\|\mathbf{P}\mathbf{M}\mathbf{Q}\|_F \le \|\mathbf{M}\|_F$:}  
Applying the contraction properties derived in steps 1 and 2 sequentially:
$$
\|\mathbf{M}\|_F \ge \|\mathbf{P}\mathbf{M}\|_F \ge \|(\mathbf{P}\mathbf{M})\mathbf{Q}\|_F,
$$
which completes the proof.
\end{proof}

\section{Simulation study and sensitivity study}\label{SM:re}

\subsection{Nominal Response Model (NRM)}\label{sec:NRM}
In educational and psychological research, polytomous data can often be collected with relative ease. To analyze such multi-category item responses, a range of complex nonlinear item response theory (IRT) models have been developed. When the categories of the response variable are unordered, the Nominal Response Model (NRM) is a suitable choice \citep{bock1972estimating}. 

Suppose there are $n$ individuals responding to $q$ items, where each response $Y_{ij}$ takes values in a set of $L$ unordered categories $\{1, \dots, L\}$. The NRM models the probability that the $i$-th individual chooses category $l$ for the $j$-th item as a softmax function of the latent positions and item-category parameters:

\begin{equation*}
\Pr(Y_{ij} = l \mid \boldsymbol{z}_i, \boldsymbol{\beta}_{jl}, \gamma_{jl}) = \frac{\exp(\boldsymbol{z}_i^T \boldsymbol{\beta}_{jl} + \gamma_{jl})}{\sum_{m=0}^L \exp(\boldsymbol{z}_i^T \boldsymbol{\beta}_{jm} + \gamma_{jm})},
\end{equation*}
where $\boldsymbol{z}_i$ is the $k$-dimensional latent vector of individual $i$, $\boldsymbol{\beta}_{jl}$ is the discrimination vector for category $l$ of item $j$, and $\gamma_{jl}$ is the intercept. To ensure model identification, we typically set a baseline category (e.g., category 1) such that $\boldsymbol{\beta}_{j1} = \mathbf{0}$ and $\gamma_{j1} = 0$. In Section~\ref{sec:Multinomial_MHgibbs}, we describe how the MCMC algorithm  for the node variable satisfy the NRM. 

\subsection{Graded Response Model (GRM)}\label{sec:GRM}
When the multi-category item responses are ordered, the Graded Response Model (GRM) is the most widely adopted IRT model in psychological measurement \citep{zhang2022gibbs}. 
While the GRM does not belong to the exponential family, it can be seamlessly integrated into the JLSM framework.

Suppose the response variable $Y_{ij}$ is ordered, i.e., $Y_{ij} \in \{1, \dots, L\}$. The GRM characterizes the response process through cumulative probabilities. The probability that the response $Y_{ij}$ is larger than or equal to a certain threshold $l$ is defined as:

\begin{equation*}
\Pr(Y_{ij} \ge l \mid \boldsymbol{z}_i, \boldsymbol\beta_j, \gamma_{jl}) = \frac{\exp(\boldsymbol z_i^T \boldsymbol\beta_j - \gamma_{jl})}{1 + \exp(\boldsymbol z_i^T \boldsymbol\beta_j - \gamma_{jl})}, \quad l = 0, 1, \dots, L,
\end{equation*}
where $\boldsymbol\beta_j$ represents the discrimination parameters and $\gamma_{jl}$ are the threshold parameters for item $j$. By convention, we define $\Pr(Y_{ij} \ge 1) = 1$ and $\Pr(Y_{ij} \ge L+1) = 0$. Then, the probability that the response variable is equal to a specific category $l$ is the difference between two adjacent cumulative probabilities:

\begin{equation}\label{eq:GRM}
\begin{aligned}
\Pr(Y_{ij} = l \mid \boldsymbol{z}_i, \boldsymbol\beta_j, \gamma_{j,l-1},\gamma_{jl}) &= \Pr(Y_{ij} \ge l \mid \boldsymbol{z}_i, \boldsymbol\beta_j, \gamma_{jl}) - \Pr(Y_{ij} \ge l+1 \mid \boldsymbol{z}_i, \boldsymbol\beta_j, \gamma_{j,l+1}) \\
&= \frac{\exp(\boldsymbol z_i^T \boldsymbol\beta_j - \gamma_{jl})}{1 + \exp(\boldsymbol z_i^T \boldsymbol\beta_j - \gamma_{jl})} - \frac{\exp(\boldsymbol z_i^T \boldsymbol\beta_j - \gamma_{j,l+1})}{1 + \exp(\boldsymbol z_i^T \boldsymbol\beta_j - \gamma_{j,l+1})}.
\end{aligned}    
\end{equation}
Since cumulative probabilities must be non-decreasing, i.e., $\Pr(Y_{ij} \ge l) \ge \Pr(Y_{ij} \ge l+1)$, the intercept parameters for item $j$ must satisfy the following monotonicity constraint:
$$-\infty  = \gamma_{j0} \le \gamma_{j1} \le \dots \le \gamma_{j,L-1}\le \gamma_{j,L} = +\infty.$$

Based on the result of \cite{zhang2022gibbs}, we propose a Gibbs sampling algorithm for node variables that satisfy the GRM in Section~\ref{sec:ordinal_gibbs}.

We then conducted an additional simulation study to evaluate the performance of the JLSM when node variables follow Eq.~(\ref{eq:GRM}). We set $L = 5$, corresponding to the conventional Likert-type scoring format widely used in educational and psychological measurement. The threshold parameters $\gamma_{j1}$, $\gamma_{j2}$, $\gamma_{j3}$, and $\gamma_{j4}$ were independently generated from $U[-3,-1.5]$, $U[-3,-1.5]$, $U[-3,-1.5]$ and $U[-3,-1.5]$, respectively. All other data-generating  parameters and MCMC settings were identical to those in Section 6.1.

\begin{sidewaystable}[ph]
    \centering
    \small
    \caption{Parameter recovery simulation results. Values are means over 100 replications, with standard errors in parentheses. The values in the parentheses of the `Acc' column are the mean absolute bias (MAB) calculated only when $\hat{K} \neq k_0$.}
    \label{tab:GRM}
    \begin{tabular}{llcccccccc}
    \toprule
    $(n,q)$ & Model & $\Delta_\alpha$ & $\Delta_{\gamma_1}$ & $\Delta_{\gamma_2}$ & $\Delta_{\gamma_3}$ & $\Delta_{\gamma_4}$ & $\Delta_B$ & $\Delta_Z$ & Acc (MAB) \\
    \midrule
    \multirow{3}{*}{(50,10)}
    & Network  & $0.461 (0.091)$ & -- & -- & -- & -- & -- & $1.053 (0.121)$ & $0.680 (1.000)$ \\
    & Node LVM & -- & $0.314 (0.073)$ & $0.242 (0.052)$ & $0.234 (0.058)$ & $0.315 (0.068)$ & $0.363 (0.079)$ & $1.170 (0.207)$ & $0.000 (2.210)$ \\
    & JLSM     & $0.430 (0.073)$ & $0.310 (0.073)$ & $0.234 (0.051)$ & $0.229 (0.056)$ & $0.302 (0.074)$ & $0.383 (0.098)$ & $1.003 (0.107)$ & $0.690 (1.000)$ \\
    \cline{2-10}
    \multirow{3}{*}{(100,20)}
    & Network  & $0.317 (0.051)$ & -- & -- & -- & -- & -- & $0.747 (0.063)$ & $1.000 (0.000)$ \\
    & Node LVM & -- & $0.284 (0.047)$ & $0.194 (0.036)$ & $0.198 (0.035)$ & $0.278 (0.042)$ & $0.380 (0.058)$ & $1.645 (0.137)$ & $0.000 (2.370)$ \\
    & JLSM     & $0.294 (0.037)$ & $0.261 (0.047)$ & $0.189 (0.033)$ & $0.189 (0.030)$ & $0.249 (0.040)$ & $0.259 (0.040)$ & $0.701 (0.056)$ & $1.000 (0.000)$ \\
    \cline{2-10}
    \multirow{3}{*}{(150,30)}
    & Network  & $0.269 (0.042)$ & -- & -- & -- & -- & -- & $0.624 (0.047)$ & $1.000 (0.000)$ \\
    & Node LVM & -- & $0.242 (0.041)$ & $0.164 (0.022)$ & $0.162 (0.024)$ & $0.245 (0.040)$ & $0.307 (0.065)$ & $1.506 (0.134)$ & $0.100 (1.689)$ \\
    & JLSM     & $0.241 (0.042)$ & $0.225 (0.035)$ & $0.156 (0.021)$ & $0.157 (0.023)$ & $0.223 (0.029)$ & $0.198 (0.023)$ & $0.573 (0.035)$ & $1.000 (0.000)$ \\
    \cline{2-10}
    \multirow{3}{*}{(300,60)}
    & Network  & $0.191 (0.034)$ & -- & -- & -- & -- & -- & $0.452 (0.034)$ & $1.000 (0.000)$ \\
    & Node LVM & -- & $0.172 (0.021)$ & $0.121 (0.012)$ & $0.122 (0.012)$ & $0.174 (0.020)$ & $0.156 (0.032)$ & $1.131 (0.099)$ & $0.850 (1.067)$ \\
    & JLSM     & $0.164 (0.016)$ & $0.169 (0.017)$ & $0.120 (0.013)$ & $0.122 (0.012)$ & $0.172 (0.017)$ & $0.133 (0.012)$ & $0.404 (0.019)$ & $1.000 (0.000)$ \\
    \bottomrule
    \end{tabular}
\end{sidewaystable}

The results of the parameter recovery study are summarized in Table~\ref{tab:GRM}. Consistent with the findings in the main text, the JLSM outperforms both the Network LSM and the Node LVM across all configurations and evaluation metrics, despite the fact that the GRM does not belong to the exponential family. Since the polytomous data contains more information than binary data but less than continuous data, the model's performance here lies between its performance with Bernoulli and Gaussian attributes.

\subsection{Mix-type node variables}\label{sec:mix}
In practice, node attributes $\mathbf{Y}$ often consist of multiple types of variables. Accordingly, we partition $\mathbf{Y}$ into $R$ blocks, that is, $\mathbf{Y} = [\mathbf{Y}_1 \mid \cdots \mid \mathbf{Y}_R]$, where each sub-block contains variables of the same type and is associated with its own link function. In this section, we focus on the setting in which node attributes consist of a mixture of Gaussian, Bernoulli, and ordinal variables. Specifically, we consider the case $R = 3$ and write $\mathbf{Y} = [\mathbf{Y}_g, \mathbf{Y}_b, \mathbf{Y}_o]$, where $\mathbf{Y}_g \in \mathbb{R}^{n \times q_g}$, $\mathbf{Y}_b \in \mathbb{R}^{n \times q_b}$, and $\mathbf{Y}_o \in \mathbb{R}^{n \times q_o}$. The corresponding Gibbs sampling algorithm is described in Section~\ref{sec:mix_gibbs}. It follows directly from the algorithms developed earlier and can be obtained through straightforward modifications.

We further conduct an additional simulation study to evaluate the performance of the JLSM under mixed-type node attributes. The sample size and attribute dimensions $(n, q_g, q_b, q_o)$ are set to $(50,5,5,5)$, $(100,10,10,10)$, $(150,15,15,15)$, and $(300,30,30,30)$. All remaining data-generating parameters and MCMC settings are identical to those used in Section 6.1 and Section~\ref{sec:GRM}.

\begin{sidewaystable}[ph]
    \centering
    \scriptsize
    \setlength{\tabcolsep}{3pt}
    \caption{Parameter recovery simulation results for different settings. Values are means over 100 replications, with standard errors in parentheses.}
    \label{tab:MIX}
    \begin{tabular}{llcccccccccccc}
    \toprule
    $(n, q_g, q_b, q_o)$ & Model & $\Delta_\alpha$ & $\Delta_{\gamma_g}$ & $\Delta_{\gamma_b}$ & $\Delta_{\gamma_{o,1}}$ & $\Delta_{\gamma_{o,2}}$ & $\Delta_{\gamma_{o,3}}$ & $\Delta_{\gamma_{o,4}}$ & $\Delta_{B_g}$ & $\Delta_{B_b}$ & $\Delta_{B_o}$ & $\Delta_{Z}$ & Acc (MAB) \\
    \midrule
    \multirow{3}{*}{(50,5,5,5)}
    & Network  & $0.471 (0.086)$ & -- & -- & -- & -- & -- & -- & -- & -- & -- & $1.033 (0.117)$ & $0.790 (1.000)$ \\
    & Node\_LV & -- & $0.170 (0.056)$ & $0.378 (0.182)$ & $0.313 (0.090)$ & $0.219 (0.077)$ & $0.225 (0.070)$ & $0.315 (0.085)$ & $0.431 (0.139)$ & $0.450 (0.136)$ & $0.454 (0.149)$ & $1.586 (0.203)$ & $0.000 (2.200)$ \\
    & JLSM     & $0.461 (0.077)$ & $0.173 (0.056)$ & $0.384 (0.179)$ & $0.303 (0.093)$ & $0.214 (0.073)$ & $0.223 (0.069)$ & $0.309 (0.079)$ & $0.187 (0.061)$ & $0.295 (0.090)$ & $0.275 (0.090)$ & $0.949 (0.110)$ & $0.800 (1.000)$ \\
    \cline{2-14}
    \multirow{3}{*}{(100,10,10,10)}
    & Network  & $0.326 (0.056)$ & -- & -- & -- & -- & -- & -- & -- & -- & -- & $0.762 (0.061)$ & $1.000 (0.000)$ \\
    & Node\_LV & -- & $0.124 (0.032)$ & $0.244 (0.067)$ & $0.269 (0.074)$ & $0.194 (0.048)$ & $0.184 (0.046)$ & $0.265 (0.061)$ & $0.364 (0.090)$ & $0.365 (0.076)$ & $0.377 (0.095)$ & $1.519 (0.161)$ & $0.010 (1.960)$ \\
    & JLSM     & $0.314 (0.052)$ & $0.125 (0.033)$ & $0.258 (0.072)$ & $0.251 (0.070)$ & $0.187 (0.048)$ & $0.181 (0.045)$ & $0.249 (0.053)$ & $0.119 (0.024)$ & $0.238 (0.041)$ & $0.201 (0.036)$ & $0.678 (0.058)$ & $1.000 (0.000)$ \\
    \cline{2-14}
    \multirow{3}{*}{(150,15,15,15)}
    & Network  & $0.257 (0.039)$ & -- & -- & -- & -- & -- & -- & -- & -- & -- & $0.612 (0.043)$ & $1.000 (0.000)$ \\
    & Node\_LV & -- & $0.099 (0.022)$ & $0.211 (0.049)$ & $0.236 (0.048)$ & $0.157 (0.032)$ & $0.166 (0.030)$ & $0.232 (0.044)$ & $0.370 (0.064)$ & $0.305 (0.062)$ & $0.284 (0.078)$ & $1.278 (0.181)$ & $0.190 (1.519)$ \\
    & JLSM     & $0.244 (0.032)$ & $0.098 (0.021)$ & $0.213 (0.050)$ & $0.223 (0.041)$ & $0.155 (0.032)$ & $0.162 (0.030)$ & $0.215 (0.039)$ & $0.099 (0.013)$ & $0.207 (0.029)$ & $0.169 (0.024)$ & $0.533 (0.033)$ & $1.000 (0.000)$ \\
    \cline{2-14}
    \multirow{3}{*}{(300,30,30,30)}
    & Network  & $0.184 (0.030)$ & -- & -- & -- & -- & -- & -- & -- & -- & -- & $0.441 (0.029)$ & $1.000 (0.000)$ \\
    & Node\_LV & -- & $0.069 (0.014)$ & $0.149 (0.025)$ & $0.172 (0.020)$ & $0.120 (0.018)$ & $0.119 (0.017)$ & $0.172 (0.026)$ & $0.338 (0.066)$ & $0.174 (0.026)$ & $0.142 (0.025)$ & $0.811 (0.108)$ & $0.910 (1.000)$ \\
    & JLSM     & $0.173 (0.023)$ & $0.070 (0.013)$ & $0.148 (0.024)$ & $0.171 (0.021)$ & $0.120 (0.017)$ & $0.118 (0.017)$ & $0.169 (0.026)$ & $0.067 (0.007)$ & $0.154 (0.016)$ & $0.121 (0.012)$ & $0.379 (0.025)$ & $1.000 (0.000)$ \\
    \bottomrule
    \end{tabular}
\end{sidewaystable}

The results of the parameter recovery study are summarized in Table~\ref{tab:MIX}. Consistent with previous findings, the JLSM consistently outperforms both the Network LSM and the Node LVM across all experimental configurations and evaluation metrics. Moreover, the parameters associated with Gaussian, Bernoulli, and ordinal variables are all accurately recovered by the JLSM, demonstrating its effectiveness in handling mixed-type node attributes.

\subsection{Sensitivity Study}\label{sec:ss}
In this section, we investigate the sensitivity of the proposed method to key hyperparameters of the COSS prior, namely the stick-breaking parameter $a$ and the slab variance parameters $(a_\theta, b_\theta)$. All other data-generating mechanisms and MCMC settings are held identical to those in Section~6.1.1 to ensure a fair and controlled comparison. As shown in Tables~\ref{tab:gaussian_results} and~\ref{tab:bernoulli_results}, the empirical performance remains stable across a range of moderate and practically reasonable hyperparameter choices. These findings suggest that the proposed method is robust to different specifications of the COSS prior. Notably, this robustness persists even when $a_\theta = 1.5$, which is samller than its theoretical lower bound, despite the theoretical requirement that $a_\theta > 2$.

\begin{table}[ht]
    \centering
    \small
    \caption{Parameter recovery results for the JLSM with Gaussian node variables. Values are means over 100 replications, with standard errors in parentheses.}
    \label{tab:gaussian_results}
    \begin{tabular}{ccccccc}
    \toprule
    $(n,q)$ & $(a, a_\theta, b_\theta)$ & $\Delta_\alpha$ & $\Delta_\gamma$ & $\Delta_B$ & $\Delta_Z$ & Acc (MAB) \\
    \midrule
    \multirow{6}{*}{(50,10)} 
    & (2.5, 3, 3)   & 0.435 (0.076) & 0.177 (0.044) & 0.221 (0.059) & 0.930 (0.101) & 0.810 (1.053) \\
    & (5, 3, 3)     & 0.450 (0.086) & 0.192 (0.052) & 0.224 (0.061) & 0.942 (0.110) & 0.760 (1.042) \\
    & (10, 3, 3)    & 0.427 (0.072) & 0.177 (0.041) & 0.224 (0.049) & 0.917 (0.107) & 0.650 (1.057) \\
    & (5, 1.5, 3)   & 0.440 (0.079) & 0.179 (0.049) & 0.213 (0.063) & 0.926 (0.108) & 0.790 (1.143) \\
    & (5, 3, 1.5)   & 0.452 (0.089) & 0.183 (0.049) & 0.224 (0.062) & 0.926 (0.123) & 0.750 (1.000) \\
    & (5, 1.5, 1.5) & 0.440 (0.076) & 0.180 (0.044) & 0.218 (0.059) & 0.938 (0.109) & 0.670 (1.000) \\
    \midrule
    \multirow{6}{*}{(100,20)} 
    & (2.5, 3, 3)   & 0.313 (0.050) & 0.127 (0.029) & 0.135 (0.021) & 0.648 (0.062) & 1.000 (0.000) \\
    & (5, 3, 3)     & 0.320 (0.057) & 0.134 (0.031) & 0.134 (0.023) & 0.661 (0.065) & 1.000 (0.000) \\
    & (10, 3, 3)    & 0.307 (0.045) & 0.129 (0.027) & 0.129 (0.018) & 0.644 (0.056) & 1.000 (0.000) \\
    & (5, 1.5, 3)   & 0.311 (0.044) & 0.128 (0.027) & 0.132 (0.021) & 0.645 (0.059) & 1.000 (0.000) \\
    & (5, 3, 1.5)   & 0.312 (0.052) & 0.128 (0.031) & 0.131 (0.019) & 0.652 (0.060) & 1.000 (0.000) \\
    & (5, 1.5, 1.5) & 0.318 (0.051) & 0.132 (0.029) & 0.133 (0.020) & 0.659 (0.062) & 1.000 (0.000) \\
    \midrule
    \multirow{6}{*}{(150,30)} 
    & (2.5, 3, 3)   & 0.252 (0.036) & 0.108 (0.021) & 0.103 (0.012) & 0.525 (0.042) & 1.000 (0.000) \\
    & (5, 3, 3)     & 0.253 (0.032) & 0.105 (0.021) & 0.100 (0.012) & 0.524 (0.039) & 1.000 (0.000) \\
    & (10, 3, 3)    & 0.250 (0.039) & 0.106 (0.024) & 0.103 (0.013) & 0.535 (0.039) & 1.000 (0.000) \\
    & (5, 1.5, 3)   & 0.254 (0.043) & 0.109 (0.023) & 0.098 (0.010) & 0.532 (0.049) & 1.000 (0.000) \\
    & (5, 3, 1.5)   & 0.251 (0.037) & 0.106 (0.021) & 0.103 (0.012) & 0.528 (0.041) & 1.000 (0.000) \\
    & (5, 1.5, 1.5) & 0.253 (0.036) & 0.106 (0.021) & 0.101 (0.012) & 0.531 (0.041) & 1.000 (0.000) \\
    \midrule
    \multirow{6}{*}{(300,60)} 
    & (2.5, 3, 3)   & 0.180 (0.026) & 0.077 (0.014) & 0.068 (0.005) & 0.376 (0.029) & 1.000 (0.000) \\
    & (5, 3, 3)     & 0.187 (0.030) & 0.080 (0.016) & 0.069 (0.005) & 0.381 (0.031) & 1.000 (0.000) \\
    & (10, 3, 3)    & 0.180 (0.025) & 0.076 (0.015) & 0.068 (0.004) & 0.377 (0.026) & 1.000 (0.000) \\
    & (5, 1.5, 3)   & 0.181 (0.028) & 0.077 (0.014) & 0.070 (0.005) & 0.374 (0.030) & 1.000 (0.000) \\
    & (5, 3, 1.5)   & 0.185 (0.034) & 0.080 (0.017) & 0.069 (0.005) & 0.382 (0.036) & 1.000 (0.000) \\
    & (5, 1.5, 1.5) & 0.185 (0.031) & 0.080 (0.016) & 0.069 (0.005) & 0.378 (0.033) & 1.000 (0.000) \\
    \bottomrule
    \end{tabular}
\end{table}

\begin{table}[ht]
    \centering
    \small
    \caption{Parameter recovery results for the JLSM with Bernoulli node variables. Values are means over 100 replications, with standard errors in parentheses.}
    \label{tab:bernoulli_results}
    \begin{tabular}{ccccccc}
    \toprule
    $(n,q)$ & $(a, a_\theta, b_\theta)$ & $\Delta_\alpha$ & $\Delta_\gamma$ & $\Delta_B$ & $\Delta_Z$ & Acc (MAB) \\
    \midrule
    \multirow{6}{*}{(50,10)} 
    & (2.5, 3, 3)   & 0.448 (0.084) & 0.481 (0.282) & 0.625 (0.177) & 1.027 (0.095) & 0.760 (1.042) \\
    & (5, 3, 3)     & 0.449 (0.080) & 0.460 (0.136) & 0.633 (0.174) & 1.041 (0.107) & 0.780 (1.045) \\
    & (10, 3, 3)    & 0.457 (0.079) & 0.458 (0.136) & 0.636 (0.173) & 1.034 (0.088) & 0.730 (1.000) \\
    & (5, 1.5, 3)   & 0.451 (0.081) & 0.526 (0.339) & 0.615 (0.163) & 1.033 (0.116) & 0.740 (1.000) \\
    & (5, 3, 1.5)   & 0.430 (0.061) & 0.450 (0.135) & 0.636 (0.181) & 1.031 (0.106) & 0.630 (1.027) \\
    & (5, 1.5, 1.5) & 0.447 (0.081) & 0.476 (0.181) & 0.631 (0.181) & 1.052 (0.108) & 0.630 (1.135) \\
    \midrule
    \multirow{6}{*}{(100,20)} 
    & (2.5, 3, 3)   & 0.308 (0.049) & 0.351 (0.182) & 0.402 (0.069) & 0.732 (0.064) & 1.000 (0.000) \\
    & (5, 3, 3)     & 0.308 (0.047) & 0.320 (0.083) & 0.400 (0.075) & 0.732 (0.062) & 1.000 (0.000) \\
    & (10, 3, 3)    & 0.318 (0.047) & 0.312 (0.089) & 0.384 (0.075) & 0.730 (0.055) & 1.000 (0.000) \\
    & (5, 1.5, 3)   & 0.313 (0.051) & 0.315 (0.091) & 0.397 (0.083) & 0.720 (0.063) & 1.000 (0.000) \\
    & (5, 3, 1.5)   & 0.314 (0.050) & 0.317 (0.070) & 0.396 (0.069) & 0.743 (0.056) & 1.000 (0.000) \\
    & (5, 1.5, 1.5) & 0.319 (0.047) & 0.316 (0.075) & 0.388 (0.082) & 0.742 (0.057) & 1.000 (0.000) \\
    \midrule
    \multirow{6}{*}{(150,30)} 
    & (2.5, 3, 3)   & 0.256 (0.043) & 0.245 (0.044) & 0.291 (0.048) & 0.606 (0.053) & 1.000 (0.000) \\
    & (5, 3, 3)     & 0.260 (0.038) & 0.252 (0.058) & 0.290 (0.043) & 0.605 (0.047) & 1.000 (0.000) \\
    & (10, 3, 3)    & 0.264 (0.047) & 0.248 (0.042) & 0.302 (0.050) & 0.609 (0.054) & 1.000 (0.000) \\
    & (5, 1.5, 3)   & 0.254 (0.039) & 0.246 (0.051) & 0.295 (0.041) & 0.596 (0.043) & 1.000 (0.000) \\
    & (5, 3, 1.5)   & 0.251 (0.032) & 0.236 (0.041) & 0.289 (0.041) & 0.601 (0.041) & 1.000 (0.000) \\
    & (5, 1.5, 1.5) & 0.255 (0.043) & 0.242 (0.041) & 0.284 (0.042) & 0.608 (0.047) & 1.000 (0.000) \\
    \midrule
    \multirow{6}{*}{(300,60)} 
    & (2.5, 3, 3)   & 0.179 (0.025) & 0.163 (0.027) & 0.183 (0.018) & 0.425 (0.026) & 1.000 (0.000) \\
    & (5, 3, 3)     & 0.181 (0.028) & 0.167 (0.023) & 0.183 (0.017) & 0.429 (0.031) & 1.000 (0.000) \\
    & (10, 3, 3)    & 0.186 (0.030) & 0.164 (0.023) & 0.184 (0.017) & 0.434 (0.029) & 1.000 (0.000) \\
    & (5, 1.5, 3)   & 0.183 (0.024) & 0.164 (0.023) & 0.188 (0.019) & 0.429 (0.023) & 1.000 (0.000) \\
    & (5, 3, 1.5)   & 0.178 (0.027) & 0.163 (0.018) & 0.187 (0.020) & 0.428 (0.027) & 1.000 (0.000) \\
    & (5, 1.5, 1.5) & 0.185 (0.027) & 0.169 (0.023) & 0.186 (0.018) & 0.431 (0.027) & 1.000 (0.000) \\
    \bottomrule
    \end{tabular}
\end{table}

\newpage

\section{Convergence Diagnostics}\label{sec:CD}
To evaluate the convergence of the proposed Gibbs sampler, we examine trace plots and Effective Sample Sizes (ESS) for key model parameters. Specifically, we analyze the posterior samples of the latent dimension $K^*$ and representative entries from the matrices $\boldsymbol\Theta^A$ and $\boldsymbol\Theta^Y$. Given the high dimensionality of these matrices, visualizing all elements is both impractical and redundant. To ensure an objective and representative assessment, we implemented a systematic selection strategy: we first identified a candidate pool of entries with posterior means within the range of $[-3, 3]$, representing the most statistically relevant effects. From this subset, a single entry was stochastically selected for visualization. For the symmetric matrix $\boldsymbol\Theta^A$, the selection was further restricted to the strictly upper triangular elements to avoid structural redundancy.

\begin{figure}[ht]
    \centering
    \includegraphics[width=0.99\linewidth]{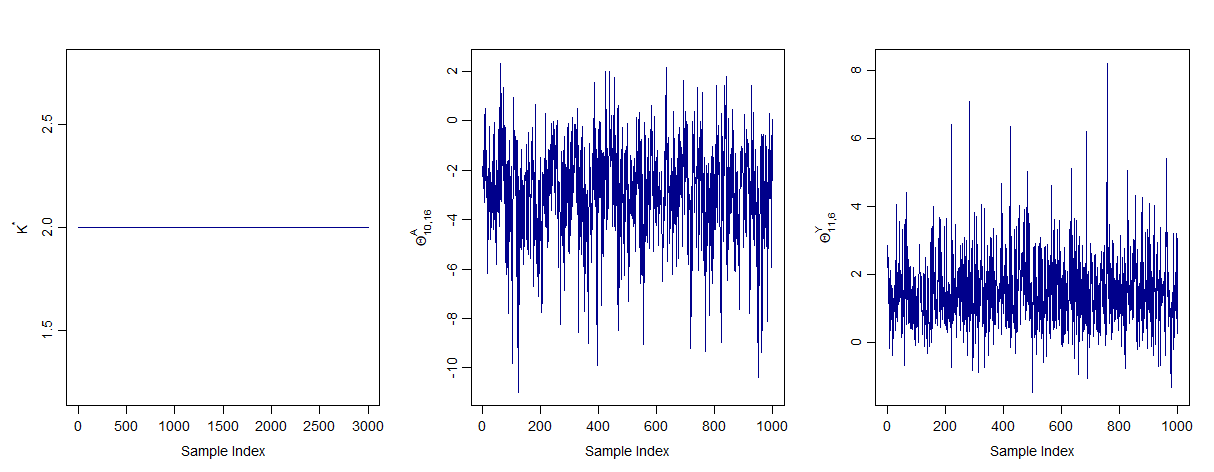}
    \caption{Trace plots of $K^*$, $\Theta^A_{10,16}$, and $\Theta^Y_{11,6}$ for the French Financial Elites dataset obtained from the Gibbs sampler, after discarding the first 10000 iterations as burn-in and retaining every 5th sample. The corresponding ESS values for $\Theta^A_{10,16}$ and $\Theta^Y_{11,6}$ are 441.7421 and 792.9565, respectively.}
    \label{fig:traceFE}
\end{figure}

\begin{figure}[ht]
    \centering
    \includegraphics[width=0.99\linewidth]{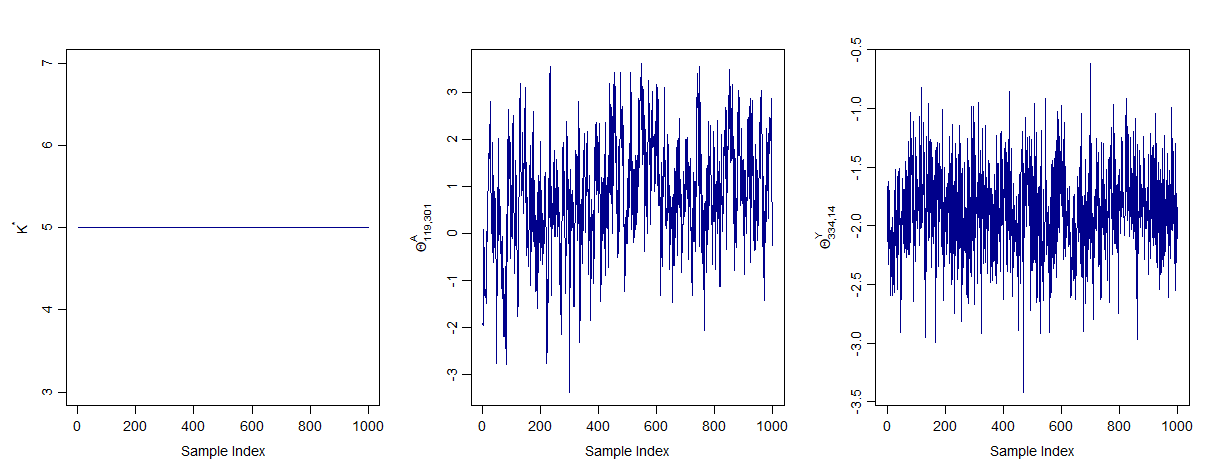}
    \caption{Trace plots of $K^*$, $\Theta^A_{119,301}$, and $\Theta^Y_{334,14}$ for Circle~1 of the Facebook ego-network dataset, obtained from the Gibbs sampler after discarding the first 10000 iterations and retaining every 5th sample. The corresponding ESS values for $\Theta^A_{119,301}$ and $\Theta^Y_{334,14}$ are 126.1373 and 382.9351, respectively.}
    \label{fig:trace1}
\end{figure}

As illustrated in Figures \ref{fig:traceFE} and \ref{fig:trace1}, the posterior trajectories for both datasets exhibit well-mixed chains with no discernible trends or computational pathologies. Furthermore, the substantial ESS values across these representative parameters indicate that the thinned samples provide a reliable and nearly independent characterization of the posterior distribution. Collectively, these diagnostics confirm that the Gibbs sampler has achieved satisfactory convergence and performed an effective exploration of the posterior space.

\section{Teenage Friends and Lifestyle Study}
\subsection{Loading Matrix}

\begin{figure}[ht]
    \centering
    \includegraphics[width=0.99\linewidth]{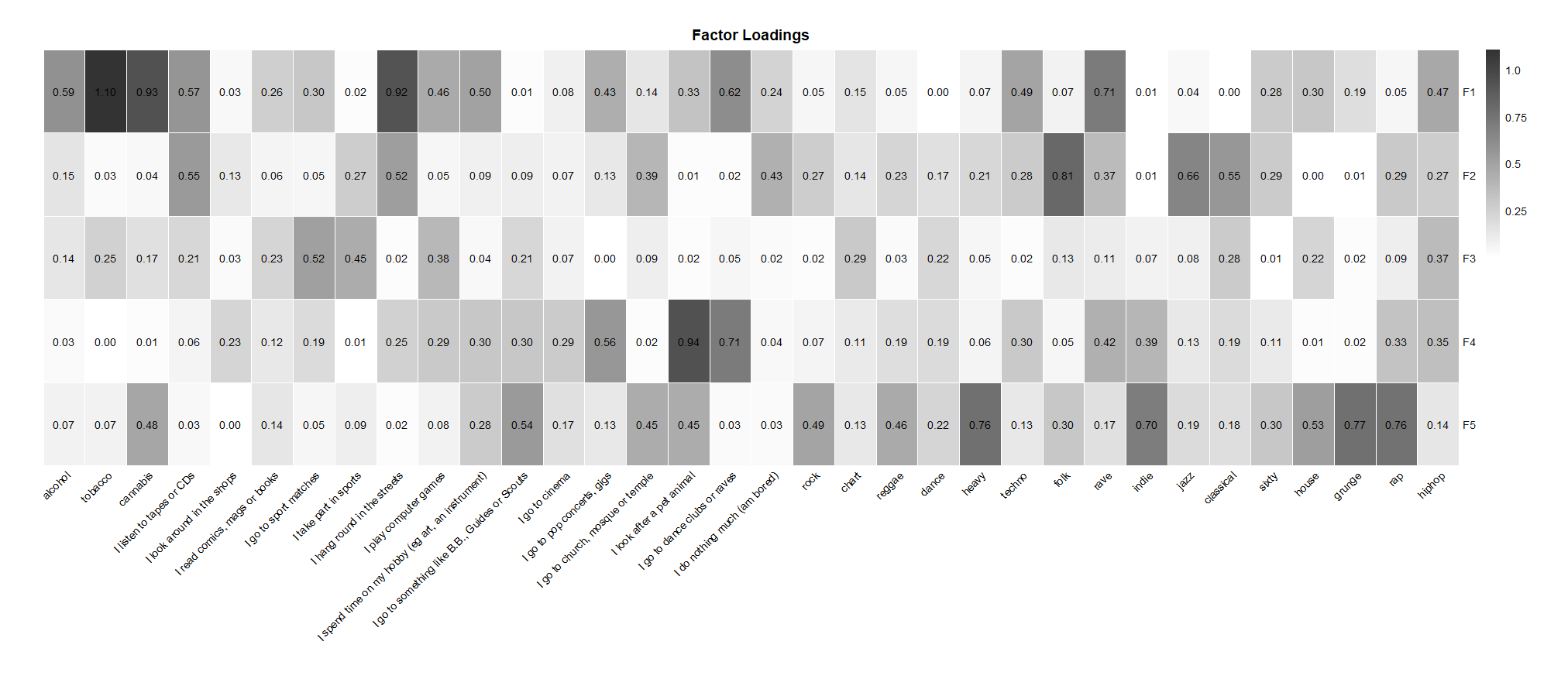}
    \caption{Heatmap of the estimated loading matrix $\mathbf{B}$ in absolute value under Geomin rotation for the Teenage Friends and Lifestyle Study dataset.}
    \label{fig:loadingdata3}
\end{figure}
In this section, we provide the detailed statistical summaries of the latent dimension selection and a comprehensive interpretation of the factor loading matrix $\hat{\mathbf{B}}$.  Figure~\ref{fig:loadingdata3} displays a heatmap of the absolute values of the estimated loading matrix following a Geomin rotation \citep{yates1987multivariate}. The rotated solution reveals a five-dimensional structure that effectively disentangles substance use from distinct leisure and musical subcultures.
\begin{itemize}
    \item \textbf{Factor 1 (Risk and Street Culture):} This dimension represents the primary axis of deviance. It shows substantial positive loadings for all substance use indicators, ``tobacco'' (1.10), ``cannabis'' (0.93), and ``alcohol'' (0.59), alongside unstructured socialization activities such as ``hanging round in the streets'' (0.92) and listening to ``rave'' music (0.71).
    \item \textbf{Factor 2 (Cultural Appreciation):} This factor captures a passive or high-culture musical profile, characterized by strong loadings for ``folk'' (0.81), ``jazz'' (0.66), and ``classical'' (0.55).
    \item \textbf{Factor 3 (Sports Activity):} The third dimension is strictly defined by physical engagement, dominated by ``going to sport matches'' (0.52) and ``taking part in sports'' (0.45).
     \item \textbf{Factor 4 (Social Engagement):} This factor reflects a mixed profile of domestic and social leisure, characterized by high loadings on ``looking after a pet'' (0.94) and ``going to dance clubs'' (0.71).
    \item \textbf{Factor 5 (Alternative Subculture):} The final dimension identifies a specific musical identity, exhibiting high loadings for ``hard'' genres including ``grunge'' (0.77), ``heavy metal'' (0.76), and ``rap'' (0.76). Notably, this factor correlates moderately with ``cannabis'' use (0.48), suggesting a subculture-specific pattern of use distinct from the general risk profile.
\end{itemize}
Collectively, $\hat{\mathbf{B}}$ demonstrates that adolescent substance use in this cohort is not random but structurally embedded within specific lifestyle clusters, particularly the ``street culture'' identified by the first latent factor.

\subsection{Social sub-groups}
\begin{figure}[ht]
\centering
\includegraphics[width=0.99\linewidth]{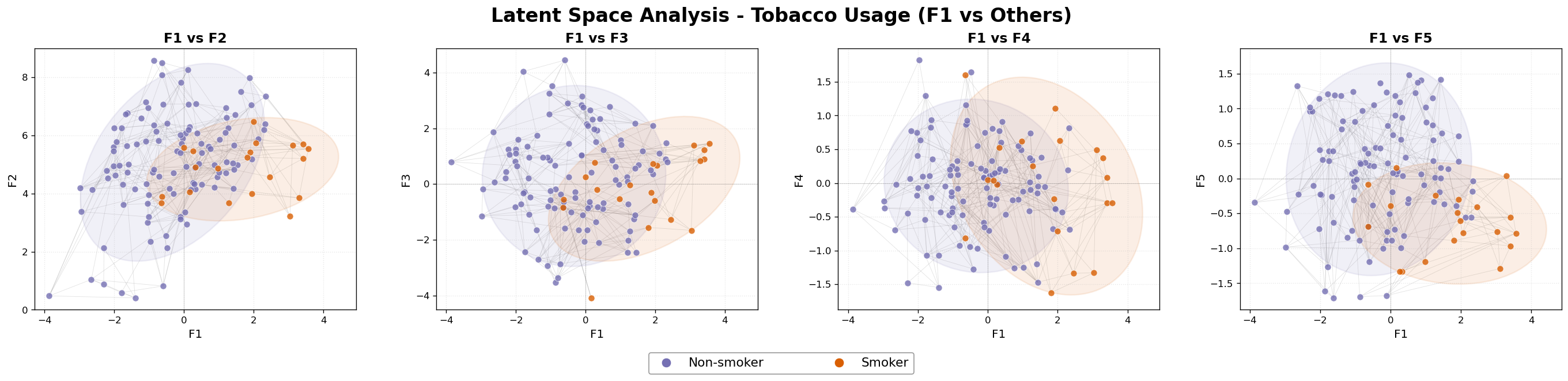}
\caption{Bivariate projections of estimated student positions in the latent space, overlaid with the friendship network (gray lines). Students colored by tobacco use.}
\label{fig:tobacco}
\end{figure}

\begin{figure}[ht]
\centering
\includegraphics[width=0.99\linewidth]{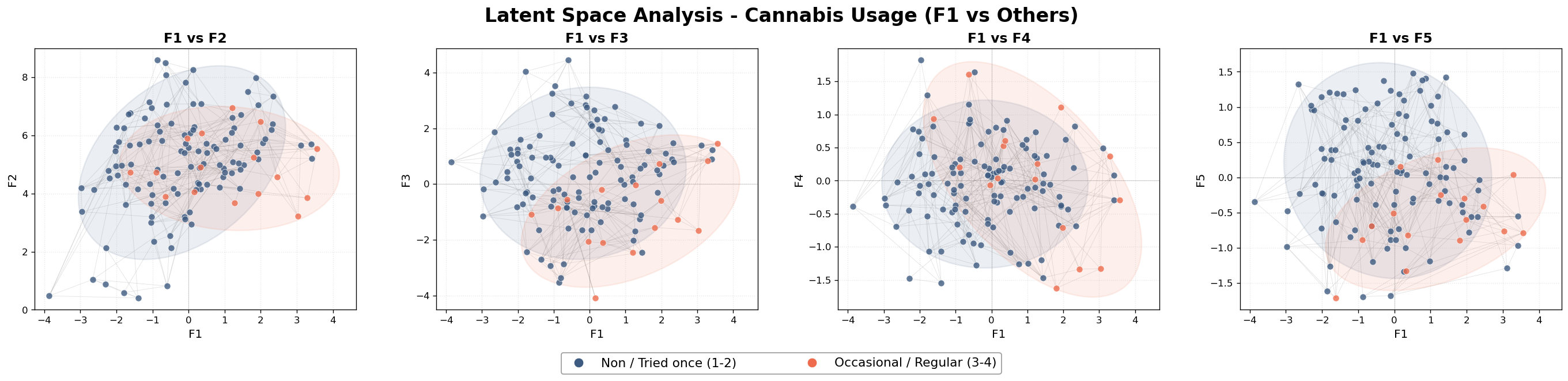}
\caption{Bivariate projections of estimated student positions in the latent space, overlaid with the friendship network (gray lines). Students colored by cannabis use.}
\label{fig:cannabis}
\end{figure}

Figures~\ref{fig:tobacco} and \ref{fig:cannabis} illustrate the latent space projections of students' tobacco and cannabis use, specifically contrasting Factor 1 against the remaining dimensions. The visualization reveals a distinct clustering pattern: students who engage in smoking or cannabis use are predominantly concentrated in regions characterized by high scores on Factor 1 (Risk and Street Culture). This observation confirms that Factor 1 effectively delineates a social subgroup defined by shared high-risk behaviors, a result that aligns with homophilous selection processes where social ties are strengthened among individuals with similar lifestyle choices.

\begin{figure}[ht]
\centering
\includegraphics[width=0.99\linewidth]{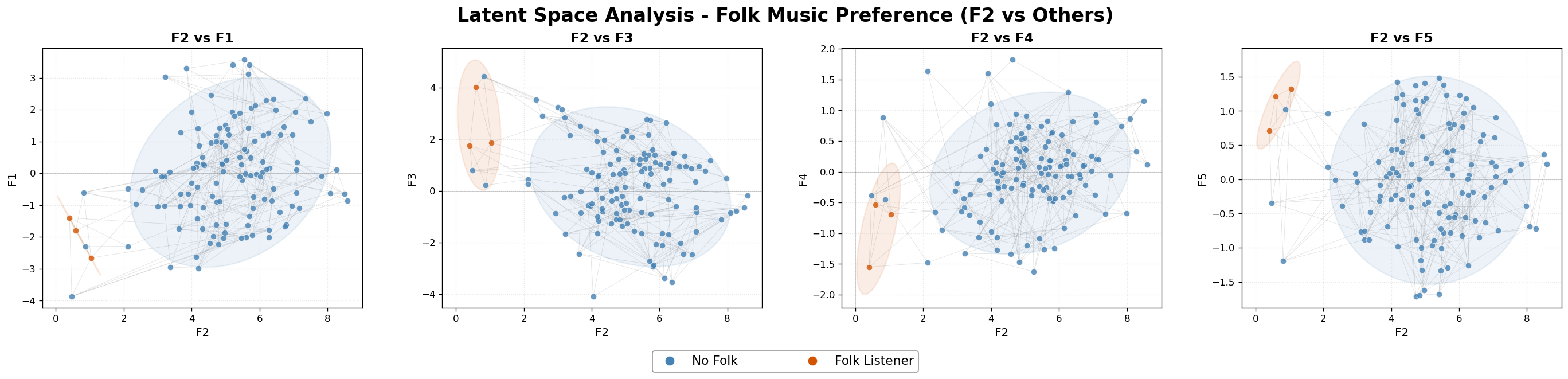}
\caption{Bivariate projections of estimated student positions in the latent space, overlaid with the friendship network (gray lines). Students colored by folk music preference.}
\label{fig:folk}
\end{figure}

\begin{figure}[ht]
\centering
\includegraphics[width=0.99\linewidth]{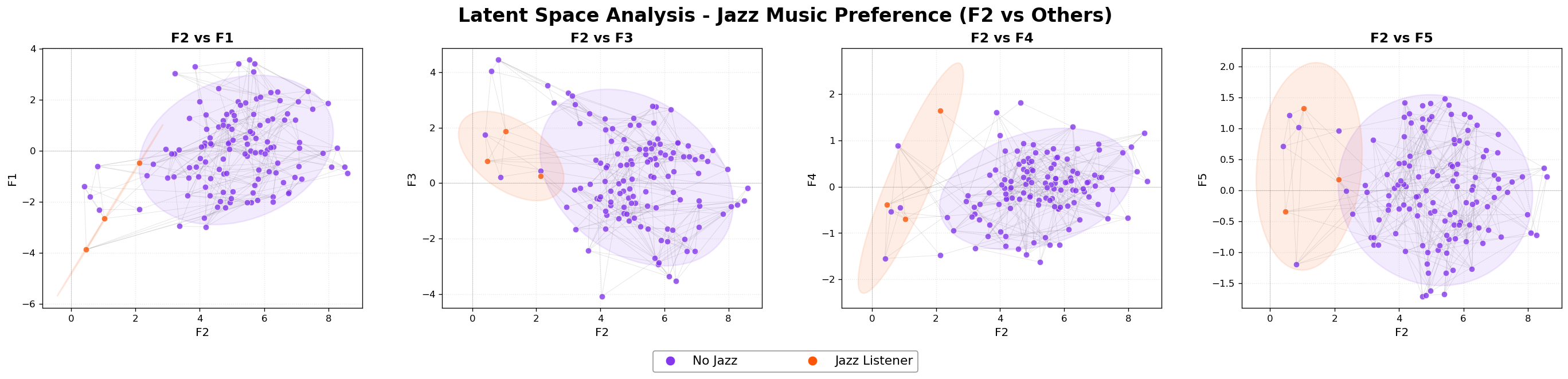}
\caption{Bivariate projections of estimated student positions in the latent space, overlaid with the friendship network (gray lines). Students colored by jazz music preference.}
\label{fig:jazz}
\end{figure}

Similarly, Figures~\ref{fig:folk} and \ref{fig:jazz} present  the latent space projections of students’ musical preferences, focusing on the relationship between Factor 2 and the remaining dimensions. The scatter plots of Factor 2 against other factors reveal a striking pattern of social stratification: a small cohort of students who enjoy folk and jazz forms a dense, isolated cluster, primarily concentrated in regions with low scores on Factor 2 (Cultural Appreciation). This pronounced separation from the majority suggests that these specific musical tastes act as a significant social identifier. Consequently, Factor 2 successfully identifies a niche subculture of students distinguished by their appreciation for folk, jazz, and related genres, further validating the model’s ability to uncover subtle social boundaries.

\clearpage

\bibliographystyle{apalike}
\bibliography{refs}